\documentclass[a4paper,11pt]{article}
\pdfoutput=1
\usepackage{jcappub}[=2021-11-22]
\keywords{cosmological parameters from LSS, galaxy clusters, power spectrum, redshift  surveys}
\arxivnumber{2106.07641}

\usepackage{microtype}

\usepackage{physics}
\usepackage{multirow}
\usepackage[dvipsnames]{xcolor}
\usepackage{tikz}
\usetikzlibrary{shapes,arrows}

\title{ShapeFit: extracting the power spectrum shape information in galaxy surveys beyond BAO and RSD}

\makeatletter
\def\thickhline{%
5  \noalign{\ifnum0=`}\fi\hrule \@height \thickarrayrulewidth \futurelet
   \reserved@a\@xthickhline}
\def\@xthickhline{\ifx\reserved@a\thickhline
               \vskip\doublerulesep
               \vskip-\thickarrayrulewidth
             \fi
      \ifnum0=`{\fi}}
\makeatother

\makeatletter
\tikzset{circle split part fill/.style  args={#1,#2}{%
 alias=tmp@name,
  postaction={%
    insert path={
     \pgfextra{%
     \pgfpointdiff{\pgfpointanchor{\pgf@node@name}{center}}%
                  {\pgfpointanchor{\pgf@node@name}{east}}%
     \pgfmathsetmacro\insiderad{\pgf@x}
      \fill[#1] (\pgf@node@name.base) ([xshift=-\pgflinewidth]\pgf@node@name.east) arc
                          (0:180:\insiderad-\pgflinewidth)--cycle;
      \fill[#2] (\pgf@node@name.base) ([xshift=\pgflinewidth]\pgf@node@name.west)  arc
                           (180:360:\insiderad-\pgflinewidth)--cycle;
         }}}}}
 \makeatother

\newcommand{\colored}[1]{\textcolor{black}{#1}}
\newcommand{\cred}[1]{\textcolor{black}{#1}}

\newcommand{\underlying}{}
\newcommand{\reference}{^\mathrm{ref}}
\newcommand{\om}{\omega_\mathrm{m}}
\newcommand{\ob}{\omega_\mathrm{b}}
\newcommand{\ocdm}{\omega_\mathrm{cdm}}
\newcommand{\ocb}{\omega_\mathrm{cb}}
\newcommand{\og}{\omega_\gamma}
\newcommand{\Om}{\Omega_\mathrm{m}}
\newcommand{\Ob}{\Omega_\mathrm{b}}
\newcommand{\Ocdm}{\Omega_\mathrm{cdm}}
\newcommand{\Omr}{\Omega_\mathrm{r}}
\newcommand{\Og}{\Omega_\gamma}
\newcommand{\Ol}{\Omega_\Lambda}
\newcommand{\rd}{r_\mathrm{d}}

\newcommand{\Mpc}{\mathrm{Mpc}}
\newcommand{\mpcoh}{\,h^{-1}\,{\rm Mpc}}

\author[a,b]{Samuel Brieden}

\author[a]{H\'ector Gil-Mar\'in}

\author[a,c]{Licia Verde}

\emailAdd{sbrieden@icc.ub.edu}\emailAdd{hectorgil@icc.ub.edu}\emailAdd{liciaverde@icc.ub.edu}

\affiliation[a]{ICC, University of Barcelona,\\
 IEEC-UB, Mart\'i i Franqu\`es, 1, E-08028 Barcelona, Spain}
\affiliation[b]{Dept.~de  F\'isica Qu\`antica i Astrof\'isica, Universitat de Barcelona,\\
 Mart\'i  i Franqu\`es 1, E-08028 Barcelona, Spain}
\affiliation[c]{ICREA,\\
 Pg. Llu\'is Companys 23, Barcelona, E-08010, Spain}

\abstract{In the standard (classic) approach, galaxy clustering measurements from spectroscopic surveys are compressed into baryon acoustic oscillations  and redshift space distortions measurements, which in turn can be compared to cosmological models. Recent works have shown that avoiding this intermediate step and fitting directly the full power spectrum signal (full modelling)  leads to much tighter constraints on cosmological parameters. Here we show where this extra information is coming from and extend the classic approach  with one additional effective parameter,  such that it captures, effectively,  the same amount of information as the full modelling approach, but in a model-independent way. We validate this new method (\emph{ShapeFit}) on mock catalogs, and compare its performance to the full modelling approach finding both to deliver equivalent results. The \emph{ShapeFit} extension of the classic approach promotes the standard analyses at the level of full modelling ones in terms of information content, with the advantages of \emph{i)} being more model independent; \emph{ii)} offering an  understanding of  the origin of the extra cosmological information; \emph{iii)} allowing a robust control on  the impact of  observational~systematics.}
\providecommand\inspire[1]{\href{https://inspirehep.net/search?p=find+#1}{{\tiny IN}{\footnotesize SPIRE}}}

\providecommand\erratum[4][ibid.\ ]{\emph{Erratum #1}{\bf #2} (#3) #4}

\providecommand{\jhep}[3] {\ifnum#2>2009%
\href{https://doi.org/10.1007/JHEP#1(#2)#3}{\emph{JHEP} {\bf #1} (#2) #3}%
\else%
\href{https://doi.org/10.1088/1126-6708/#2/#1/#3}{\emph{JHEP} {\bf #1} (#2) #3}%
\fi}
\providecommand{\jcap}[3] {\href{https://doi.org/10.1088/1475-7516/#2/#1/#3}{\emph{JCAP} {\bf #1} (#2) #3}} 
\def\issueFromCounter.#1#2#3#4#5#6.{#2#3}
\providecommand{\jstat}[2]{\PackageWarningNoLine{\jname}{The macro \protect\jstat\space is obsolete!\MessageBreak Please typeset JSTAT as any other journal}%
  \href{https://doi.org/10.1088/1742-5468/#1/\issueFromCounter.#2./#2}{\emph{J.\ Stat.\ Mech.\ }(#1) #2}} 

\providecommand{\astroph}[1]{\href{https://arxiv.org/abs/astro-ph/#1}{\tt astro-ph/#1}}

\providecommand{\arXivid}[1]{\href{https://arxiv.org/abs/#1}{\tt arXiv:#1}}
\providecommand{\Math}[2]{%
\if!#1!%
\href{https://arxiv.org/abs/math/#2}{\tt math/#2}%
\else%
\href{https://arxiv.org/abs/math.#1/#2}{\tt math.#1/#2}%
\fi}

\begin{document}
\maketitle\flushbottom

\section{Introduction}

Observations of the Cosmic Microwave Background (CMB, e.g., \citep{2013ApJS..208...19H,Aghanim-ml-2018eyx}) radiation have been pivotal in establishing the standard cosmological model (the so-called $\Lambda$CDM model) and to open the doors to precision cosmology. The CMB has, however, the fundamental limitation of originating from a 2D surface at a given cosmic epoch. Observations of the Large Scale Structure (LSS) over large 3D volumes can yield a dramatic increase in the number of  accessible modes  and  trace the evolution of clustering across cosmic times.

The three-dimensional clustering of galaxies is rapidly becoming one of the most promising  avenue to study cosmology from the late-time Universe.  
Spectroscopic galaxy redshift surveys have witnessed a spectacular success covering increasing larger volumes:  2-degree Field (2dF, \cite{2016MNRAS.462.4240B}), Sloan Digital Sky Survey II (SDSS-II, \citep{adsabs-ml-2000AJ....120.1579Y}), SDSS-III Baryon Oscillation spectroscopic Survey (BOSS, \citep{adsabs-ml-2011AJ....142...72E,2013AJ....145...10D,alamdr12}), SDSS-IV  extended BOSS (eBOSS, \citep{ebossELG_catalogue,ebossLRG_catalogue,ebossQSO_catalogue}); and this trend is set to continue: the on-going Dark Energy Spectroscopic Instrument (DESI, \cite{aghamousa_desi_2016,aghamousa_desi_2016-1}) and up-coming Euclid satellite mission \cite{ laureijs_euclid_2011}, as just two examples.   

Baryon Acoustic Oscillations (BAO) is an imprint in the power spectrum of sound waves in the pre-recombination Universe offering a ``standard ruler'' observable \cred{through} galaxy clustering \cite{2005ApJ...633..560E, cole05, PeeblesYu1970, BondEfstathiou1984, Holtzmann1989, HuSugiyama1996, EH_TransferFunction,2014PhRvL.113x1302H}. 
 The standard  approach to \cred{analyse} galaxy redshift clustering, used extensively and part of official surveys' pipelines,  has used the standard ruler signature in the galaxy power spectrum to obtain determinations of the distance-redshift relation at the effective redshift of the surveys' samples exploiting the Alcock-Paczynski effect \cite{1979Natur.281..358A}. The process of density-field reconstruction, e.g., \cite{Eis2007}, is widely adopted to reduce the information loss induced by non-linearities. In this approach, the geometric information extracted from the BAO peak position is largely model-independent: the physical quantity constrained is directly related to the expansion history and independent of the parametrization of the expansion history given by specific cosmological models. The reconstruction step induces some model-dependence but this has been shown to be very weak \cite{Carter_BAOtest}. 
Redshift Space Distortions (RSD, pioneered by \cite{kaiser_clustering_1987}) arise from the non-linear relation between cosmological distances-- natural input to the theory \cred{modelling}-- and the (observed) redshifts. They enclose information about the velocity field and have been used to extract constraints on the amplitude of velocity fluctuations times the dark matter amplitude fluctuations, characterized by the parameter combination $f\sigma_8$.  

BAO and RSD results and their cosmological interpretation for state-of-the-art surveys have been presented e.g., in \cite{alam_clustering_2017}, for the SDSS-III BOSS survey, and in \cite{eboss_collaboration_dr16}, for the SDSS-IV eBOSS survey, and the success of this approach is behind much of the science case for forthcoming surveys.
 \colored{From now on we refer to this, now standard, approach as ``classic"\footnote{Classic in the Merriam Webster dictionary: serving as a standard of excellence, of recognized value. We use here the word classic as  ``of high quality standard in its respective genre based on judgement over a period of time" and "can be considered as standard".}}
The classic approach is conceptually different from the way, for example, CMB data are interpreted and from the analysis of LSS data pre-BAO \colored{era} \colored{(see e.g., \cite{2dFGRS-ml-2001csf,SDSS-ml-2003tbn,Efstathiou-ml-2000fk,2003ApJS..148..195V}}\cred{)}. \colored{When the BAO detection in galaxy redshift surveys became of high enough signal to noise, it was quickly recognized that it carried most of the interesting signal, see e.g., \cite{2010JCAP...07..022H,Song-ml-2008qt}, and the community then adopted the, now classic, BAO and RSD approach.}
BAO and RSD analyses, with the help of a template of the power spectrum,  compress the power spectrum data into few physical observables which are sensitive only to late-time physics, and it is these observables that are then interpreted in light of a cosmological model. CMB data analyses, on the other hand, compare directly the measured  power spectrum to the model prediction, requiring the choice of a cosmological model to be done {\it ab initio}.
Recently, the development of \colored{high performance codes \cred{based on the} \textsc{FFTLog} algorithm \citep{fftlog_paper} giving rise to fast model evaluations of, for instance,} the so-called ``Effective Field Theory of Large Scale Structure",  e.g., \cite{2012JHEP...09..082C,2020PhRvD.102l3541N,2020PhRvD.102f3533C,2020JCAP...06..001C, 2020JCAP...05..042I,2020JCAP...05..032P,2018PhRvD..97f3526L,DAmico-ml-2019fhj},  has prompted part of the community to \cred{analyse} galaxy redshift clustering in a similar way as CMB \colored{and pre-BAO era LSS} data \colored{(see e.g., \cite{Efstathiou-ml-2000fk,2dFGRS-ml-2001csf,SDSS-ml-2003tbn} and references therein)}, by comparing directly the observed power spectrum, including the BAO signal, the RSD signal, as well as the full shape of the broadband power to the model's prediction. A full Markov Chain Monte Carlo (MCMC) exploration of the cosmological parameter space can then be performed obtaining cosmological constraints significantly tighter than in the standard analysis. For example \colored{\cite{2020JCAP...05..042I, 2020JCAP...05..032P}}, imposing a Big Bang Nucleosynthesis (BBN) prior, obtain a 1.6\%  constraint on the Hubble constant, which is instead very mildly \colored{($\sim 10\%$)} constrained in the standard approach (see e.g., \colored{red contours in} figure 5 of \cite{eboss_collaboration_dr16}). \colored{In what follows, we refer to this approach as ``full \cred{modelling}", to highlight the fact that  while currently Effective Field Theory of Large Scale Structure is the theoretical \cred{modelling} of choice for this approach,   other choices are also possible.}

The additional constraining power afforded by the second approach must arise, at least in part, from the broadband shape of the power spectrum, but a full physical interpretation of the origin \colored{of the extra constraining power} is still lacking (but see e.g.,~\cite{Philcox_rsmeas} and refs. therein). 

This paper  serves three main objectives:
1) identify clearly where the additional information comes from and what physical processes it corresponds to, 2) bridge the classic and new analyses in a transparent way and 3) extend the classic analysis in a simple and effective manner to capture the bulk of this extra information.  In passing, we also present a new  definition and interpretation of the physical parameter describing amplitude of velocity fluctuations which further reduce the model-dependence of the traditional RSD analysis.
\colored{ We stress here that the theoretical models for the power spectra adopted by the published works of the ``classic" and ``full \cred{modelling}" approaches are different. The main motivation of this paper is not to do a first principles comparison including all combinations of theoretical models and fitting methodologies. In this paper we stick to the theoretical models and fitting methodologies as adopted in the literature, using as much as possible codes made publicly available by the authors of the relevant papers. When extending the classic approach we will take care in introducing as minimal modifications as possible.   }
The rest of the paper is organized as follows: in section~\ref{sec:theory} we review the known approaches for the cosmological interpretation of galaxy clustering. While this is background material it serves the purpose of highlighting differences and similarities across approaches and make explicit their dependence on (or independence of) assumptions about a cosmological model. Section~\ref{sec:shapefit} introduces the phenomenological extension of the classic approach, which we call {\it ShapeFit}, an executive summary of it in the form of a flowchart is presented in figure~\ref{fig:cheatsheet}, and section~\ref{sec:Practical} presents our setup for its application to mock catalogs. In section \ref{sec:results} we show a direct comparison between the different analysis approaches and perform additional systematic tests of the proposed {\it ShapeFit} in section \ref{sec:additional_tests}.
The conclusions are presented in section~\ref{sec:conclusions}. The appendices present technical details and relevant systematic tests.

\section{Theoretical background}
\label{sec:theory}
In this section we provide an overview of the most common LSS analysis strategies to date. To understand how to compare them directly with each other, and how to interpret the resulting parameter constraints as done in section~\ref{sec:theory-extract-info}, it is important to spell out clearly what physical processes, what observational features and what model ingredients are relevant for each of the approaches. This background section serves for this purpose.  

\subsection{The $\Lambda$CDM model: \cred{notation} and definitions} \label{sec:theory-lcdm-model}

If not stated otherwise, we work in the flat $\Lambda$CDM model with the following parameter basis
\begin{align} \label{eq:parambase}
    \mathbf{\Omega} = \left\lbrace \Omega_i \right\rbrace = \left\lbrace \colored{\ocdm}, \ob, h, \sigma_8, n_s, M_\nu \right\rbrace ~,
\end{align}
where $\colored{\ocdm}$ and $\ob$ are the physical density parameters of the cold \cred{dark} matter and baryons respectively. In addition, we use the subscript `cb' to refer to the cold \cred{dark matter} + baryon component, m for the total matter including non-relativistic neutrinos, and $\nu$ for neutrinos. When the dimensionless  Hubble-\cred{Lemaître} parameter, $h$,  is introduced $H_0=h\times 100$ km s$^{-1}$ Mpc$^{-1}$, they are related to the energy density fractions $\Omega_X$ for any species $X \in \lbrace \mathrm{b},\mathrm{cdm},\mathrm{m},\mathrm{r},\gamma,\nu,\Lambda, \dots \rbrace$ for baryons, cold dark matter, matter, relativistic species, photons, \cred{neutrinos, cosmological constant, etc., as} 
\begin{align}
    \omega_X = \Omega_X h^2.
\end{align}
Within the flat $\Lambda$CDM model these quantities fulfill the budget equation 
\begin{align}
    \underbrace{\Og + \Omega_{\nu,r}}_{\Omr} + \underbrace{\Ocdm + \Ob + \Omega_{\nu,m}}_{\Om} +  \Ol = 1
\end{align}
at all times. The physical photon densiy $\og$ is effectively fixed by the precise COBE measurement of the CMB temperature $T_0 = 2.7255 \pm 0.0006$ \cite{Fixsenetal-ml-1996}, via
\begin{align}
    \og = \frac{8\pi^3T_0^4}{45\left(H_0/h\right)^2M_P^2} = (2.472\pm0.002) \times 10^{-5}~,
\end{align}
where $M_P$ is the Planck mass in natural units. This measurement of $\omega_\gamma$ is commonly used as a prior, and its central value is implicitly  adopted within the term ``flat $\Lambda$CDM". In the following we stick to this convention, although one could in principle allow $T_0$ to be a free parameter \cite{Ivanov_Alihamoud_Lesgourgues-ml-2020}.

We  also include the sum of the neutrino masses $M_\nu$ as a free parameter, where we choose 2 massless states (counted as radiation) and 1 massive state (counted as matter).  

On the background level, assuming homogeneity and isotropy on large scales, the geometry of the universe is fully described by the Hubble expansion rate as function of redshift, $z$, 
\begin{align}\label{eq:Hubble}
    H(z) = H_0 \sqrt{(1+z)^4\Omr + (1+z)^3\Om + \Ol}~,
\end{align}
where the Hubble distance $D_H$ and the comoving angular distance $D_M$ are given as
\begin{align} \label{eq:dist}
    D_H(z) = \frac{c}{H(z)}~, \qquad D_M(z) = \int_0^z \,  \frac{c dz^\prime}{H(z^\prime)}.
\end{align}

Linear perturbations in the energy density of a given species $X$ from the homogeneous background are encoded in the power spectrum $P_X(k)$ describing the \cred{2-point} statistics as a function of wavevector $k$ in Fourier space. It is written as the product of the primordial power spectrum $\mathcal{P}_\mathcal{R}(k)$ and the squared transfer function $\delta_X(k)$ obtained from solving the perturbed coupled Boltzmann equations for each species $X$,
\begin{align} \label{eq:Theory-PX}
    P_X(k,z) = \frac{2\pi^2}{k^3} \mathcal{P}_\mathcal{R}(k) \delta_X^2(k,z) = \frac{2\pi^2}{k^3}  A_s \left(\frac{k}{0.05 \, \mathrm{Mpc}^{-1}} \right) ^{n_s-1} \delta_X^2(k,z)~,
\end{align}
where the standard inflationary model assumes the primordial power spectrum to be nearly scale-invariant, with global amplitude $A_s$ and scalar tilt $n_s$.

Since the global amplitude is modulated by the transfer function as well, it is common in LSS analyses to replace $A_s$ by the total amplitude $\sigma_8$ at redshift \cred{zero} as a free parameter. In general the redshift-dependent $\sigma_8$ is defined as the matter density fluctuation at a given redshift $z$ smoothed on spheres of 8 $\mathrm{Mpc}/h$,
\begin{equation} \label{eq:Theory_sigma8}
\begin{aligned} 
    \sigma_8^2(z) \equiv \int_0^\infty \!\mathrm{d}(\ln k) \, k^3 P_\mathrm{m}(k,z) W_\mathrm{TH}^2(k \cdot 8h^{-1}\Mpc) ~,
\end{aligned}
\end{equation}
 where $W_\mathrm{TH}$ is the spherical top hat filter.
 The ``amplitude parameters", $A_s$ and $\sigma_8$, are defined on different scales and at different epochs; in particular, while $A_s$ is a primordial quantity with direct interpretation in terms of early-time physics, $\sigma_8$  is a late-time quantity, with more direct interpretation from observations of LSS clustering. \colored{Also note that the primordial amplitude $A_s$ is defined with respect to a certain pivot scale given in $\mathrm{Mpc}$ units, while the scale of interest for $\sigma_8$ has $h^{-1}\mathrm{Mpc}$ units. Therefore, the wave-number $k$ in eq.~\eqref{eq:Theory-PX} is given in $\mathrm{Mpc}^{-1}$ units, while in eq.~\eqref{eq:Theory_sigma8} and in what follows we write $k$ in units $h\mathrm{Mpc}^{-1}$.}

 The reference cosmology $\mathbf{\Omega}^\mathrm{ref}$ used throughout this work, if not stated otherwise, is given in table~\ref{tab:cosmoparams} for the parameter base introduced in eq.~\eqref{eq:parambase} and other derived parameters. We refer to this set of parameters  as ``Planck".
 
 \begin{table}[h]
    \centering
    \begin{tabular}{c||c|c|c|c|c|c|c||c|c}
        Cosmology & $\colored{\ocdm}$ & $\ob$ & $h$ & $\sigma_8$ & $n_s$ & $M_\nu\,[\mathrm{eV}]$ & $\Om$ & $r_{\rm d} \,[\mathrm{Mpc}]$ \\ \hline
       Planck & \colored{0.1190} & 0.022  & 0.676 & 0.8288 & 0.9611 & 0.06 & 0.31  & 147.78\\
    \end{tabular}
    \caption{Reference values of cosmological parameters for the $\Lambda$CDM base of eq.~\eqref{eq:parambase} and its derived parameters, the matter density parameter, $\Om$, and the sound horizon at radiation drag, $\rd$. \colored{As it is customary, we report $\rd$ in units Mpc, although we use $h^{-1}{\rm Mpc}$ units throughout the rest of this work. The reported} parameter values are close to the Planck best-fit cosmology \cite{Aghanim-ml-2018eyx}.}
    \label{tab:cosmoparams}
\end{table}

\subsection{Parameter dependence of the (real space) linear matter power spectrum} \label{sec:theory-param-dependence}
The main quantity needed to model the observable galaxy power spectrum multipoles is the real-space, linear matter power spectrum, $P_m(k)$.
\cred{Here,} we illustrate its dependence on key cosmological and physical parameters.  
Real world effects such as galaxy bias are discussed in section~\ref{sec:theory-galaxies-redshift-space}.
As eq.~\eqref{eq:Theory-PX} indicates, the primordial power spectrum is assumed to be a power law with an amplitude $A_s$  and a spectral slope $n_s$. These quantities are set by the mechanism that generated the initial conditions, but not by the subsequent evolution of the Universe. The late-time linear matter power spectrum  is not a power law; this is  encoded by the transfer function, which captures linear physics relevant after the end of inflation. \cred{As such,} it depends on the content of the Universe, and its early-time expansion history.

The primordial power spectrum is always defined for $k$ in units of Mpc$^{-1}$. The LSS power spectrum on the other hand is usually defined for $k$ in units of $h$Mpc$^{-1}$. This is intimately related to the fact that observations measure angles and redshifts and not distances directly. \cred{Hence,} distances are obtained assuming a specific theoretical model, with specific parameters values, as \cred{given by} the reference model. The model-dependence of this step can be  made more transparent  by defining distances in units of a theoretical quantity (or ``ruler")  and then making explicit the scaling of the power spectrum and the wave vector with the theory ruler. The use of ``little $h$" is the classic example, with $h=H_0/H_0^{\rm ref}$ where the reference model has  $H_0^{\rm ref}=100$ km s$^{-1}$ Mpc$^{-1}$. Moreover, as it will become clear below, it is useful to go beyond $h$ and also consider scaling of distances  with respect to the BAO standard ruler, the sound horizon at radiation drag, $r_d$, yielding $s\equiv r_{\rm d}/r_{\rm d}^{\rm ref}$;  $s$ enters the normalization of the power spectrum and the scaling of the wave vector in much the same way as $h$.  

\begin{figure}[t]
    \centering
    \includegraphics[width=0.83\textwidth]{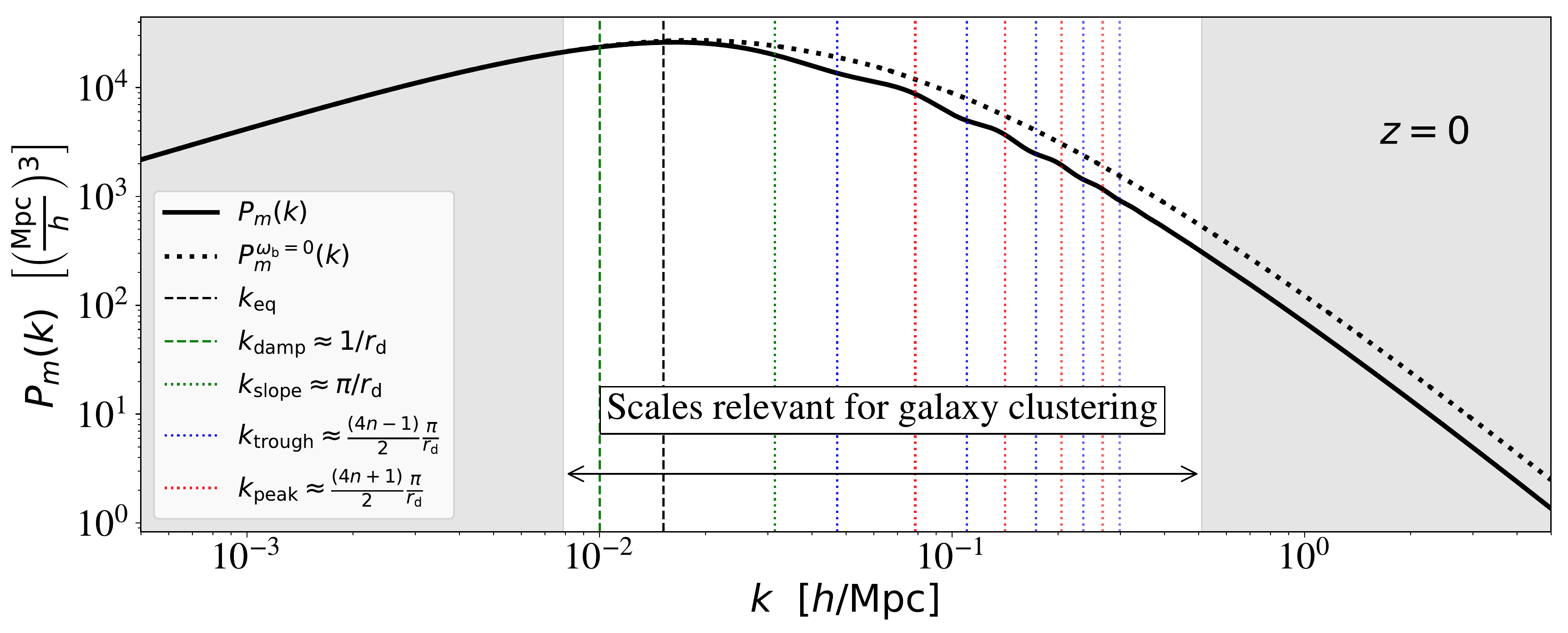} \\
    \includegraphics[width=0.83\textwidth]{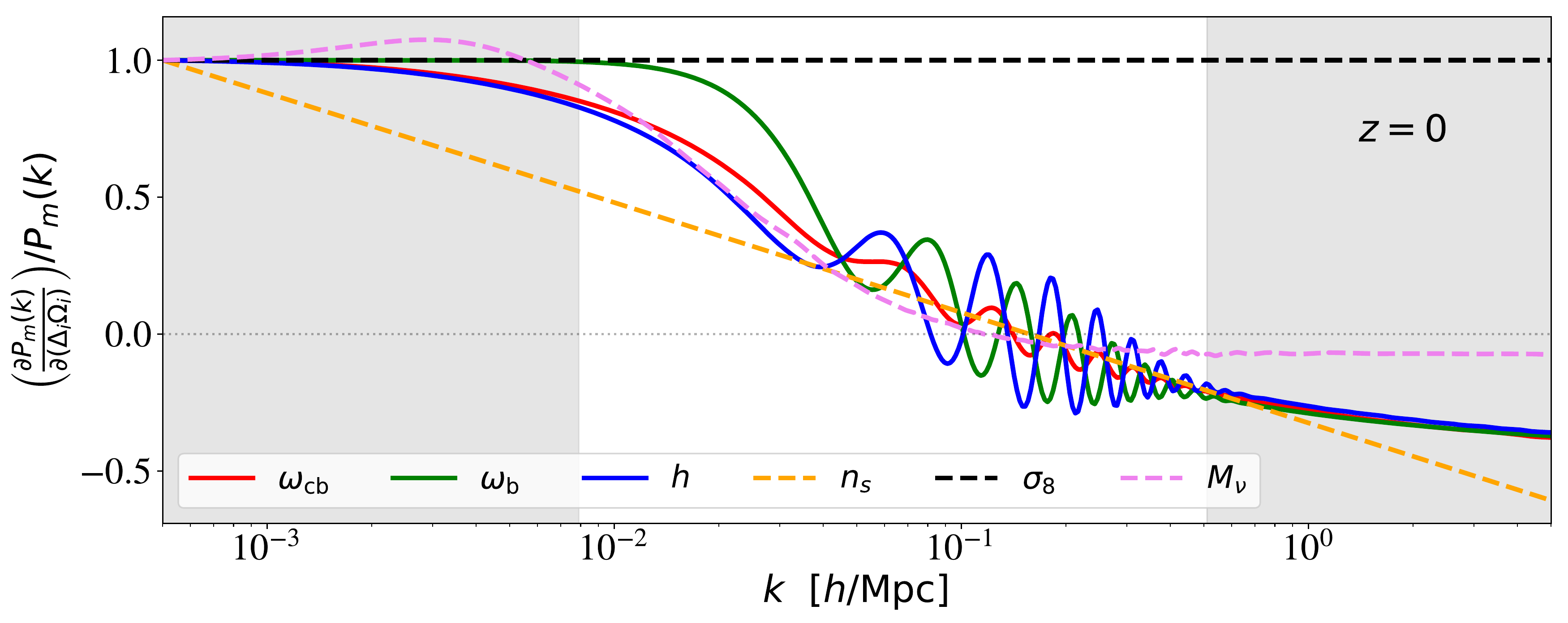} \\
    \includegraphics[width=0.83\textwidth]{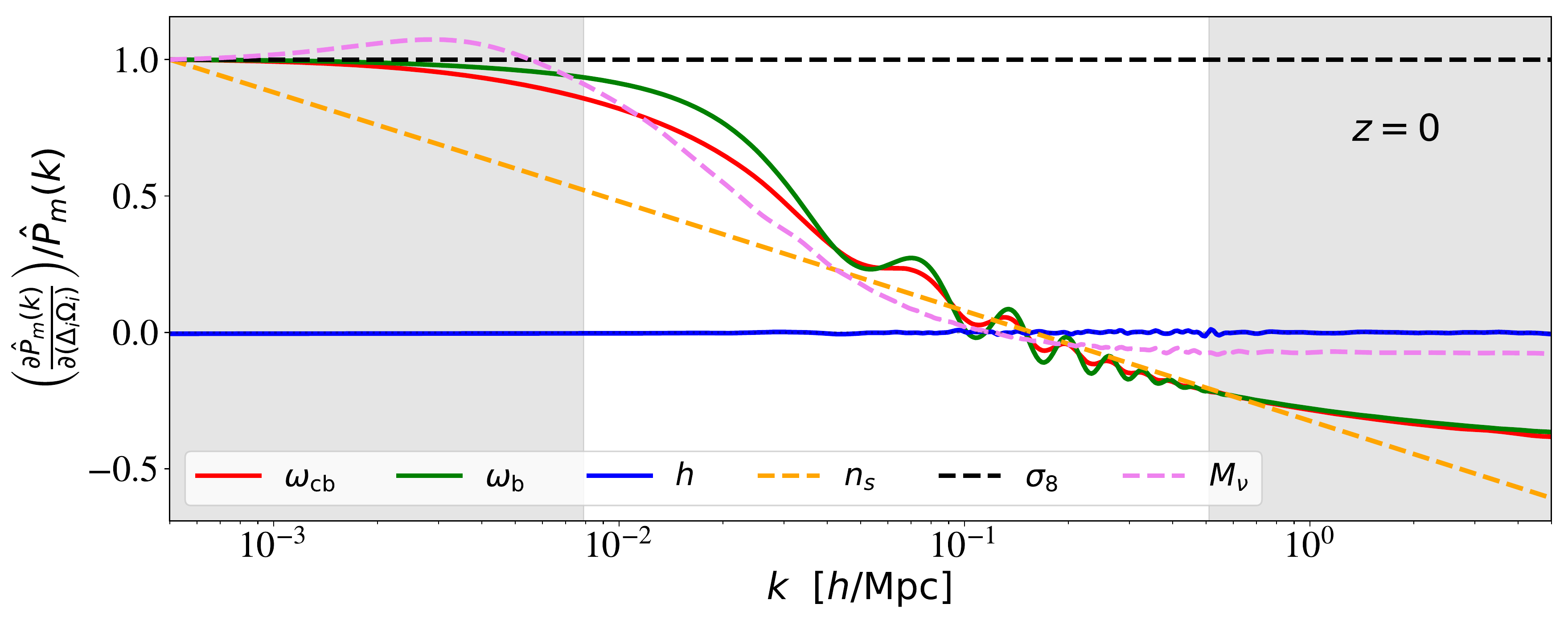}
    \caption{{\it Top panel:} Linear matter \colored{(cold dark matter + baryons+ neutrinos)} power spectrum $P_m(k)$ \colored{at $z=0$} and its characteristic scales: the equality between matter and radiation $k_\mathrm{eq}$ (black dashed  vertical line), the turn-around, which coincides with $k_\mathrm{eq}$, the  BAO-peaks  and -troughs (red and blue vertical dotted lines, respectively) estimated as function of $\rd$, and the scale where the baryon suppression reaches its maximal slope, $k_\mathrm{slope}=\pi/\rd$ (green dotted vertical line). To highlight the baryon suppression effect, the black dotted line shows $P_m(k)$ in the zero-baryon case for comparison. {\it Middle panel:}  \colored{$z=0$} $P_m(k)$ parameter dependence for varying $\Omega_i \in \left\lbrace \colored{\ocb}, \ob, h, n_s, \sigma_8, M_\nu \right\rbrace$. The normalization factors $\Delta_i$ are chosen such that \colored{all parameters} have the same impact on the power spectrum as $\sigma_8$ in the large scale limit.
    {\it Bottom panel:} same lines as in the middle panel, but  after rescaling by $\rd$ according to eq.~\eqref{eq:Theory-rsrescaled}, so that the BAO wiggle positions overlap. In all panels, the non-shaded $k$-range highlights the usual observed range for spectroscopic galaxy surveys, $0.008\leq k\,[h{\rm Mpc}^{-1}]\leq0.5$. \colored{This figure should remind the reader of the landmark  works \cite{BBKS, Sugiyama95, EH_TransferFunction} and references therein.}}
    \label{fig:pk-paramderivatives}
\end{figure}

In the top panel of figure~\ref{fig:pk-paramderivatives} we show the matter power spectrum for the ``Planck" cosmology (solid black line), and the one corresponding to a universe without baryons, where all the matter consists of dark matter (dotted black line). The latter can be described by only one characteristic scale, the scale of matter-radiation equality,
\begin{align}
    k_\mathrm{eq} = a_\mathrm{eq} H_\mathrm{eq} = 7.46 \times10^{-2}\, \frac{\colored{\omega_\mathrm{cb}}}{h} \left(\frac{T_0}{2.7\,\mathrm{K}}\right)^{-2}~,
\end{align}
which corresponds to the modes entering the Hubble horizon at the redshift of equality, $z_\mathrm{eq}$, \colored{which depends on the physical density $\ocb=\ocdm+\ob$ comprising cold dark matter and baryons. The numerical calculation of the matter power spectrum is carried out with the Boltzmann code CLASS \cite{2011JCAP...07..034B}, which uses $\ocdm$ as input parameter. However, for some applications it is more instructive to show the parameter $\ocb$.}

The effect of baryons on the matter power spectrum is characterized by an additional scale, the sound horizon at baryon radiation drag epoch,
\begin{equation} \label{eq:Theory_bao_standard_ruler}
    \begin{aligned}
        r_\mathrm{d} = \int_{z_\mathrm{d}}^{\infty} \! d\tilde{z} \, \frac{c_s(\tilde{z})}{H(\tilde{z})}~,
    \end{aligned}
\end{equation}
where $c_s(z)$ is the sound speed of the tightly coupled photon-baryon fluid, and $z_d$ the epoch of baryon drag. 
This is the maximum scale over which baryon pressure waves could have \cred{travelled} from initial times until the baryon-photon decoupling. The sound horizon has two major effects on the power spectrum, which can be seen in figure~\ref{fig:pk-paramderivatives} by comparing the dotted to the solid black line. First, it acts as a Jeans scale, damping the power spectrum for modes $k_\mathrm{damp}>1/\rd$ (green dashed line). Second, it introduces the BAO, whose peaks and troughs locations are given by the red and blue dotted lines. Interestingly, the slope of the baryon suppression reaches its maximum at the same scale that corresponds to the zero-crossing before the first BAO trough at  $k_\mathrm{slope}=\pi/\rd$ (green dotted line). This shows that the scale of the suppression and the BAO wiggle position are indeed directly linked to each other by $\rd$. We anticipate here that we will make use of this important fact in section~\ref{sec:shapefit-pk}. 

In the middle panel of figure~\ref{fig:pk-paramderivatives} we show the normalized derivative of the matter power spectrum with respect to the base $\Lambda$CDM parameters introduced in section~\ref{sec:theory-lcdm-model}. The effect of varying $\sigma_8$ (black dashed line) and $n_s$ (orange dashed line) is trivial, they just change the power spectrum  global amplitude and tilt respectively. The sum of neutrino masses $M_\nu$ (magenta dashed line) acts as a step-like suppression at the scales at which neutrino free streaming occurs $k\approx 0.01\,h^{-1}\Mpc$. 
\colored{Note that the differential of each parameter $\partial \Omega_i$ is normalized by a factor $\Delta_i$ as to match the effect of $\sigma_8$ at large scales. While varying one  parameter, all other parameters are fixed to their fiducial value in table \ref{tab:cosmoparams}. In effect, the normalized derivative with respect to each parameter is the same at the smallest wavevectors and the zero-crossing occurs at the same characteristic wavevector $k\approx 1/(8 \,h^{-1}\mathrm{Mpc})$ for the cases where $\sigma_8$ is fixed.} 

We can appreciate  that the parameters $\left\lbrace \colored{\ocb}, \ob, h \right\rbrace$ have the same effect on the small and  large scale limit, but show differences at  intermediate scales: the onset of suppression on scales $0.01<k\,[h{\rm Mpc}^{-1}]<0.05$ and the oscillation amplitude and position on scales $0.05<k\,[h {\rm Mpc}^{-1}]<0.5$. However, from the plot it is not clear whether different apparent amplitudes are related to a pure change in amplitude or to the shift of BAO position.

Therefore, in the bottom panel of figure~\ref{fig:pk-paramderivatives} we show the same cases after rescaling by $s$, the shift in sound horizon corresponding to the shift in cosmological parameter, such that the BAO position overlaps for all the lines,

\begin{align}\label{eq:Theory-rsrescaled}
    \colored{P_m(k) \longrightarrow \hat{P}_m(k) = \frac{1}{s^3} P_m \left(\frac{k}{s}\right)}, \qquad s= \frac{r_\mathrm{d}(\Omega_i + \partial \Delta_i\Omega_i)}{r_\mathrm{d}(\Omega_i)} ~.
\end{align}
Lines corresponding to a shift in parameters that leave $\rd$ unchanged (dashed lines) are, of course, unaffected by the rescaling. For $\sigma_8$ this is  not  strictly correct, there is a residual dependence on the scale in the filter function (see eq.~\eqref{eq:Theory_sigma8}), as  $\sigma_8$ does change after the transformation \eqref{eq:Theory-rsrescaled}. Here we actually used a redefinition of $\sigma_8$ introduced and motivated in section~\ref{sec:shapefit-scaling}. 

On the other hand, for the parameters that have an impact on $\rd$ (solid lines), we observe a systematically different behaviour.  First, the \colored{effect of the} parameter $h$ \colored{is completely absorbed by the rescaling, because we express the sound horizon $r_\mathrm{d}$ in $h^{-1}{\rm Mpc}$ units}. Second, $\colored{\ocb}$ and $\ob$ have \cred{a} nearly identical effect on the slope, with only a small offset coming from $k_\mathrm{eq}$. In fact, $k_\mathrm{eq}$ and $\rd$ are closely related within standard $\Lambda$CDM, as the relevant physical effects leading to these scales occur at relatively adjacent times, not allowing for much freedom to change one without changing the other. Third, while their effect on the slope is \colored{qualitatively} similar \colored{in the $k$-range of interest}, $\ob$ has a larger impact on the BAO wiggle amplitude than $\colored{\ocb}$. This is expected, as the amplitude depends on the ratio $\ob/\ocb$. Note that, to reduce the dynamic range to display in figure~\ref{fig:pk-paramderivatives} \colored{(and to normalize their effect on large scales)}, $\Delta_{\colored{\ocb}}$ has a different sign than $\Delta_{\ob}$. 
\colored{Although the effect of the parameters $\colored{\ocb}$, $\ob$ and $n_s$ on the shape of the matter power spectrum is expected to be somewhat degenerate, the change in slope by $\colored{\ocb}$, $\ob$ is scale-dependent, while for $n_s$ it is scale-independent by definition. We will come back to this point later.}

\colored{Of course what the figure shows and the discussion refers to is the effect of the $\Gamma$ parameter ($\Gamma \sim \Omega_m h$ in $\Lambda$CDM where however the $\sim$ sign is key as there is a rich dependence on early-time physics in the shape of the matter transfer function  see e.g., \cite{BondSzalay83, BBKS, Sugiyama95} and the extensive discussion in \cite{EH_TransferFunction}). These references, especially \cite{EH_TransferFunction} as will be clear later, are   key to offer a physical interpretation of the information provided by the power spectrum and transfer function shape.}

From this purely theoretical investigation of the linear matter power spectrum 
we conclude that when trying to measure even the base $\Lambda$CDM parameters directly from clustering data, without external priors or data-sets, the resulting constraints are expected to be highly degenerate.

In particular, we have shown that the effect on the power spectrum slope of $\left\lbrace \colored{\ocb},\ob,h,n_s \right\rbrace$ (or of $\left\lbrace \colored{\ocb},\ob,n_s \right\rbrace$ when removing the $\rd$ dependence) is \colored{qualitatively} similar. The situation is further complicated by the fact that we observe galaxies, which are biased tracers of the \colored{cold dark matter + baryon} power spectrum in redshift space, and with non-linear corrections playing an important role.
\colored{It is well known that using the matter power spectrum or the cold dark matter + baryon power spectrum  as an input for \cred{modelling} the galaxy clustering in redshift space can make a difference in the constraints of  the sum of neutrino masses \cite{Raccanelli-ml-2017kht, Vagnozzi-ml-2018pwo, Castorina-ml-2013wga}. We refer the reader to these references for more details. This is, however, beyond the scope of this paper. In the following we stick to the convention and nomenclature of the CLASS code.}

\subsection{From dark matter in real space to galaxies in redshift space} \label{sec:theory-galaxies-redshift-space}

We start by writing the density and velocity real space spectra for biased tracers at 1-loop standard perturbation theory (SPT) as in \cite{Beutler-ml-2013yhm}:
\begin{equation}
\begin{aligned}
P_{g,\delta\delta}(k) ~=~ &b_1^2 P_{m,\delta\delta}(k) +  2b_2 b_1 P_{m,b2\delta}(k) +  2b_{s2} b_1 P_{bs2,\delta}(k) + b_2^2 P_{m,b22}(k) ~ + \\ & 2 b_2 b_{s2} P_{m,b2s2}(k) + b_{s2}^2P_{bs22}(k) + 2 b_1 b_{\rm 3nl} \sigma_3^2(k) P_\mathrm{m,lin}(k) \\
P_{g,\delta\theta}(k) ~=~ &b_1 P_{m,\delta\theta}(k) + b_2 P_{m,b2\theta} (k) + b_{s2} P_{m,bs2\theta}(k) + b_{\rm 3nl} \sigma_3^2(k) P_\mathrm{m,lin}(k) \\
P_{g,\theta\theta}(k) ~=~ &P_{\theta\theta}~,
\end{aligned}
\label{eq:Pmodel}
\end{equation}
where $P_{xy}$ with $x,y = \delta$ or $\theta$ are the auto and cross power spectra of non-linear density ($\delta$) and velocity ($\theta$) perturbations, $P_{m,{\rm lin}}$ denotes the linear matter power spectrum and $P_{b2,x} P_{bs2,x}$ represent 1-loop corrections to the linear bias expansion. The exact expressions for these terms and $\sigma_3$ can be found in eq. B2- B7 of \cite{gil-marin_power_2015}.
Biasing is parametrized by four bias parameters, the first and second order biases $b_1, b_2$ \cite{Fry_Gazta-ml-93}, and the non-local biases $b_{s2}, b_{\rm 3nl}$ \citep{McDonald_2009}. Under the assumption of local Lagrangian conditions these two non-local biases can be written as a function of $(b_1-1)$ and are not independent parameters. Some studies have shown that in general this local condition holds for dark-matter haloes \citep{Baldauf_2012,Saito-ml-2014qha,kwanetal-ml-2012}, but is not necessarily true for galaxies with an arbitrary halo occupation distribution e.g., \cite{Barreira-ml-2021ukk}. We follow the usual assumption that, at the scales of interest, the galaxy velocity field is unbiased. 

Going from real space to redshift space introduces an additional dependence on the angle $\vartheta$ of wavevectors with respect to the line-of-sight (LOS), which is usually parametrized by $\mu = \cos(\vartheta)$. It is widespread to adopt the redshift space formulation from \cite{scoccimarro_redshift-space_2004} and extended by \cite{Taruya-ml-2010mx},
\begin{equation} \label{eq:TNS}
\begin{aligned}
P_\mathrm{RSD} (k,\mu) = \left(1+ \left[ k\mu\sigma_P \right]^2/2\right)^{-2} &\left[ \,P_{g,\delta\delta}(k) + 2 f \mu^2 P_{g,\delta\theta}(k) + f^2 \mu^4 P_{g,\theta\theta}(k) ~+ \right. \\
&\left. ~~b_1^3 A^\mathrm{TNS}(k,\mu,f/b_1)  + b_1^4 B^\mathrm{TNS}(k,\mu,f/b_1) \,\right] ~,
\end{aligned}
\end{equation}
where the Lorentzian damping term in front incorporates the effect of non-linear RSD, also called Fingers-of-God  effect. Here $\mu$ is the cosine of the angle to the LOS, $\sigma_P$ is a phenomenological incoherent velocity dispersion parameter, and $f$ denotes the linear growth rate $d D/d\ln a$ where $D$ is the linear growth factor and $a$ the scale factor.  \cred{Eq.} ~\eqref{eq:TNS} describes the so-called TNS model (see the definition of the coefficients $ A^\mathrm{TNS},  B^\mathrm{TNS}$ in \cite{Taruya-ml-2010mx}). We follow the usual approach of expanding the power spectrum $\mu$-dependence in the Legendre-polynomials orthonormal base. This procedure allows us to describe the LOS dependence through a series of multipoles. Although the multipole-expansion requires an infinite set of multipoles, in practice just the first 2 or 3 non-null multipoles are used.\footnote{In the same fashion an infinite $\mu$-binning is required to extract the full available information, but in practice signal-to-noise arguments limit this to just 2 or 3 bins in $\mu$ (see for e.g., \citep{wedges})}
The power spectrum multipoles are thus constructed by integrating $P_\mathrm{RSD}$ times the corresponding Legendre polynomials over $\mu$
\begin{align} \label{eq:Prsdmultipoles}
P_\mathrm{RSD}^{(\ell)} (k) =  (2\ell+1) \int_{-1}^1 \! P_\mathrm{RSD} (k,\mu) \mathcal{L}_\ell(\mu) \, d\mu~.
\end{align}
Combining the signal from the monopole ($\ell=0$) and quadrupole ($\ell=2$) allows to break the usual large-scale degeneracy between linear bias and growth of structure. Adding the hexadecapole ($\ell=4$) helps in breaking degeneracies between the AP effect and redshift space distortions. Although the non-linear terms $\{A,B\}^{\rm TNS}$ of eq.~\eqref{eq:TNS} include $\mu^6$ and $\mu^8$ contributions, the amount of information of these in the scales of interest is very small, and so, the information contained in the higher-order multipoles ($\ell>4$). For this reason all the cosmological analysis up-to-date stop at the hexadecapole level. We do not consider the odd-multipoles such as the dipole ($\ell=1$) and octopole ($\ell=3$) in our standard cosmological analyses. These are, by definition, zero under the flat-sky approximation and in the absence of selection effects, and do not contain cosmological information. However, some recent studies have shown that these measurements may be useful for an accurate modelling of the window function at very large-scales (wide-angle effects) on real surveys \cite{wide_angle}.

\subsection{Extracting cosmological information from the galaxy power spectrum: \\ An overview of BAO, RSD and FM analyses} \label{sec:theory-extract-info}

A spectroscopic galaxy survey measures the redshifts of a large number of targeted galaxies at a given angular position. The galaxies are grouped in redshift bins with different effective redshift. For each bin the summary statistics are measured, these are the 2-point correlation function and power spectrum; the 3-point correlation function and bispectrum, and even higher order moments if needed.
These statistics may contain several spurious signals related to how the observations have been performed: the angular and radial selection function \cite{Wilson-ml-2015lup}; the effect of imaging observational systematics \cite{ebossLRG_catalogue}; the effect of redshift failures or collisions \cite{Bianchi_Percival17,PIPeboss}; which need to be corrected either in the catalogue (usually by weighting the galaxies, or down-sampling the random catalogue) or by accounting them in the modelling part.

In a nutshell, the standard approach,  (e.g., BAO and RSD analyses, which from now on we will refer to as ``classic" approach) relies in compressing the data into physical observables that, {\it i}) represent the universe's late-time dynamics; {\it ii}) are as much as possible model-independent; and {\it iii}) can be in turn interpreted in light of the cosmological model of choice.

In the case of the classic BAO analysis the physical observable is the position of the BAO peak in the  clustering signal along and across the LOS. Thus, in this approach a power spectrum or correlation function template (computed once for a reference cosmological model) is used to fit the data, that is separated into a wiggle or oscillatory component containing the BAO information, and a broadband component (also referred to as non-wiggle or smooth), which does not contain any BAO information. The smooth component is marginalized over and the BAO position is measured by rescaling the wiggle component\footnote{The BAO amplitude is also damped in the wiggle component in order to account for the bulk-flow motions.} by the following free (physical) parameters,
\begin{equation}\label{eq:Theory_alphas}
\begin{aligned}
\alpha_\perp(z) &= \frac{D_M\underlying(z)\, r_{\rm d}\reference}{D_M\reference(z)\, r_{\rm d}\underlying}~, \quad
&
\quad \alpha_\parallel(z) &= \frac{H\reference(z)\, r_{\rm d}\reference}{H\underlying(z)\, r_{\rm d}\underlying}~.
\end{aligned}
\end{equation}
These are used as rescaling variables and correspond to the ratios between the underlying and the reference distances\footnote{The reference (sometimes referred to as ``fiducial") distances depend on the chosen model used to convert redshifts into distances. On the other hand, the reference sound horizon is the theory prediction of the reference model (for fixed-template approaches). Although one could choose two different reference models, for the sound horizon and the distances, is of common practice to use the same, which is the approach we follow in this work.} across and along the LOS in units of the sound horizon at baryon drag epoch defined in equation~\eqref{eq:Theory_bao_standard_ruler}.

In practice, the combined scaling is applied to the model via a coordinate transformation of wavevector $k$ and cosine of angle with respect to the LOS $\mu$

\begin{align} \label{eq:APtrafo}
k \longrightarrow \widetilde{k} &= \frac{k}{\alpha_\perp} \left[ 1 + \mu^2 \left( \frac{\alpha_\perp^2}{\alpha_\parallel^2} -1 \right)\right]^{1/2}, \\
\mu \longrightarrow \widetilde{\mu} &= \mu \frac{\alpha_\perp}{\alpha_\parallel} \left[ 1 + \mu^2 \left( \frac{\alpha_\perp^2}{\alpha_\parallel^2} -1 \right)\right]^{-1/2}. 
\end{align}

Finally, the modeled power spectrum multipoles of eq.~\eqref{eq:Prsdmultipoles} can then be written in terms of the transformed coordinates as

\begin{align} \label{eq:Prsdmultipoles-transformed}
P_\mathrm{RSD}^{(\ell)} (k) = \frac{(2\ell+1)}{2\alpha_\perp^2 \alpha_\parallel} \int_{-1}^1 \! P_\mathrm{model} (\widetilde{k}(k,\mu), \widetilde{\mu}(\mu)) \mathcal{L}_\ell(\mu) \, d\mu~.
\end{align}

Hence, the classic BAO analysis compresses the measured galaxy power spectrum multipoles in a given redshift bin, into $\alpha_\parallel$, $\alpha_\perp$, which are interpreted as the BAO peak position information, along and across the LOS, at  that redshift. These quantities describe the geometry and  expansion history of the Universe in a  model-independent  way. Under the umbrella of $\Lambda$CDM,
they can be interpreted in terms of the $\Om$ and $H_0\rd$ variables. However, the scaling parameters do not capture the effect that $\Om$, $H_0$ and the matter-radiation equality scale
have on the matter transfer function which contains extra, non-BAO-based, cosmological information \cite{Pradaetal-ml-2011,Philcox_rsmeas,EH_TransferFunction}.

A widely used approach to enhance the BAO signal and obtain more stringent constraints on cosmological parameters, is the reconstruction algorithm e.g., \cite{Eis2007,2009PhRvD..79f3523P, burden_efficient_2014}, that uses the measured overdensity field to sharpen the BAO peak by partially undoing non-linear evolution. Although it involves weak model assumptions (such as GR, linear bias, homogeneity, etc ...), the bulk of information obtained after reconstruction is still purely geometric and model-independent. In a recent work, \cite{Carter_BAOtest} show how some of these assumptions have a negligible impact on the final results.

In the case of the classic RSD analysis,  the physical observable is not only the BAO position, but also the anisotropy signal generated by redshift space distortions, mainly at linear and quasi-linear scales. The analysis follows a similar strategy as the BAO analysis, with the difference that the scaling parameters are applied to the full $P^{(\ell)}(k)$ template (i.e., the $P^{(\ell)}(k)$ for a reference cosmological model)  without any wiggle-broadband decomposition. Due to the inclusion of the broadband signal, the RSD analysis is sensitive to the monopole-to-quadrupole ratio which is  parametrized by $f\sigma_8$.\footnote{To be precise, the ratio is only parametrized by $f$, while the absolute amplitude is given by $\sigma_8$, which is fixed by the template. In practice, both parameters are very degenerate and the combination $f\sigma_8$ is template-independent.} The growth rate of structures $f$, is responsible for the large-scale bulk velocity component along the LOS, that induces an enhanced clustering signal in this direction. Unlike the anisotropic signal generated by the AP effect, the enhanced clustering caused by the RSD does not modify the BAO peak position: this \cred{makes it} possible to disentangle \cred{the RSD from the AP effect}, that otherwise would appear very degenerate. 
\colored{Note that, what we call ``classic RSD analysis" has been called Full Shape analysis in earlier works, as it includes both the BAO and the broadband. However, this name is too easy to confuse with what we call ``Full Modelling analysis". Hence the name RSD analysis, which can be thought of as an enhanced BAO (or `BAO-plus' as in the SDSS-IV official release\footnote{\href{https://svn.sdss.org/public/data/eboss/DR16cosmo/tags/v1_0_1/likelihoods/}{https://svn.sdss.org/public/data/eboss/DR16cosmo/tags/v1\_0\_1/likelihoods/}}) \cred{analysis,} that also includes the amplitude part of the broadband and its anisotropy, induced by RSD.}

The classic RSD analysis compresses the power spectrum multipoles into $\alpha_\parallel$, $\alpha_\perp$, $f\sigma_8$, similarly to what the \cred{classic BAO analysis} does, but with the additional growth of structures information.
It is well known that in GR $f$ is  determined by $\Omega_m$ (the growth history being completely determined by the expansion history).\footnote{For models where dark energy has an equation of state parameter different from $w=-1$, this parameter also appears with $\Omega_m$ in the expression for $f$ but  introduces only small corrections} The classic approach however, does not make this connection and treats $f\sigma_8$ as an independent quantity to be measured directly.

\colored{Therefore, in the case we assume a flat $\Lambda$CDM model and GR as the theory of gravity, $f\sigma_8$ is effectively a measurement of $\sigma_8$, as within $\Lambda$CDM $\Om(z)$ is obtained from $\alpha_\parallel$ and $\alpha_\perp$ and within GR the growth rate evolution $f(z)$ is completely fixed by $\Om(z)$. Crucially, the constraining power on $\sigma_8$ comes from the effect that $\Om(z)$ has on the background, not  the effect of the matter density on the epoch of matter-radiation equality and thus on the shape of  the transfer function.} 
In summary, the classic RSD analysis is only sensitive to the effect that $\Om$, $H_0\rd$ and $\sigma_8$ have at the level of BAO peak position and the relative amplitude of the isotropic and anisotropic signals, but not on their effects on the matter transfer function itself. 
This is an important point to bear in mind: the shape of the matter transfer function is set by the physics of the early Universe ($z>1000$);  on the other hand, the expansion history and growth history probed by the ``classic" BAO/RSD approach is only sensitive to late-time physics ($z\sim z_{\rm sample}\lesssim 1$ where $z_{\rm sample}$ denotes the typical redshift of the galaxy sample used to measure the power spectrum multipoles).

``Classic" BAO and RSD analyses have in common a key aspect: the attempt to compress, in a lossless way, the robust part of $P^{(\ell)}(k)$ signal into physical observables, that only depend on the late-time geometry and kinematic in a model-independent way, and not on other physics relevant to processes at play at a different epoch in the Universe evolution such as equality scale, sound horizon scale, primordial power spectrum or other quantities that enter in the matter transfer function.

In practice, this is achieved by fixing the power spectrum template:  the information contained in the transfer function does not propagate into $\alpha_\parallel$, $\alpha_\perp$ and $f\sigma_8$. It can be demonstrated that this assumption actually holds by testing the universality of $f\sigma_8$ and  the radial and angular distances in units of $r_d$,
\begin{equation} \label{eq:compression-universal}
\begin{aligned}
    \frac{D_M(z)}{\rd} &= \alpha_\perp(z) \frac{D_M\reference(z)}{\rd\reference} \\
    \frac{D_H(z)}{r_d}\equiv \frac{c}{H(z) \rd} &= \alpha_\parallel(z)  \left[ H\reference(z) \cdot \rd\reference\right]^{-1}c \\
\end{aligned}
\end{equation}
 when performing the fits with different power spectrum templates. In the case of BAO analysis, this universality has been demonstrated to hold impressively well even for exotic Early Dark Energy (EDE) and $\Delta N_\mathrm{eff}$ models \cite{Bernal_BAObias}. The template independence for the RSD analysis has been studied for eBOSS \cite{Gil-Marin-ml-2020bct} yielding reassuring results.\footnote{While the small residual template dependence has been small enough (a factor 5 smaller than the statistical errors) for eBOSS data, improvements might be needed for future data.
 }

While BAO fits are very mildly affected by non-linear corrections (the reconstruction step  removes the bulk of the non-linear effects on the BAO signal and the small scales non-linear corrections are marginalized), for RSD fits it is important to model the $P^{(\ell)}(k)$ up to 1-loop or 2-loop order in Perturbation Theory (PT). \colored{ In the classic approach} these are \colored{ usually }computed once for the reference cosmology of the template, and scaled by $\alpha_\parallel$, $\alpha_\perp$ and $f\sigma_8$ accordingly during the fit. It has been shown that the PT kernels have a very weak dependence on cosmology \cite{Catelanetal-ml-1995}, so that \colored{ the amplitude parameters ($f$, $\sigma_8$) can just be a re-scaling} and this is a valid assumption. 

Recently, there have been a series of works following a fundamentally different route than the classic BAO and RSD analyses and very close to the way CMB data are \cred{analysed} and interpreted.\footnote{ \colored{As already mentioned, and for historical completeness, this is more a going back to the  way galaxy surveys were analysed before circa 2010  rather than a radically new idea see e.g., \cite{Efstathiou-ml-2000fk,2dFGRS-ml-2001csf,SDSS-ml-2003tbn,Sanchez-ml-2005pi,wedges,Reid-ml-2009xm}}} This approach avoids the compression step and directly fits cosmological models to the $P^{(\ell)}(k)$ signal. We do not review the technical details of this approach here, we direct the reader to the references for that, but highlight important similarities and  differences  with the ``classic" approach.
 As the parameter space is explored (usually via a MCMC), the likelihood evaluation involves calculating for every choice of cosmological parameter values the model prediction of the transfer function and the non-linear correction to the power spectrum corresponding to perturbation theory  frameworks such as EFT e.g., \cite{2020JCAP...05..042I, DAmico-ml-2019fhj} or gRPT \cite{Trosteretal-ml-2020}. We call this approach the ``Full modelling" (FM) analysis/fit in what follows. In this approach the parameter dependence of the transfer function and the geometry are not kept separated; in this way the information carried  by the shape of the transfer function improves constraints on cosmological parameters that are usually interpreted as purely geometrical (e.g., $\Omega_m$, $h$).
 
 This connection between early-time transfer function and late-time background dynamics \colored{ in-built in the FM approach} can be seen as an ``{\it internal model prior}". The classic fixed template methods do not invoke a prior of that kind, as they do not establish this link. \colored{Such approaches are not  taking advantage of a model prior and are thus recognized as ``model-independent". While it is true that the template is fixed, it has been extensively demonstrated\cred{, that} this choice does not introduce biases nor affect the error-bars. e.g.,\cite{Bernal_BAObias} and refs therein.}
 For this reason in the ``classic" approach the choice of the cosmological model matters only at the stage of interpreting the constrains on the physical (compressed) parameters as constrains on cosmological parameters, \cred{with} the physical (compressed) parameters \cred{being} effectively model-independent. \colored{Of course this compression is not lossless, but, as extensively shown in the literature (see e.g. \cite{ross_information_2015}), it captures fully the relevant information and there is conscious control on the information loss \cite{Eisenstein-ml-2006nj,Eisenstein-ml-2004an,Song-ml-2008qt,Brieden-ml-2021cfg}}. In the FM approach on the other hand, the cosmological model must be chosen {\it ab initio}. \colored{Figure 7 of \cite{2020JCAP...05..042I} drives this point home: in the FM approach for simple extensions of the $\Lambda$CDM that change \cred{late}-time physics assumptions, the resulting error-bars increase to almost match those of the classic approach. }

 In practice, the FM approach must ``undo" the effect of the reference model assumed to transform \cred{redshifts} and angles into distances. 
 \colored{This is achieved (see section 2.2. of \cite{DAmico-ml-2019fhj}) by rescaling the modeled power spectrum multipoles from the model in consideration $\mathbf{\Omega}$ to the reference $\mathbf{\Omega\reference}$. This is similar to eq.~\eqref{eq:Prsdmultipoles-transformed} with the only difference that the ``$\alpha$ scaling parameters" are replaced by the so-called late-time scaling parameters defined as}

\begin{equation}\label{eq:AP_scaling_parameters}
\begin{aligned}
q_\perp(z) &= \frac{D_M\underlying(z)}{D_M\reference(z)}~, \quad
&
\quad q_\parallel(z) &= \frac{D_H\underlying(z)}{D_H\reference(z)}~, \\
q_0(z) &= \left[ q_\perp^2(z) q_\parallel (z)\right]^{1/3} = \frac{D_V\underlying(z)}{D_V\reference(z)} ~, \quad &
D_V &= \left[D_M^2(z) D_H(z)\right]^{1/3} ~,
\end{aligned}
\end{equation}
and $q_0$ is the late-time scaling associated to the power spectrum monopole. \colored{It describes the volume-averaged, isotropic distance scaling and is of integral importance as we will discuss later.} 
The main differences between the FM analysis and the classic BAO/RSD analyses are summarized in table~\ref{tab:overview_BAO_RSD_FM}.

 \begin{table}[t]
    \centering
    \makebox[\textwidth][c]{
    \begin{tabular}{l||c|c||c}
       \multirow{2}{*}{\textbf{Fit type}} &
       \multicolumn{2}{c||}{\textbf{Classic}} & \textbf{Full Modelling} \\ \cline{2-4}
         & \textbf{BAO Fit} & \textbf{RSD Fit} & \textbf{FM fit}  \\ \hline \hline
       \textbf{Information Source}  & BAO Wiggles only & BAO + $P^{(\ell)}(k)$ amp. & Full $P^{(\ell)}(k)$ \\
       $\mathbf{P_\mathrm{lin}(k)}$ \textbf{template} & fixed & fixed & varies with model \\
       \textbf{\cred{Non}-linear correction} & marginalized over & computed once & varies with model \\
       \textbf{Scaling parameters} & free $\alpha_\parallel$, $\alpha_\perp$ & free $\alpha_\parallel$, $\alpha_\perp$ & $\alpha_\parallel$, $\alpha_\perp$ derived by model \\
       \textbf{Linear RSD} & marginalized over & free $f$ & $f$ derived by model \\
       \textbf{Global amplitude} & marginalized over & $\sigma_8$ fixed or free & free $\sigma_8$ or $A_s$ \\ \hline
        & $\alpha_\parallel$, $\alpha_\perp$ can be &  $\alpha_\parallel$, $\alpha_\perp$, $f\sigma_8$ can be & done in a single step, \\
       \textbf{Cosmological}  & compared to any & compared to any & but whole fit needs \\
       \textbf{interpretation} & model, sensitive & model, sensitive to & to be repeated \\
        & to $\Omega_\mathrm{m},H_0 r_\mathrm{d}$, & $\Omega_\mathrm{m},H_0 r_\mathrm{d}$, $A_s$, $D(z)$ & for each model \\ \hline
    \end{tabular}}
    \caption{Overview of the three main approaches to extract cosmological information from galaxy surveys to date. 
    }
    \label{tab:overview_BAO_RSD_FM}
\end{table}

\begin{figure}[t]
    \centering
    \begin{minipage}[h]{0.495\textwidth}
    \centering
    BOSS DR12 \\
    \includegraphics[width=\textwidth]{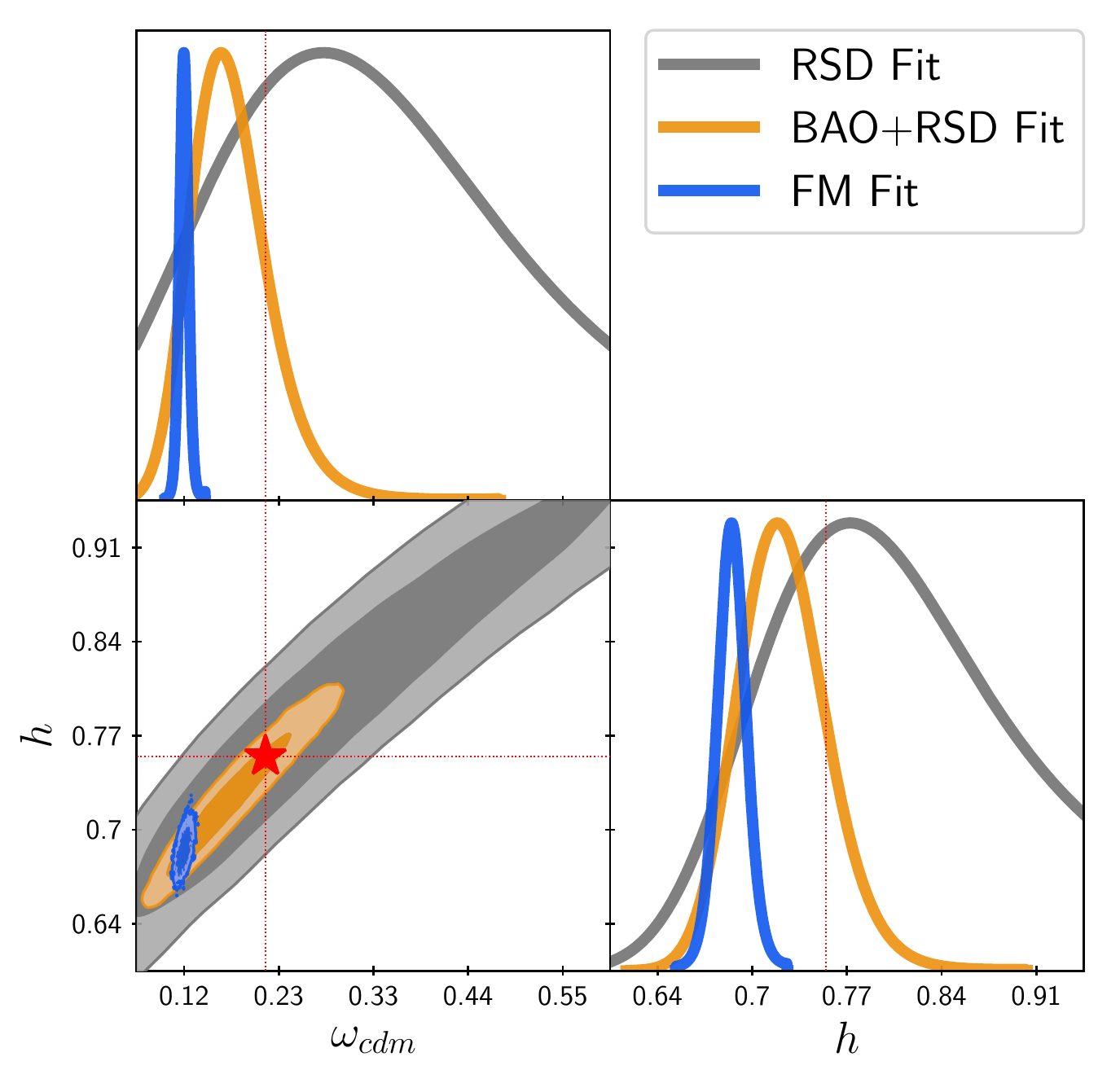}
    \end{minipage}
    \begin{minipage}[h]{0.495\textwidth}
    \centering
    Planck 2018 + BOSS DR12 \\
    \includegraphics[width=\textwidth]{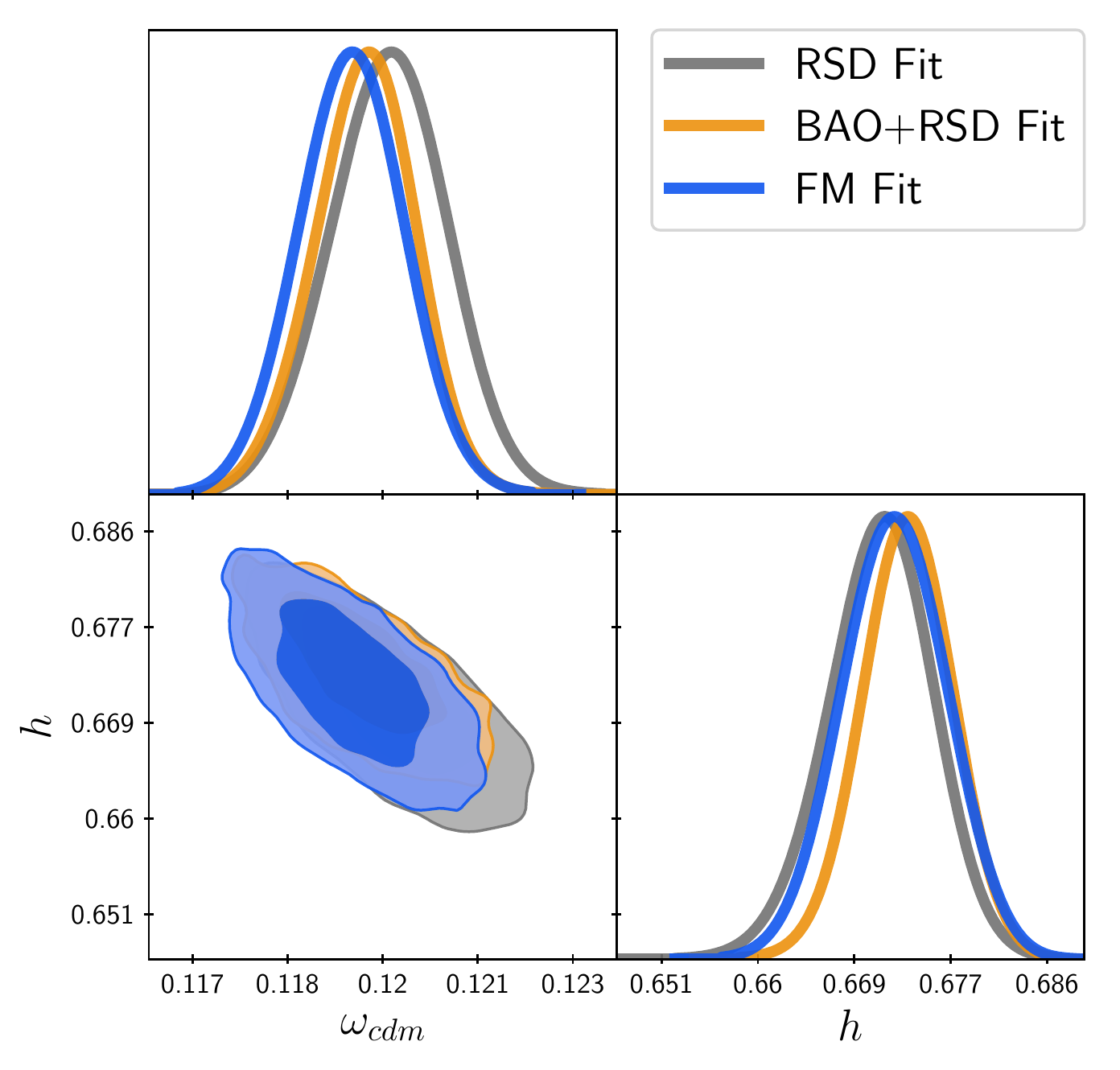}
    \end{minipage}
    \caption{Posterior results of the base $\Lambda$CDM runs -- \colored{where, following \cite{2020JCAP...05..042I} $h, A_s$ and $\omega_\mathrm{cdm}$ are varied, tight Gaussian priors are imposed  on $\omega_\mathrm{b}$ and $M_\nu$ and  $n_s$ is fixed} on BOSS DR12 data alone (left panel) and in combination with Planck (right panel) in the $(\omega_{\mathrm{cdm}}-h)$ plane. Grey contours correspond to the 68\% and 95\% confidence levels of the \cred{classic} RSD-fit from \cite{Beutler-ml-2016arn}, orange contours to the BOSS consensus result combining RSD and BAO analyses on pre- and post-reconstructed catalogues, respectively, \cite{alam_clustering_2017} and blue contours to the FM-fit using the EFT approach from \cite{2020JCAP...05..042I}. The red star corresponds to a trial model close to the RSD bestfit and still within 1-$\sigma$ of the BAO+RSD constraints, but completely excluded by the FM constraints.}
    \label{fig:trial-model-omcdm-h-plane}
    \centering
    \includegraphics[width=0.33\textwidth]{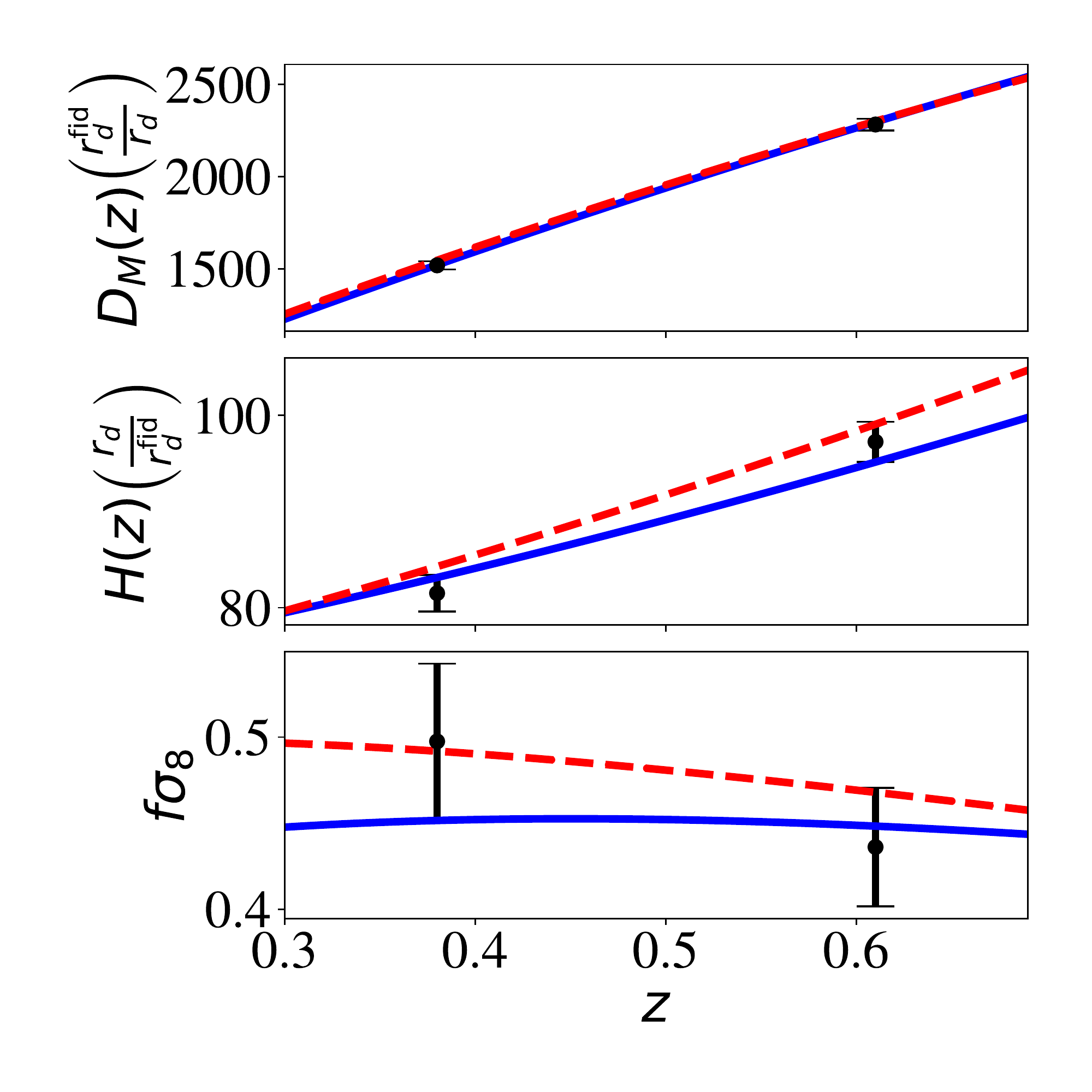}
     \includegraphics[width=0.66\textwidth]{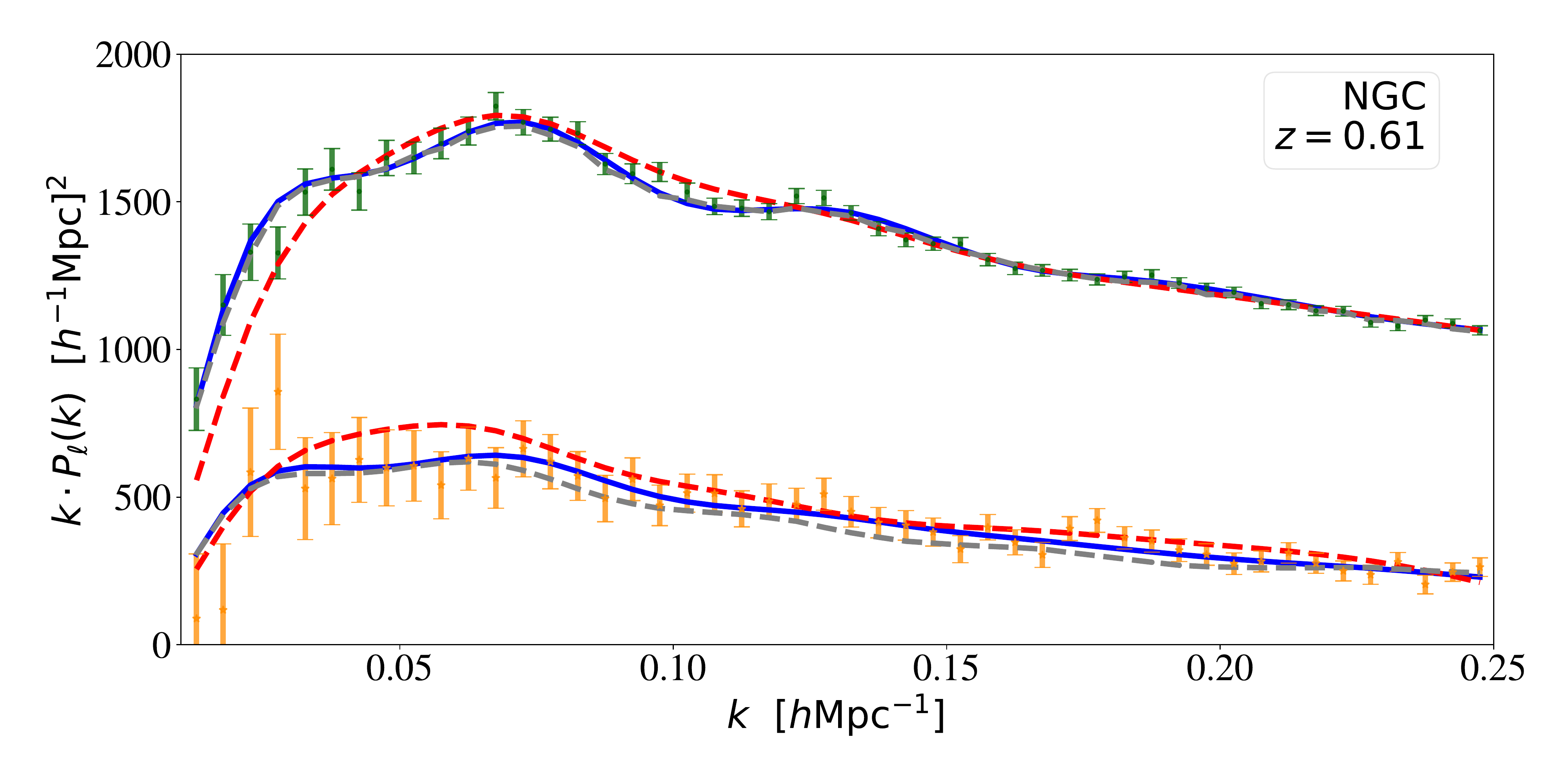}
     \includegraphics[width=\textwidth]{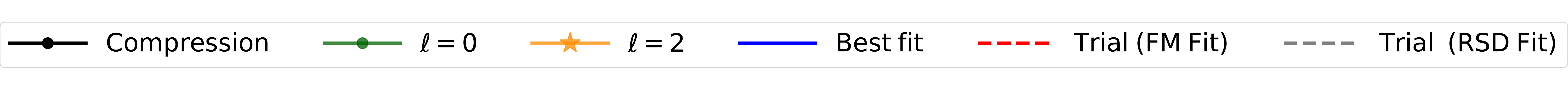}
    \caption{Left panel: the datapoints correspond to the BOSS DR12 BAO+RSD consensus results from \cite{alam_clustering_2017}. The blue line is the model prediction from the FM bestfit, the red dashed line from the trial model (red star in figure~\ref{fig:trial-model-omcdm-h-plane}). Right panel: here both models are compared to the monopole and quadrupole measurements from CMASS NGC. The grey dashed line corresponds to the trial model evaluated with the \cred{classic} RSD method, while the red dashed line is computed with the FM method (with refitted nuisance in both cases). }
    \label{fig:datat-bf-vs-trial}
\end{figure}

\subsection{BAO, RSD, and FM analyses: \cred{Direct} comparison on data} \label{sec:Theory-direct-comparison}
How do the differences between the FM and the classic approach described above translate into differences in cosmological parameter constraints? In the left panel of figure~\ref{fig:trial-model-omcdm-h-plane} we show the 1-$\sigma$ and 2-$\sigma$ confidence intervals in the $\ocdm-h$ plane obtained from fitting the flat $\Lambda$CDM model to BOSS DR12~\cite{Reid-ml-2015gra} data using the Boltzmann code CLASS \cite{2011JCAP...07..034B} within the cosmological Sampler MontePython\footnote{The code can be found at \href{https://github.com/brinckmann/montepython_public}{https://github.com/brinckmann/montepython\_public}} \cite{Brinckmann-ml-2018cvx} for three cases as follows. We fit the model to the compressed variables $\left\lbrace \alpha_\parallel, \alpha_\perp, f\sigma_8 \right\rbrace$ obtained from the Fourier Space RSD fit \cite{Beutler-ml-2016arn} (grey contours) and from the consensus BAO (post-reconstruction) + RSD fit \cite{alam_clustering_2017} (orange contours). Additionally, we show the constraints of the FM fit using the EFT approach and the publicly available code with the standard settings as in \cite{2020JCAP...05..042I} \footnote{We use their publicly available code \href{https://github.com/Michalychforever/CLASS-PT}{https://github.com/Michalychforever/CLASS-PT} from \cite{2020JCAP...05..042I} and its interface with MontePython \href{https://github.com/Michalychforever/lss_montepython}{https://github.com/Michalychforever/lss\_montepython}} (blue contours). \colored{Recall that,  as in the baseline set up of \cite{2020JCAP...05..042I} $h, A_s$ and $\omega_\mathrm{cdm}$  are varied with a flat uninformative prior, tight priors are imposed on $\omega_\mathrm{b}$ ($\ob = 0.02268 \pm 0.00038 $, Gaussian) and $M_\nu$ ($0.06\,\mathrm{eV} < M_\nu < 0.18 \mathrm{eV}$, flat) and  $n_s$ is fixed to its Planck 2018 base \cred{$\Lambda$CDM} value.} It is evident that the ``internal model prior" of the FM fit leads to substantially more precise constraints than the classic method.  

In past and present data releases of spectroscopic galaxy surveys, cosmological results are almost never presented for galaxy clustering data alone, but usually in combination with other datasets, especially with CMB data such as Planck \cite{Aghanim-ml-2018eyx}. This effectively fixes the sound horizon scale and the shape of the transfer function, so that the remaining galaxy clustering information beneficial for cosmological constraints is mostly captured by the geometrical information alone. In this particular case, as we see in the right panel of figure~\ref{fig:trial-model-omcdm-h-plane}, the FM and RSD fits deliver effectively equivalent results. One may argue that the classic template-based fits have hence been designed to constrain cosmology in combination with Planck, which justifies fixing the template to Planck's cosmology in the first place. We stress here that this is not the case. Crucially, the agreement between the FM and classic fits is independent of the template used for the classic analysis, e.g., even for a template very different from the Planck cosmology, the obtained geometrical information would be the same. This is shown in appendix \ref{sec:appendixtempldep}, see also \cite{Gil-Marin-ml-2020bct} for reference.

However, in this work we are especially interested in constraining cosmology with LSS data alone. In order to further understand, also visually,  where the difference in constraining power between the fitting approaches arises, let us  compare two suitably chosen models directly to the measurements.

From figure~\ref{fig:trial-model-omcdm-h-plane} (left panel) we select two models: the bestfit model from the FM fit located at the center of the blue contours, and a trial model displayed with the red star, selected such that it is still within the joint 1-$\sigma$ region in the $\omega_{cdm}-h$ plane of the BAO+RSD fit and close to the bestfit value of the RSD fit. In figure~\ref{fig:datat-bf-vs-trial} we compare the FM-bestfit model (blue solid line) and the trial model (red dashed line) both evaluated within the EFT framework to the data. In the left panel they are compared to the compressed variables corresponding to the BOSS DR12 consensus values (corresponding to the orange contours in figure~\ref{fig:trial-model-omcdm-h-plane}). None of the models seems to be a particularly better fit than the other. In fact, both reside at the $1-\sigma$ boundary of the orange contour in figure~\ref{fig:trial-model-omcdm-h-plane} within the same degeneracy direction between $\ocdm$ and $h$.  This is why they are basically indistinguishable in $D_M(z)/\rd$. The right panel shows the two models in comparison with the measured $P^{(\ell)}(k)$ signal; for conciseness we only show the BOSS NGC sample at $z=0.61$, as the picture does not change qualitatively for the other samples. Now it is possible to appreciate that the trial model (with refitted nuisance parameters) is a much worse fit, in fact it is completely excluded by the FM method. So why it is still a good fit within the classic method? The grey dashed line shows the trial model $P^{(\ell)}(k)$ evaluated within the RSD framework as follows. We use the values of $\left\lbrace \alpha_\parallel, \alpha_\perp, f\sigma_8 \right\rbrace$ calculated from the trial model, apply them to the reference template and refit the nuisance parameters to the data. Since the transfer function is not altered during that process, the difference between the solid blue and the dashed grey line is purely geometrical (see left panel). This is why the trial model monopole is basically identical to the one of the bestfit model (the gray dashed line is indistinguishable from the blue line) and only the quadrupole shows some (small and statistically insignificant) residual differences due to the AP and RSD anisotropies.

\colored{It should be noted that the perturbation theory models implemented in this comparison are  different between the FM and the classic RSD methods. Later we show, that  the differences are unimportant in practice, as the agreement between the methods in the right panel of figure~\ref{fig:trial-model-omcdm-h-plane} indicates.} 
To understand the meaning and relevance of the  extra information that the FM fit captures, in the next section we show how to encode this extra information with a  simple phenomenological extension of the classic fit which will enable one to bridge the two approaches in a transparent way.

\section{Connecting FM analysis and classic RSD analysis: \cred{\textit{ShapeFit}}} \label{sec:shapefit}

We now proceed to present a way to connect the two ``classic" and FM approaches which, for reasons which will become clear later, we call ``{\it ShapeFit}". We will demonstrate that two ingredients are needed to bridge the two approaches:  the correct definition, application and interpretation of the scaling parameters and the ability to model the signatures of  early-time physics  in the  large-scales broadband shape of the real-space matter power spectrum. 

\subsection{Connection: \cred{scaling} parameters interpretation} \label{sec:shapefit-scaling}
The ``late-time scaling" used in the FM approach \colored{(described at the end of section \ref{sec:theory-extract-info})} takes into account that the data is measured for a certain redshift-distance mapping corresponding to the reference model.
\colored{Here, for purely pedagogical purpose, we review this late-time rescaling from a different point of view: What if, instead of scaling the model in consideration to the measured data, we correct the data in order to match the model at each step. For simplicity, 
we now focus on the real-space monopole data $P_\mathrm{data}^{(0)}$ (without loss of generality) and write conceptually\footnote{Eq. (\ref{eq:Theory_Pdata_rescaling}) is presented only for illustrative purposes, in reality one needs to take into account the full angle dependence as done in eq.~\eqref{eq:Prsdmultipoles-transformed} for example.} how to scale it from the reference $\mathbf{\Omega\reference}$ to the model in consideration $\mathbf{\Omega}$,}
\begin{equation} \label{eq:Theory_Pdata_rescaling}
    \begin{aligned}
       P_\mathrm{data}^{(0)}(k, \mathbf{\Omega}) = q_0^3 \, P_\mathrm{data}^{(0)}(q_0 k, \mathbf{\Omega\reference})~.
 \end{aligned}
\end{equation}

\colored{It is important to note that this operation involves the ``average late-time scaling parameter" $q_0$ defined in \eqref{eq:AP_scaling_parameters} at two places: Inside the argument of $P_\mathrm{data}^{(0)}$ and as an overall amplitude factor in units of volume. While this is a well known fact, we find it important to stress the dependence on the units here in order to motivate the next steps.}

\colored{Crucially}, in contrast \colored{to this ``late-time scaling"}, we can identify an ``early-time scaling" that takes into account that the linear power spectrum template is computed for the reference cosmology. The classic RSD analysis assumes\footnote{\colored{This is a very crude approximation as it just what is needed to shift the BAO bump to the equivalent location. So it is useful pedagogically but should not be applied as is. }} that all the early-time cosmology dependence is captured by the sound horizon scale $\rd$ (defined in eq.~\eqref{eq:Theory_bao_standard_ruler}). We can apply this rescaling to the model in a similar fashion as to the data (see eq.~\eqref{eq:Theory_Pdata_rescaling}) by 
\begin{equation} \label{eq:Theory_Pmodel_rescaling}
    \begin{aligned}
        P_\mathrm{model}^{\mathrm{lin}}(k, \mathbf{\Omega}) = s^3 \, P_\mathrm{model}^{\mathrm{lin}}(s k, \mathbf{\Omega\reference}) \quad \mathrm{with} \quad s = \frac{\rd}{\rd\reference}~,
    \end{aligned}
\end{equation}
where, again, we need to introduce a volume rescaling $s^3$ taking into account that the power spectrum has units of volume. In this way the power spectrum amplitude is preserved when changing $s$.  
One can see, that the early-time rescaling on the model and the late-time rescaling on the data are very similar. The only difference is that the scaling with $s$ is purely isotropic and redshift independent, while 
a rescaling that involves $q_\parallel$ and $q_\perp$ 
allows for an additional anisotropic degree of freedom and redshift dependence.
But the isotropic components of both scalings at a given redshift, $q_0$ and $s$, are indistinguishable in practice. This is the motivation for combining both scalings into the scaling parameters
\begin{equation}
    \begin{aligned}
    \alpha_\perp(z) &= \frac{q_\perp(z)}{s}~, \quad
&
\quad \alpha_\parallel(z) &= \frac{q_\parallel(z)}{s}~, \quad
    \end{aligned}
\end{equation}
already introduced in eq.~\eqref{eq:Theory_alphas}. Thus in the ``classic" approach, instead of rescaling the data and the model separately, $\alpha_\perp$ and $\alpha_\parallel$ are applied to the model only, for reasons of practicality. This simply means that the data does not account for the  arbitrary choice of a "fiducial"  cosmology adopted to convert observed redshifts in distances to provide the input data-catalog,  but the model is transformed into the ``fiducial" coordinate system of the data instead. Although both ways of coordinate transformation are completely equivalent, we stress the difference in physical meaning here, as it is important later for the cosmological interpretation.

\subsection{New scaling for the fluctuation amplitude} \label{sec:shapefit-s8}

Having described the scaling parameters that change the modeled power spectra horizontally, in this section we look at the parameter that captures the ``vertical" information, the matter fluctuation amplitude smoothed on spheres with radius of $8\,h^{-1}\Mpc$ (see eq.~\eqref{eq:Theory_sigma8}),
\begin{equation} \label{eq:Theory_sigmaR}
\begin{aligned} 
    \sigma_8 &\equiv \sigma(R=8h^{-1}\Mpc, \mathbf{\Omega} )~, \\
    \sigma^2(R,\mathbf{\Omega}) &= \int \!\mathrm{d}(\ln k) \, k^3 P_\mathrm{lin}(k,\mathbf{\Omega}) W_\mathrm{TH}^2(kR) ~,
\end{aligned}
\end{equation}
where $W_\mathrm{TH}(kR)$ is the spherical top-hat filter. In the classic RSD analysis the amplitude of the matter fluctuations is usually fixed to the reference cosmology. The logic  is that a change in $\sigma_8$ can be seen, in a very good approximation, as being completely absorbed into the scale-independent growth rate $f$ and the bias parameters. In this sense, it is possible to obtain template-independent quantities just by multiplying $f, b_1, b_2, ...$ by $\sigma_8$. 

\colored{However, as explained in section \ref{sec:shapefit-scaling}, the classic RSD analysis implicitly assumes the ``early-time rescaling" , which induces a change in the interpretation of $\sigma_8$ via $\alpha_\parallel$ and $\alpha_\perp$. Therefore, $\sigma_8$ as defined in eq.~\eqref{eq:Theory_sigmaR} is actually \textit{not} kept fixed while exploring parameter space during the RSD fit}.

This can easily be accounted for by defining the fluctuation amplitude in such a way that it does not change during the fitting process, i.e., such that it is independent of changes in $s$:
\begin{equation} \label{eq:news8}
\begin{aligned} 
    \sigma_{s8} &\equiv \sigma(R=s\cdot 8h^{-1}\Mpc, \mathbf{\Omega} )~. 
\end{aligned}
\end{equation}
We can show, that this quantity is indeed uniquely defined for a given reference template independent of the value of $s$ by plugging into the $\sigma(R)$ definition
\begin{equation} \label{eq:news8-derivation}
\begin{aligned} 
    \sigma_{s8}^2(\mathbf{\Omega}) &= \int_0^\infty \!\mathrm{d}(\ln k) \, k^3 P_\mathrm{lin}(k,\mathbf{\Omega}) W_\mathrm{TH}^2(k s\cdot 8h^{-1}\Mpc) \qquad &\Big| \mathbf{\Omega} \rightarrow \mathbf{\Omega}\reference \\
    &= \int_0^\infty \!\mathrm{d}(\ln k) \, k^3 s^3 P_\mathrm{lin}(sk,\mathbf{\Omega}\reference) W_\mathrm{TH}^2(k s\cdot 8h^{-1}\Mpc) \qquad &\Big| k^\prime = ks \\
    &= \int_0^\infty \!\mathrm{d}(\ln k^\prime) \, k^{\prime 3} P_\mathrm{lin}(k^\prime,\mathbf{\Omega}\reference) W_\mathrm{TH}^2(k^\prime \cdot 8h^{-1}\Mpc) \\
    &= \sigma_{8}^2(\mathbf{\Omega}\reference)~.
\end{aligned}
\end{equation}
To conclude,  in the classic RSD analysis  the  fixed template fit allows for a dependence on early-time physics to be  parametrized by $\rd$. Therefore  it does not actually  measure the velocity fluctuation amplitude $f\sigma_8$ defined at an absolute smoothing scale, but the quantity $f\sigma_{s8}$, where the smoothing scale is defined relative to the sound horizon scale. This fact has been ignored in recent clustering data releases, mainly because cosmological constraints were presented in combination with Planck data, which implies $\sigma_{s8} = \sigma_8$. But for the scope of constraining cosmology from galaxy clustering alone, we emphasize that the following statement is of particular importance and an integral part of the {\it ShapeFit} presented in this work.   The three physical parameters that the classic RSD analysis actually measures at a given redshift bin,
    $D_M/s$,  $H \cdot s$, and $f\sigma(s \cdot 8 h^{-1}\Mpc)$
are \textit{all} given in units of the sound horizon ratio $s$, whenever units of length are involved. This holds for cosmological distances and smoothing scales in particular. It should be noted that by using this convention the question whether to use length units of Mpc or $h^{-1}$Mpc (see \cite{Sanchez-ml-2020vvb}) does not need to be posed.  For this reason we recommend to slightly modify the interpretation of the classic RSD parametrization of the perturbations amplitude,  to use $\sigma_{s8}$ as a parameter and have $\sigma_8$ as a derived parameter instead. \colored{We stress here  that  our proposed  redefinition of $f\sigma_8$ does not involve any changes on how to carry out the fit, but becomes important at the level of interpretation (see sections \ref{sec:shapefit-cosmo} and \ref{sec:recipe} for details).}

\subsection{\cred{Modelling} the linear power spectrum shape} \label{sec:shapefit-pk}
\begin{figure}[t]
    \centering
    \includegraphics[width=\textwidth]{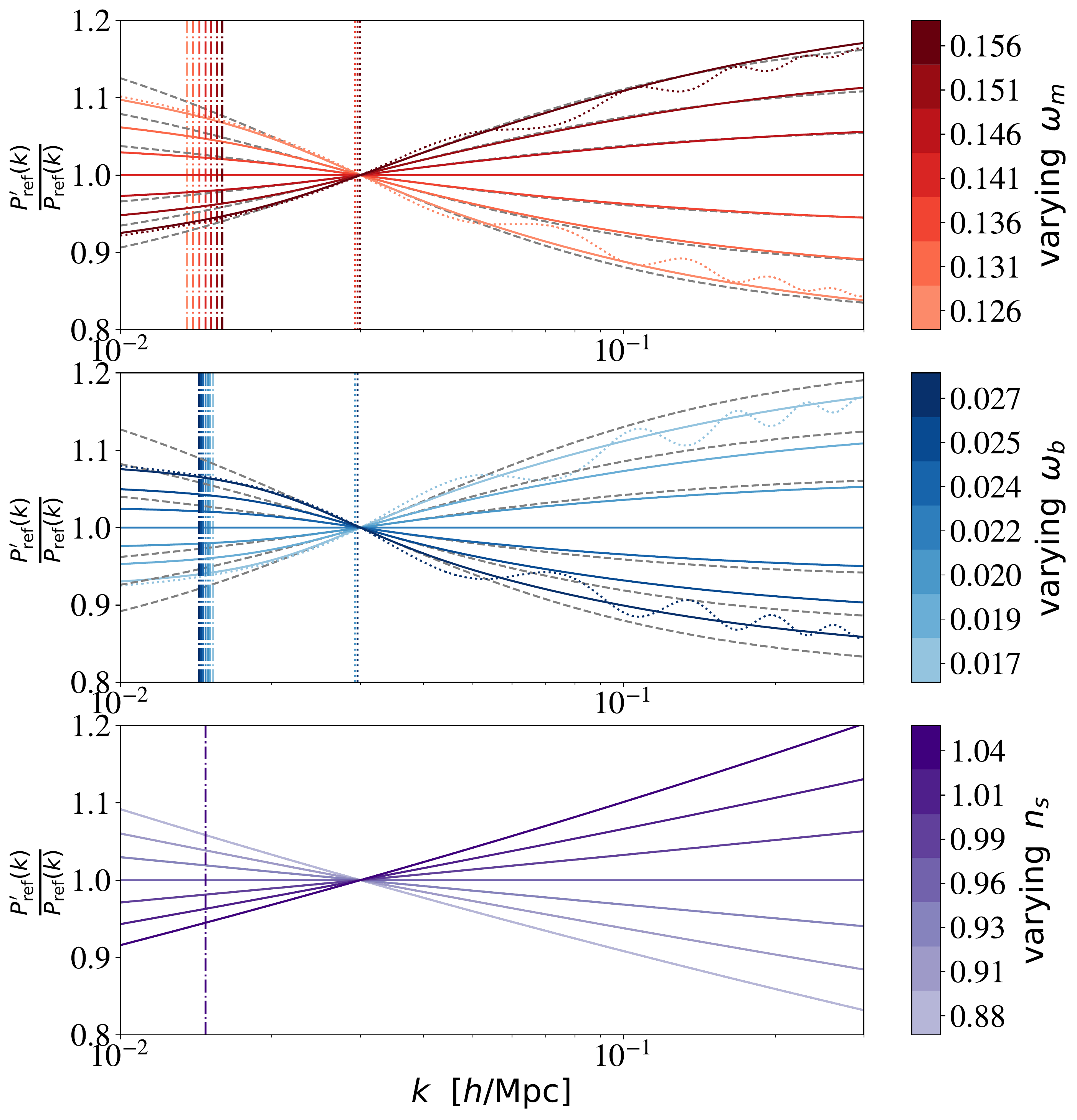}
 \caption{Rescaled EH98 prediction of the power spectrum shape (colored sold lines) comparison with the parameterization of eq.~\eqref{eq:shapefit-m} for $a=0.6, k_p=0.03\,h\mathrm{Mpc}^{-1}$. This choice fits the response to $\om$ (upper panel) very well and to $\ob$ (middle panel) less well but still sufficient for our purposes. Dashed-dotted vertical lines show the rescaled location of $k_\mathrm{eq}$ for each model and dotted vertical lines highlight the positions where the scale-dependent slope reaches a maximum. This position is constant with varying cosmological parameters and very close to the expectation $k_p=\pi/\rd^\mathrm{ref}\approx 0.03\,h\mathrm{Mpc}^{-1}$. The scale independent slope fits the prediction of varying $n_s$ (bottom panel) perfectly by definition. For the most extreme shifts in parameters we also show in dotted lines the CLASS prediction, whose shape is matched very well by the EH98 formula.}
    \label{fig:m-param-dependence}
\end{figure}
The classic BAO and RSD approaches assume that all early-time physics is captured by the free parameter $\rd$.\footnote{In fact, it is treated more as a unit rather than a free parameter at this step. However, when we interpret the unit in terms of cosmological parameters we actually constrain the parameter $\rd$.} Yet, as discussed in section~\ref{sec:theory-param-dependence}, there is additional early-time physics signal in the power spectrum. First, information on the primordial power spectrum independent of $\rd$ is present about the primordial amplitude $A_s$ (which is completely absorbed by $\sigma_8$) and the primordial tilt $n_s$, which is not captured in any way within the classic approach. Moreover, the broadband is shaped by  the transfer function encoding the evolution (scale and time dependence) of the initial fluctuations from inflation until the time of decoupling of the photon-baryon fluid, which in a $\Lambda$CDM model, depends on the physical baryon and matter densities $\om, \ob$ and  $h$ (see section~\ref{sec:theory-param-dependence} and the middle panel of figure~\ref{fig:pk-paramderivatives}). The bottom panel of this figure clearly shows that even after absorbing the dependence on $\rd$ (and hence aligning the BAO wiggle position), there is an additional dependence mostly visible in the slope and the BAO wiggle amplitude.

This additional signal is ignored in classic BAO and RSD approaches for two main reasons.  On one hand, the BAO wiggles are the most prominent feature in the power spectrum and \cred{their} position provide the most robust standard ruler to infer the universe's expansion history. On the other hand, this approach decouples the early-time information from the late-time information, that encodes the dynamics of the universe during the matter and dark energy dominated epochs (without an internal model prior).

We present here a simple, phenomenological extension of the classic RSD fit that is able to capture the bulk of the  information coming from the early-time transfer function. We propose to compress this additional signal into 1 or 2  effective parameters in such a way that  early-time and late-time information is still decoupled, but can be  easily and consistently combined at the interpretation stage when constraining cosmological parameters (i.e., the internal model prior can be imposed at the cosmological parameters inference step, but not before). 
Our goal is get the best of both approaches: on one hand to preserve the model-independent nature of the compressed physical variables of the classic approach; and on the other hand match the constraining power of the FM approach when interpreted  within the cosmological model parametrization of choice. 

As the bottom panel of figure~\ref{fig:pk-paramderivatives} shows, the classic RSD fit already takes into account the change in the global amplitude due to $\sigma_8$ and $h$ through $f\sigma_{s8}$ and of course the BAO position through $\rd$. As mentioned before, the additional degrees of freedom are the slope of the power spectrum (in a $\Lambda$CDM model depending on $\om, \ob, n_s$) and the BAO wiggle amplitude depending on $\ob, \om$.  Within $\Lambda$CDM both effects are directly coupled, this is what we refer to as the internal model prior. 
As we aim to find a model-independent parametrization, we should keep both effects separate. Moreover, we focus only on the slope and do not model the BAO wiggle amplitude, keeping it to the prescription provided by the perturbation theory model at a given fixed template. We adopt this approach for two reasons.
First, we expect the bulk of the additional  signal to come from the variation of the slope, not the BAO amplitude. Second, the BAO amplitude signal is not as robust as its position. Some bias models for example can change the BAO amplitude (see e.g., \cite{Verde-ml-2014nwa}) and the amount of non-linear BAO damping is somewhat model-dependent \cite{Hinton-ml-2019nky}.

Our {\it Ansatz} for \cred{modelling} the slope of the linear power spectrum template is as follows. We assume that the logarithmic slope consists of two components: the overall scale-independent slope $n$ (this is completely degenerate with $n_s$) and a scale-dependent slope $m$, that follows the transition of the linear power spectrum from the large scale to the small scale limit (in a $\Lambda$CDM model this is driven by the combined effect of $\ob \cred{and} \om$). To do so, we transform the reference power spectrum template, $P_\mathrm{ref}(k)$, into a new reference template, $P^\prime_\mathrm{ref}(k)$, via a slope rescaling
\begin{align} \label{eq:shapefit-m}
   \ln \left( \frac{P^\prime_{\rm ref}(k)}{P_{\rm ref}(k)} \right)=\frac{m}{a} \tanh{\left[a \ln\left(\frac{k}{k_p}\right) \right] + n \ln\left(\frac{k}{k_p}\right)  }~,
\end{align}
where the hyperbolic tangent is a generic sigmoid function reaching its maximum slope $m$ at the pivot scale $k_p$ and the amplitude $a$ controls how fast the large scale and small scale limits are reached. The pivot scale $k_p$ introduced here should not be confused with the ``primordial pivot scale" $k_\mathrm{piv}$, that is usually chosen to be $k_\mathrm{piv} = 0.05\,\mathrm{Mpc}$ (see  eq.~\eqref{eq:Theory-PX}). In contrast, the pivot scale $k_p$ is chosen to coincide  with $k_\mathrm{slope}=\pi/\rd$ introduced in figure~\ref{fig:pk-paramderivatives} corresponding to the location where the slope due to baryon suppression reaches its maximum. 

We test this expectation numerically by comparing eq.~\eqref{eq:Theory-PX} to the actual power spectrum shape (without BAO wiggles) obtained with the analytic EH98 \cite{EH_TransferFunction} formula and its response to the parameters $\om$ and $\ob$ after rescaling each curve by the corresponding value of $\rd$. The latter is important, since the transformation displayed by eq.~\eqref{eq:shapefit-m} is applied before rescaling the template, so that we need to rescale the cosmological prediction (here given by the EH98 formula) such that it matches the value of $\rd^\mathrm{ref}$. 

The comparison is shown in figure~\ref{fig:m-param-dependence} for varying $\om$ (upper panel), $\ob$ (middle panel) and $n_s$ (bottom panel). Colored solid lines are the EH98 no-wiggle power spectra ratios with color codes given by the adjacent color bars. The reference cosmology ``Planck", to which they are compared, is given by table~\ref{tab:cosmoparams}.  We also show the position of $k_\mathrm{eq}$ after rescaling by $\rd$ (dashed-dotted vertical lines). One can see that it is mildly affected by $\ob$ (through the weak $\ob$ dependence of $\rd$) and by $\om$, since $k_\mathrm{eq}$ and $\rd$ scale similarly with $\om$ as explained in section~\ref{sec:theory-param-dependence}. For the most extreme parameter shifts, we also show the CLASS prediction (dotted lines), which agrees with the shape of the EH98 formula very well. The dashed grey curves correspond to the RHS of eq.~\eqref{eq:shapefit-m}, where in the case of the $\om$ and $\ob$ sub-panels the slope $m$ is given as the derivative of the colored curves at the pivot scale $k_p$ with $n = 0$; and vice versa for the $n_s$ sub-panel. For the latter, as we see from the bottom panel, the agreement between eq.~\eqref{eq:shapefit-m} and the model is exact. This is simply because $n$ and $n_s$ are equivalent by definition, and this holds independent of the chosen pivot scale (as both describe a scale independent slope). In what follows, for simplicity,  we will
focus on the case where $n$ is fixed to $0$, which is equivalent to impose a prior $n_s = n_s^\mathrm{ref}$.

We calibrate the remaining parameters $a$ and $k_p$ with the EH98 formula for varying $\om$ and find,
\begin{align}
    a \approx 0.6, \qquad k_p \approx 0.03\,h\mathrm{Mpc}^{-1} \approx \pi/\rd^\mathrm{ref},  
\end{align}
matching the EH98 formula at  0.5\% level precision on scales $0.02 < k\,[h^{-1}\rm{Mpc}] < 0.25$. The same choice of parameters also captures very well the $\ob$-dependence, with at most 3\% deviation in the same range of scales.  

Now we have all the ingredients for the \textit{ShapeFit}, where the transformation eq.~\eqref{eq:shapefit-m}  is applied before the  rescaling by $\alpha_\perp$ and $\alpha_\parallel$. In this sense, the \textit{ShapeFit} consists of applying the classic RSD fit to a reference template $P^\prime_\mathrm{ref}(k)$, that is transformed at each step via \eqref{eq:shapefit-m} with free parameters $m$ (and $n$, if needed).
In principle this transformation should be applied also to the reference power spectra that appear in the integrand of higher-order perturbation corrections.  In our implementation, however, in order to avoid a re-evaluation of all perturbative terms at each step of the likelihood exploration, we apply this transformation as if it were independent of scale. In practice this means that we pre-compute all the loop corrections using the linear power spectrum given by $P_{\rm ref}$. During the likelihood evaluation we transform each of these terms using eq.~\eqref{eq:shapefit-m}, taking into account the power of the linear power spectrum used to compute them, which is a power of $N+1$ for the $N$-loop corrections. \colored{To be more precise, in the case of SPT we evaluate the $1$-loop correction $P_\mathrm{1-loop} = P_{13} + P_{22}$ depending on the (new) reference linear template $P_\mathrm{ref}^\prime$ and the corresponding kernels $F_i$ using the following approximations,
\begin{equation}
\begin{aligned}\label{eq:m-nonlinearities}
    P_{13}(k) &= P_\mathrm{ref}^\prime(k) \int_0^\infty \! d^3q \, P_\mathrm{ref}^\prime(q) F_3(k,q,-q) \\
    &\approx \left( \frac{P^\prime_{\rm ref}(k)}{P_{\rm ref}(k)} \right)^2 P_\mathrm{ref}(k) \int_0^\infty \! d^3q \, P_\mathrm{ref}(q) F_3(k,q,-q),  \\
    P_{22}(k) &= \int_0^\infty \! d^3q \, P_\mathrm{ref}^\prime(q) P_\mathrm{ref}^\prime(|q-k|) F_2 (k,q-k) \\
    &\approx \left( \frac{P^\prime_{\rm ref}(k)}{P_{\rm ref}(k)} \right)^2 \int_0^\infty \! d^3q \, P_\mathrm{ref}(q) P_\mathrm{ref}(|q-k|) F_2 (k,q-k).
\end{aligned}
\end{equation}}

We show in appendix \ref{sec:rescaling} that this approximation is very good and more than sufficient for our purposes. 
With this, the computational time of {\it ShapeFit} is effectively indistinguishable from that of the classic RSD approach, except at the MCMC level,  where the posterior sampling involves one (or two) extra parameters.
{\colored It is of academic interest but still instructive to consider how the discussion of sections \ref{sec:shapefit-scaling} and \ref{sec:shapefit-s8} would change if there were no BAO. In this case it would be misleading to interpret $s$ as the sound horizon ratio. Nevertheless, {\it ShapeFit} can be used in the case of zero baryons (or no-BAO) either by setting $s=1$ (which is similar to the classic method where $s$ would be set to 1 inside the $\alpha$ terms), or by interpreting $s$ not as the sound horizon ratio, but rather as the ratio of "pivot scale" $k_{p}/k_{p}^\mathrm{ref}$ (see eqs. \eqref{eq:news8-derivation}-\eqref{eq:m-nonlinearities}). Generally speaking, the {\it ShapeFit} parameterization does not rely on BAO, but rather on the notion, that there is some early-time physics scale - a ruler - that mostly defines the power spectrum shape.}

\subsection{Cosmological interpretation} \label{sec:shapefit-cosmo}
The \textit{ShapeFit} constraints on  the physical and phenomenological parameters, $\left\{ \right. \alpha_\parallel,\allowbreak \alpha_\perp,\allowbreak f\sigma_{s8},\allowbreak m,\allowbreak n \left. \right\}$ can be then interpreted in terms of cosmological parameters. This step is, naturally, very similar to the way standard RSD likelihoods are implemented already in the most common cosmological inference codes. 
In the classic RSD approach, results on $\left\lbrace \alpha_\parallel, \alpha_\perp, f\sigma_{8} \right\rbrace$ and their covariance for all redshift bins are used as input for cosmological parameters inference, where the $\chi^2$ (or log-likelihood) is  computed for the theoretical prediction for each quantity given an input cosmological model and parameters values. 
 For the \textit{ShapeFit} the relevant aspects in this step are the calculation of the scaling parameters, fluctuation amplitude and growth rate, and the power spectrum slope (the only new ingredient). 

\paragraph{Scaling parameters.} The interpretation of $\left\lbrace \alpha_\parallel, \alpha_\perp \right\rbrace$ is exactly the same as in the classic RSD approach. Therefore, any existing likelihood computing these quantities using eq.~\eqref{eq:Theory_alphas} is left unchanged.

\paragraph{Fluctuation amplitude and growth.} The interpretation of $f\sigma_{s8}$ is nearly the same as in the classic RSD analysis with the only difference that we advocate the fluctuation amplitude to be  defined as $\sigma_{s8}$ instead of $\sigma_8$, see eqs.~\eqref{eq:news8} and \eqref{eq:news8-derivation}  and section~\ref{sec:shapefit-s8}.

However, the  slope rescaling (eq.~\eqref{eq:shapefit-m} and section~\ref{sec:shapefit-pk}) changes $\sigma_{s8}$ for $m,n \neq  0$. Therefore, it is  convenient to define the fluctuation amplitude $A_p$ at the pivot scale $k_p$
\begin{align}\label{eq:shapefit-Ap}
A_p^\mathrm{ref} = P_\mathrm{no-wiggle}^\mathrm{lin}\left(k_p, \mathbf{\Omega}^\mathrm{ref}\right)~,
\end{align}
which does not change with slope by definition;\footnote{It should be noted, that the amplitude $A_{sp}$ needs to be obtained from the ``no-wiggle" power spectrum (given by the EH98 formula for instance), to ensure that the BAO wiggles do not influence the amplitude. Normally, this is ensured by using $\sigma_8$ as the amplitude, a quantity, for which the BAO wiggles are integrated over. But since eq.~\eqref{eq:shapefit-m} operates in Fourier Space, it is more convenient to define the amplitude in Fourier space as well. }  $A_p^\mathrm{ref}$ is defined for the reference template. As the analysis explores the posterior of the physical and phenomenological parameters, following section~\ref{sec:shapefit-pk} it is possible to recognize that the amplitude parameter becomes, internally to the fit,  
\begin{align}\label{eq:shapefit-Asp}
A_{sp} = \frac{1}{s^3} P_\mathrm{no-wiggle}^\mathrm{lin}\left(\frac{k_p}{s}, \mathbf{\Omega}\right)~, \quad \mathrm{with} \quad s = \frac{\rd}{\rd^\mathrm{ref}}~.   
\end{align}
This amplitude, $A_{sp}$, can be understood as the ``late-time" counterpart of the amplitude of the primordial power spectrum $A_s$, but the two quantities should not be confused. The actual velocity fluctuation amplitude measurement is then given as $f A_{sp}^{1/2}$, and thus
\begin{align}\label{eq:shapefit-interpretation-fAsp}
    f\sigma_{s8} = \frac{(f\sigma_{s8})^\mathrm{ref}}{(f A_{sp}^{1/2})^\mathrm{ref}}  f A_{sp}^{1/2}\,.
\end{align}
\colored{Eq.~\eqref{eq:shapefit-interpretation-fAsp} can be used in order to obtain the more frequently used $f\sigma_{s8}$ variable, although we advocate using $f A_{sp}^{1/2}$ for cosmological parameter inference. It should be clear (see also section~\ref{sec:recipe}) that we only propose a reinterpretation of the amplitude parameter not a change in the analysis or definitions.} 

\paragraph{Power spectrum slope.} The new ingredient of the \textit{ShapeFit} is given by the slope, parametrized by $m,\,n$
following eq.~\eqref{eq:shapefit-m}. The interpretation of the scale independent slope $n$ is trivial, as it can be directly related to the primordial scalar tilt $n_s$ via,
\begin{align}\label{eq:shapefit-interpretation-m2}
    n = n_s - n_s^\mathrm{ref}~.
\end{align}
The interpretation of the scale-dependent slope $m$ then becomes:
\begin{align} \label{eq:shapefit-interpretation-m1}
    m = \frac{d}{dk} \left( \ln \left[ \frac{P_\mathrm{no-wiggle}^\mathrm{lin}\left(\frac{k_p}{s}, \mathbf{\Omega}\right)/\colored{\mathcal{P}_\mathcal{R}\left(\frac{k_p}{s}, \mathbf{\Omega}\right)}}{P_\mathrm{no-wiggle}^\mathrm{lin}\left(k_p, \mathbf{\Omega}^\mathrm{ref}\right)/\colored{\mathcal{P}_\mathcal{R}\left(k_p, \mathbf{\Omega}^\mathrm{ref}\right)}} \right] \right) \Bigg|_{k=k_p}~.
\end{align}
{\colored where $\mathcal{P}_\mathcal{R}\left(k, \mathbf{\Omega}\right)$ denotes the primordial density power spectrum.}
In case  $n_s$ is varied during the cosmological fit, eq.~\eqref{eq:shapefit-interpretation-m1} has to be applied to the power spectrum obtained when $n_s$ is fixed to the reference value. This ensures  that a change in $n_s$ does not lead to a different prediction for $m$ but only for $n$ via eq.~\eqref{eq:shapefit-interpretation-m2}. In other words, $n$ is obtained from the primordial power spectrum, while $m$ is obtained from the transfer function squared \colored{(\cred{which is} the power spectrum divided by the primordial power spectrum)}.

In practice, the no-wiggle linear power spectrum $P_\mathrm{no-wiggle}^\mathrm{lin}$ is computed using the EH98 formula. While being computationally much faster this formula matches the Boltzmann code output \colored{formally at the 5\% level over a wide range of cosmologies. However, for the parameter range investigated here and since we are considering power spectrum ratios we find 1\% level precision,} suitable for this application. \cred{An implementation of the cosmological likelihood containing the interpretation of our \textit{ShapeFit} BOSS DR12 results within MontePython is publicly available.}\footnote{\cred{\href{https://github.com/SamuelBrieden/shapefit_montepython_code}{https://github.com/SamuelBrieden/shapefit\_montepython\_code}}}

\tikzstyle{headerleft} = [draw, diamond, fill=green!15, node distance=0.325\textwidth-1em, text width=0.1\textwidth, text centered, anchor=center, minimum height=2em]
    
\tikzstyle{headertop} = [draw, rectangle, fill=green!15, node distance=0.425\textwidth, text width=0.1\textwidth, text centered,
    minimum height=2em]
    
\tikzstyle{block-b} = [rectangle, draw, fill=blue!15, 
    text width=0.35\textwidth, minimum width = 2em, rounded corners, minimum height=3.5em]
    
\tikzstyle{block-r} = [rectangle, draw, fill=orange!15, 
    text width=0.35\textwidth, minimum width = 2em, rounded corners, minimum height=3.5em]

\tikzstyle{block-inter} = [rectangle, draw, fill=blue!15, 
    text width=0.425\textwidth, minimum width = 2em, rounded corners, minimum height=4em]
  
\tikzstyle{cloud-b} = [draw, circle,fill=blue!15, node distance=0.159\textwidth, text width=0.085\textwidth, text centered,
    minimum height=1em]
    
\tikzstyle{cloud-r} = [draw, circle,fill=orange!15, node distance=0.159\textwidth, text width=0.085\textwidth, text centered,
    minimum height=1em]
\tikzstyle{line} = [draw, line width = 0.5mm, -stealth]

\tikzstyle{elem-b} = [draw, rectangle,fill=blue!15, node distance=0.17\textwidth, text width=0.085\textwidth, text centered,
    minimum height=1em, rounded corners]
    
\tikzstyle{elem-r} = [draw, rectangle,fill=orange!15, node distance=0.17\textwidth, text width=0.085\textwidth, text centered,
    minimum height=1em, rounded corners]
    
\tikzstyle{big-elem-b} = [draw, rectangle,fill=blue!15, node distance=0.31875\textwidth, text width=0.13\textwidth, text centered,
    minimum height=4em, rounded corners]
    
\tikzstyle{big-elem-r} = [draw, rectangle,fill=orange!15, node distance=0.31875\textwidth, text width=0.13\textwidth, text centered,
    minimum height=4em, rounded corners]
    
\tikzstyle{line} = [draw, line width = 0.5mm, -stealth]

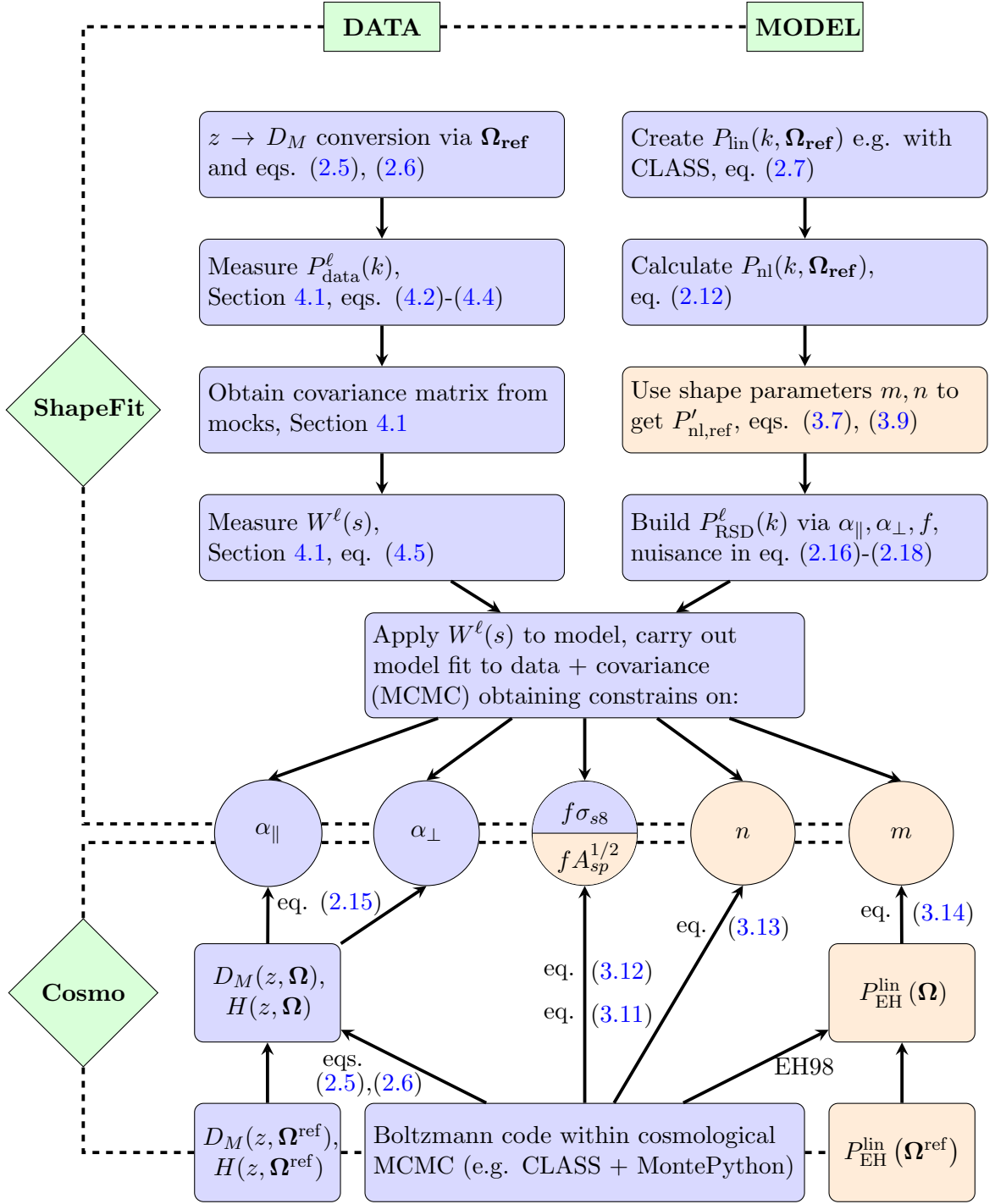
\begin{figure}[!h]
\begin{tikzpicture}[node distance = 2.0cm, auto]
    \node [headertop, node distance = 1.0cm] (D) {\textbf{DATA}};
    \node [block-b, below of=D] (D1) {$z\rightarrow D_M$ conversion via $\mathbf{\Omega}_\mathbf{ref}$ and eqs. \eqref{eq:Hubble}, \eqref{eq:dist}};
    \node [block-b, below of=D1] (D2) {Measure $P_\mathrm{data}^\ell(k)$, \par Section \ref{sec:Practical-mocks}, eqs. \eqref{eq:FKP_pse}-\eqref{eq:Pk_yama}};
    \node [block-b, below of=D2] (D3) {Obtain covariance matrix from mocks, Section \ref{sec:Practical-mocks}};
    \node [block-b, below of=D3] (D4) {Measure $W^\ell(s)$, \par Section \ref{sec:Practical-mocks}, eq. \eqref{eq:Practical-window}};   
    \node [headerleft, left of=D3] (SF) {\textbf{ShapeFit}~};
    
    \node [headertop, right of=D] (M) {\textbf{MODEL}};
    \node [block-b, below of=M] (M1) {Create $P_\mathrm{lin}(k,\mathbf{\Omega}_\mathbf{ref})$ e.g. with CLASS, eq.~\eqref{eq:Theory-PX} };
    \node [block-b, below of=M1] (M2) {Calculate $P_\mathrm{nl}(k,\mathbf{\Omega}_\mathbf{ref})$, \par eq.~\eqref{eq:Pmodel} };
    \node [block-r, below of=M2] (M3) {Use shape parameters $m,n$ to get  $P_\mathrm{nl,ref}^\prime$, eqs. \eqref{eq:shapefit-m}, \eqref{eq:m-nonlinearities}};
    \node [block-b, below of=M3] (M4) {Build $P_\mathrm{RSD}^\ell(k)$ via $\alpha_\parallel, \alpha_\perp, f,$ nuisance in eq.~\eqref{eq:APtrafo}-\eqref{eq:Prsdmultipoles-transformed}};   
    \node [block-inter, below of=M4, anchor=east] (M5) {Apply $W^\ell(s)$ to model, carry out model fit to data + covariance (MCMC) obtaining constrains on:};
   \node[minimum height=0.4em,
    shape=circle split,
    draw,node distance=0.17\textwidth, text width=0.015\textwidth, text centered,
    circle split part fill={blue!15, orange!15}, below of=M5
    ] (fs8) {$\!\!\!\!f\sigma_{s8}$ \nodepart{lower} $\!\!\!\!\!fA_{sp}^{1/2}$};
   \node [cloud-b, left of=fs8] (aperp) {$\alpha_\perp$};
   \node [cloud-b, left of=aperp] (apara) {$\alpha_\parallel$};
   \node [cloud-r, right of=fs8] (nshape) {$n$};
   \node [cloud-r, right of=nshape] (mshape) {$m$};
   \node [block-inter, below of=fs8, node distance = 5.0cm] (CLASS) {Boltzmann code within cosmological MCMC (e.g. CLASS + MontePython)};
   \node [big-elem-b, left of=CLASS] (Dref) {$D_M(z, \mathbf{\Omega}^\mathrm{ref})$, \par $H(z, \mathbf{\Omega}^\mathrm{ref})$};
   \node [big-elem-r, right of=CLASS] (EHref) {$P_\mathrm{EH}^\mathrm{lin}\left( \mathbf{\Omega}^\mathrm{ref}\right)$};
   \node [big-elem-b, above of=Dref, node distance = 2.5cm] (Dcosmo) {$D_M(z, \mathbf{\Omega})$, \par $H(z, \mathbf{\Omega})$};
   \node [big-elem-r, above of=EHref, node distance = 2.5cm] (EHcosmo) {$P_\mathrm{EH}^\mathrm{lin}\left( \mathbf{\Omega}\right)$};
   
   \node [headerleft, below of=SF, node distance=0.64\textwidth-2em] (CI) {\textbf{Cosmo}};

    \path [line] (D1) -- (D2);
    \path [line] (D2) -- (D3);
    \path [line] (D3) -- (D4);
    \path [line] (D4) -- (M5);
    \path [line] (M1) -- (M2);
    \path [line] (M2) -- (M3);
    \path [line] (M3) -- (M4);
    \path [line] (M4) -- (M5);
    
    \path [line] (M5) -- (apara.90);
    \path [line] (M5) -- (aperp.90);
    \path [line] (M5) -- (fs8.90);
    \path [line] (M5) -- (nshape.90);
    \path [line] (M5) -- (mshape.90);
    
    \path [line] (CLASS) -- (fs8) node[pos=0.4,left] {eq.} node[pos=0.4,right] {$\!$\eqref{eq:shapefit-Asp}} node[pos=0.6,left] {eq.} node[pos=0.6,right]  {$\!$\eqref{eq:shapefit-interpretation-fAsp}};
    
    \path [line] (CLASS) -- (EHcosmo) node[midway,right] {~EH98};
    \path [line] (EHcosmo) -- (mshape) node[midway,left] {eq.} node[midway,right] {\eqref{eq:shapefit-interpretation-m1}};
    \path [line] (EHref) -- (EHcosmo);
    \path [line] (CLASS) -- (nshape.270) node[pos=0.8,left] {eq.} node[pos=0.8,right] {\eqref{eq:shapefit-interpretation-m2}};
    
    \path [line] (CLASS) -- (Dcosmo) node[pos=0.55,left] {eqs.~~~~~} node[pos=0.25,left] {\eqref{eq:Hubble},\eqref{eq:dist}~~~};
    \path [line] (Dcosmo) -- (aperp.270) node[pos=0.7,left] {eq. \par \eqref{eq:Theory_alphas}~~} ;
    \path [line] (Dref) -- (Dcosmo);
    \path [line] (Dcosmo) -- (apara);
    
    \path [draw,dashed, line width = 0.5mm] (SF) |- (D);
    \path [draw,dashed, line width = 0.5mm] (D) -- (M);
    \path [draw,dashed, line width = 0.5mm] (SF) |- (apara.170);
    \path [draw,dashed, line width = 0.5mm] (apara.10) -- (aperp.170);
    \path [draw,dashed, line width = 0.5mm] (aperp.10) -- (fs8.170);
    \path [draw,dashed, line width = 0.5mm] (fs8.10) -- (nshape.170);
    \path [draw,dashed, line width = 0.5mm] (nshape.10) -- (mshape.170);
    \path [draw,dashed, line width = 0.5mm] (apara.350) -- (aperp.190);
    \path [draw,dashed, line width = 0.5mm] (aperp.350) -- (fs8.190);
    \path [draw,dashed, line width = 0.5mm] (fs8.350) -- (nshape.190);
    \path [draw,dashed, line width = 0.5mm] (nshape.350) -- (mshape.190);
    \path [draw,dashed, line width = 0.5mm] (CI) |- (apara.190);
    \path [draw,dashed, line width = 0.5mm] (CI) |- (Dref);
    \path [draw,dashed, line width = 0.5mm] (Dref) -- (CLASS);
    \path [draw,dashed, line width = 0.5mm] (CLASS) -- (EHref);
    
\end{tikzpicture}
 \caption{Executive summary of {\it ShapeFit} (upper dashed lines) from data acquisition (``DATA", see  sections~\ref{sec:theory} and  \ref{sec:Practical-mocks}) to modelling (``MODEL") and its cosmological interpretation (``Cosmo") (bottom dashed lines, sections~\ref{sec:theory} and \ref{sec:shapefit}) including all relevant equations. The purple \cred{fields} represent steps in the ``classic" approach while orange \cred{fields} represent the {\it ShapeFit} additions.  Circles represent parameters, boxes in the top part of the diagram represent analyses steps, in the bottom part of the diagram represent products of theoretical calculations.} \label{fig:cheatsheet}
\end{figure}

\subsection{{\it ShapeFit} implementation recipe}
\label{sec:recipe}
\colored{We summarize the changes to be done to the classic BAO+RSD analysis (and respective codes) to implement {\it ShapeFit} in the flowchart of figure~\ref{fig:cheatsheet}. This  chart  can be seen as an executive summary of {\it ShapeFit} (upper dashed lines) from data acquisition  to modelling  and its cosmological interpretation  (bottom dashed lines) including all pointers to  relevant equations. In this flowchart the purple \cred{fields} represent steps in the ``classic" approach while orange \cred{fields} represent the {\it ShapeFit} additions. Circles represent parameters, boxes  in the top part of the diagram represent analyses steps, in the bottom part of the diagram represent products of theoretical calculations.}

\section{Application to mocks of SDSS-III BOSS survey data} \label{sec:Practical}

We now describe our fiducial analysis setup which we use to compare the \textit{ShapeFit} introduced in section~\ref{sec:shapefit} with the FM fit. The FM application  is done following the EFT implementation by \cite{2020JCAP...05..042I}. We first present the mocks in section~\ref{sec:Practical-mocks} and describe the model choices in section~\ref{sec:Practical-priors}. 

\subsection{Mock catalogs} \label{sec:Practical-mocks}
We apply our analysis pipeline to the MultiDark-Patchy BOSS DR12 (\textsc{Patchy}) mocks created by \cite{kitaura_clustering_2016,rodriguez-torres_clustering_2016}. The fiducial $\Lambda$CDM parameters of the Multidark simulation are,
\begin{align} \label{eq:patchy_cosmo}
    \Om = 0.307115, \quad \Ob = 0.048206, \quad h = 0.6777, \quad \sigma_8 = 0.8288, \quad n_s = 0.9611\,.
\end{align}
The mock catalogs are designed to reproduce the angular and radial selection function and small scale clustering of BOSS DR12 data.
These mocks have been used extensively in the development of the analysis of the BOSS survey, and provide many realizations, which is crucial for estimating covariance matrices and for stacking to reduce statistical errors.   However it is important to keep in mind that these are not full N-body runs, but are based on Augmented Lagrangian Perturbation Theory and an exponential bias scheme. Small differences with N-body mocks are not unexpected. For this reason in section~\ref{sec:additional_tests} we also consider independently generated Nseries mocks (see section 7.2 of \cite{eboss_collaboration_dr16} as well as section 2.2.2 of \cite{Gil-Marin-ml-2020bct} for details)  based on full N-body runs, populated using halo occupation distribution parameters that match Luminous Red Galaxies (LRG) observations and with the sky-geometry of BOSS DR12 CMASS northern galactic cap sample.
In the remainder of this section we focus only on the (\textsc{Patchy}) mocks ``ngc\_z3" \cred{sample} located at the north galactic cap and covering a redshift range of $0.5<z<0.75$ with effective redshift $z_\mathrm{eff}=0.61$. We work with all 2048 realizations of the \textsc{Patchy} mocks, which are publicly available.\footnote{\href{https://fbeutler.github.io/hub/boss_papers.html}{https://fbeutler.github.io/hub/boss\_papers.html}}

In addition to angular positions and redshifts, the catalogs provide simulated close-pair weights $w_\mathrm{cp}$ to account for galaxy pairs neighboured closer than the instrument angular resolution (limited by the fiber size).
Also, the catalogs contain the angle averaged number density $\bar{n}(z)$ for each galaxy, which allows one to construct the FKP weight $w_\mathrm{FKP}(z) = 1/(1+ \bar{n}(z)P_0)$ \cite{1994ApJ...426...23F}. This weight is used to minimize the power spectrum variance at $P_0 = 10,000\, [{\rm Mpc}h^{-1}]^3$, which corresponds to the galaxy power spectrum amplitude at $k\sim0.1\,h{\rm Mpc}^{-1}$. We also use the random catalogs provided along with the mocks containing $\alpha_\mathrm{ran}^{-1}=50$ times more objects than  the individual mocks. They have the same selection function  but no intrinsic clustering.

We measure the multipole power spectra of each individual mock catalog. Then, we take the mode-weighted power spectra average of all 2048 realizations, which is used as our dataset. The error bars (including correlations between different bins) are  obtained from the covariance of the 2048 mocks. At the step of covariance matrix inversion, we apply the Hartlap correction \cite{Hartlap-ml-2006kj} taking into account the small bias due to the finite number of mock catalogs. We fit the mean of the 2048 mocks and rescale the covariance matrix by a factor $0.01$, which corresponds to the volume of $100$ stacked mocks.\footnote{We do not rescale it to a volume of 2048 \colored{mocks}, since this would decrease the error bars to a level much smaller than the model uncertainty (both of 1-loop SPT and of the semi-analytic models used to create the mocks). The corresponding effective volume of the 100 stacked mocks ($\approx 300 \, \mathrm{Gpc}^3$ assuming the ``Planck" cosmology) is still significantly larger than the effective volume of the next generation of galaxy redshift surveys.} Finally, we compute the survey window function, which is needed in order to compare theoretical models to the mock data. Our procedure of the power spectrum estimation and the window function computation (which is standard) is described in more detail below.  

\paragraph{Power spectrum estimator.} 
We place the galaxies into a cubic box with length $L_\mathrm{box}=3.6\,\mathrm{Gpc}h^{-1}$
using the reference cosmology of table~\ref{tab:cosmoparams} (with $\Om=0.31$) to convert redshifts to distances. We assign galaxies and random objects to $512^3$ grids using the triangular shape cloud (TSC) grid assignment and using the interlacing technique to mitigate aliasing effects \citep{sefusatti}. Using the obtained galaxy and random densities, $n(\mathbf{r})$ and $n_\mathrm{ran}(\mathbf{r})$, 
we \colored{ follow standard practice  and} define the FKP function as
\cite{1994ApJ...426...23F},
\begin{align}\label{eq:FKP_pse}
F(\textbf{r}) = \frac{w_\mathrm{FKP}(\mathbf{r})}{I_2^{1/2}} \left[w_\mathrm{cp}(\mathbf{r}) n(\mathbf{r}) - \alpha_\mathrm{ran} n_\mathrm{ran}(\mathbf{r})  \right]~,
\end{align}
where the normalization factor $I_2$ is given as 
\begin{align} \label{eq:Pk_I2}
I_2 = \int \! d^3\mathbf{r} \, w_\mathrm{FKP}(\mathbf{r}) \left\langle w_\mathrm{cp}(\mathbf{r})n(\mathbf{r})  \right\rangle ^2~.
\end{align}
We construct the power spectrum multipoles via Fourier transformations following the Yamamoto approximation \cite{Yamamoto-ml-2005dz,bianchi_measuring_2015}
\begin{align} \label{eq:Pk_yama}
P^{(\ell)}(k) = \frac{(2 \ell+1)}{2} \int \! \frac{d \Omega}{4 \pi} \, \left[ \int \! d \mathbf{r}_1 \, F(\mathbf{r}_1) e^{i \mathbf{k} \cdot \mathbf{r}_1}
\int \! d \mathbf{r}_2 \, F(\mathbf{r}_2) e^{- i \mathbf{k} \cdot \mathbf{r}_2} \mathcal{L}_\ell (\hat{\mathbf{k}} \cdot \hat{\mathbf{r}}_2) \right]  - P_\mathrm{sn}^{(\ell)} ~,
\end{align}
that assigns the varying LOS towards one of the galaxies of each pair. The Poisson shot noise term $P_\mathrm{sn}^{(\ell)}$ can be measured from the catalog and is subtracted from the monopole only, as for $\ell>0$ it is zero. However, the amplitude of the shot noise term is treated as a free parameter in our analyses, as described in more detail in section~\ref{sec:Practical-priors}. We measure the multipoles in bins of $\Delta k = 0.005\,h{\rm Mpc}^{-1}$  and make use of the scale-range $0.01\leq k\,[h{\rm Mpc}^{-1}]\leq 0.15$ for the analyses in this paper. 

\paragraph{Window function.}
The resulting power spectrum of eq.~\eqref{eq:FKP_pse} contains the effect of the survey selection function convolved with the actual galaxy power spectrum signal. In order to perform an unbiased analysis we need to include the effect of the survey selection in the theory model as well. We follow the formalism described in \cite{Wilson-ml-2015lup,Beutler-ml-2016arn} based on the Hankel transforms and implemented via FFT-log \cred{\cite{fftlog_paper}}, which relies on multiplying the Hankel transform of the theory-predicted power spectra multipoles by the window function pair-counts functions performed on the random catalogue,
\begin{align} \label{eq:Practical-window}
    W_\ell(s) = \frac{(2\ell+1)}{I_2\alpha_\mathrm{ran}^{-2}} \sum_{j>i}^{N_\mathrm{ran}} \frac{w_\mathrm{cp}(\mathbf{r}_i)w_\mathrm{FKP}(\mathbf{r}_i)w_\mathrm{cp}(\mathbf{r}_j+\mathbf{s})w_\mathrm{FKP}(\mathbf{r}_j+\mathbf{s})}{2\pi s^2 \Delta s} \mathcal{L}_\ell \left(\frac{\mathbf{x}_i\cdot \mathbf{s}}{x_i s} \right)~.  
\end{align}
The pair-count for each $s$-bin is normalized by the associated volume \cred{ given by $2\pi s^2 \Delta s$}, where $\Delta s$ is the binning size of the $s$-count and the $j>i$ condition prevents double counting pairs. The window function is normalized by $I_2\alpha_{\rm ran}^{-2}$ in order to account for the difference in number density between the random and data catalogue and to ensure the very same normalization as the power spectrum computed from eq.~\eqref{eq:FKP_pse}. Normalizing both eqs.~\eqref{eq:FKP_pse} and \eqref{eq:Practical-window} by the same $I_2$ factor prevents spurious leakage of the small-scale fluctuations of the random catalogue into the cosmological parameters, such as $\sigma_8$ or $A_s$, that typically could yield to systematic shifts \cite{deMattia-ml-2019vdg}. 

\subsection{Priors and likelihoods} \label{sec:Practical-priors}
Here we present our analysis choices for the two methods we aim to compare, the {\it ShapeFit} and the FM fit. The {\it ShapeFit} is performed in two steps.\footnote{Actually the {\it ShapeFit} only consists of the first step, but the second step is needed in order to compare both analysis types.} First, the physical parameters $\left\lbrace \alpha_\parallel, \alpha_\perp, f, m, n \right\rbrace$ are varied along with the nuisance parameters (compression step). Second, the results on physical parameters are treated as the new ``input data" and compared to any cosmological model of choice (cosmology inference step) {\colored As it is customary, in the cosmology inference step the full covariance between the compressed variables is included in computing the likelihood and the resulting parameters posterior is sampled via MCMC}. The FM fit consists of only one step, where the nuisance parameters are varied along with the cosmological parameters, while the physical parameters are not varied, since they are derived from the cosmological model. \colored{The fitting range in all presented runs is $0.01\leq k\,[h{\rm Mpc}^{-1}]\leq 0.15$}

\begin{table}[h]
    \centering
    \begin{tabular}{c|c|c c|c c}
      \hline \hline
       \multicolumn{2}{c}{Parameter} & \multicolumn{4}{|c}{Prior ranges} \\ \hline
       type & name & SF min & SF max & FM min & FM max \\ \hline
     \multirow{3}{*}{Cosmological} & $\omega_\mathrm{cdm}$  &  \multicolumn{2}{c|}{$[\mathrm{None}, \mathrm{None}]$} &  \multicolumn{2}{c}{$[\mathrm{None}, \mathrm{None}]$}\\
      & $h$  &  \multicolumn{2}{c|}{$[\mathrm{None}, \mathrm{None}]$} &  \multicolumn{2}{c}{$[\mathrm{None}, \mathrm{None}]$} \\
      & $\ln\left(10^{10} A_s\right)$ / $A^{1/2}$  & \multicolumn{2}{c|}{$[\mathrm{None}, \mathrm{None}]$} & \multicolumn{2}{c}{$[0.2, 2.0]$} \\ \hline
     \multirow{4}{*}{Physical}  & $\alpha_\parallel$  & \multicolumn{2}{c|}{$[0.5, 1.5]$} & \multicolumn{2}{c}{--} \\
    & $\alpha_\perp$ & \multicolumn{2}{c|}{$[0.5, 1.5]$} & \multicolumn{2}{c}{--} \\
     & $f$ & \multicolumn{2}{c|}{$[0, 3]$} & \multicolumn{2}{c}{--}  \\
     & $m$ & \multicolumn{2}{c|}{$[-3, 3]$} & \multicolumn{2}{c}{--}  \\ \hline
   \multirow{10}{*}{Nuisance}  & $b_1$ & \multicolumn{2}{c|}{$[0, 10]$} & \multicolumn{2}{c}{$[0, 10]$} \\
     & $b_2$ & \multicolumn{2}{c|}{$[\text{-} 10, 10]$} &  \multicolumn{2}{c}{$[\text{-} 10, 10]$} \\
     & $b_{s2}$ & lag.  & $[\text{-} 10, 10]$ & lag. & $[\text{-} 10, 10]$ \\
     & $b_{3\mathrm{nl}}$ & lag.  & $[\text{-} 10, 10]$ & lag. & $0$ \\
     & $c_0  \, [h^{-2} \mathrm{Mpc}^2]$ & \multicolumn{2}{c|}{--} & $0$ & $(0\pm30)$ \\
     & $c_2  \, [h^{-2} \mathrm{Mpc}^2]$ & \multicolumn{2}{c|}{--} & \multicolumn{2}{c}{$(0\pm30)$} \\
     & $c_4  \, [h^{-4} \mathrm{Mpc}^4]$ & \multicolumn{2}{c|}{--} & \multicolumn{2}{c}{$(500\pm500)$} \\
     & $\sigma_P \, [\mpcoh]$ & \multicolumn{2}{c|}{$[0, 10]$} & \multicolumn{2}{c}{--} \\
     & $A_\mathrm{noise}$ / $\Delta P_\mathrm{noise} \, [h^{-3} \mathrm{Mpc}^3]$ & \multicolumn{2}{c|}{$[\text{-} 5, 5]$} &  \multicolumn{2}{c}{$(0\pm5000)$} \\
    \end{tabular}
    \caption{Prior ranges for parameters used for the \textit{ShapeFit} and the FM fit. For each, we define a case with minimum (``min") and maximum (``max") freedom, where overlapping prior choices between the two choices are written in the center. Flat priors are given as $[\mathrm{min},\mathrm{max}]$, Gaussian priors are denoted as $(\mathrm{mean} \pm \mathrm{std})$. Parameters separated by ``/"  correspond to different conventions used between \textit{ShapeFit} and FM fit for the same physical effect, see text for details.}
    \label{tab:priors}
\end{table}

In table~\ref{tab:priors} we show the model parameters and prior choices for both methods, where the model used for the {\it ShapeFit} is based on the 1-loop SPT +TNS model introduced in section~\ref{sec:theory-galaxies-redshift-space} and the extensions described in section~\ref{sec:shapefit}. As a representative model of the FM fit approach we choose the EFT implementation of \cite{2020JCAP...05..042I}, which is also based on 1-loop SPT, but with a few differences.

It is well known that the BAO amplitude is affected by non-linear coupling to large scale displacements (bulk flows), that are hard to model within Eulerian PT (at the base of 1-loop SPT, which is used in this work).
In the FM approach this is done by implementing the so called ``Infrared (IR) resummation" effect, that can be well described within Lagrangian PT, via a phenomenological damping of the BAO amplitude.
Since there is no equivalent IR resummation correction in {\it ShapeFit} (at least not in this first implementation),  and including this effect in the FM fit broadens the constraints, we perform the {\it ShapeFit} to FM comparison by not including IR resummation in the FM fit, but we return to this point in appendix \ref{sec:IR}. {\colored It will become clear below that to see at a statistical significant level the effect of including or not the IR resummation for \textit{ShapeFit} a survey volume of $\sim 300\, [{\rm Gpc}h^{-1}]^3$ would be needed.}

The EFT model phenomenologically accounts for higher order non-linearities via the so called ``counterterms" parametrized by $\left\lbrace c_0,c_2,c_4 \right\rbrace$ (see \cite{2020JCAP...05..042I} for the explicit equations). In summary, $c_0$ effectively corrects for dark matter behaving differently than a perfect fluid on small scales (monopole only) and $\left\lbrace c_2, c_4 \right\rbrace$ take into account non-linear RSD (quadrupole only). While in the {\it ShapeFit} we use the non-linear RSD prescription of \cite{Taruya-ml-2010mx} (TNS model) in combination with a phenomenological Lorentzian damping parametrized by $\sigma_P$ (see eq.~\eqref{eq:TNS}), the counterterms $\left\lbrace c_2, c_4 \right\rbrace$ are coefficients of a $2^\mathrm{nd}$ order Taylor expansion of the phenomenological damping describing the non-linear redshift space distortions. Hence, the EFT implementation of non-linear RSD is equivalent to our {\it ShapeFit} implementation, but with more freedom (2 parameters instead of 1). 

Another difference is the interpretation of bias parameters, that in the \textit{ShapeFit} incorporate an implicit scaling with $\sigma_{s8}$, while in the EFT fit they scale with the primordial fluctuation amplitude as $A^{1/2}$, where
\begin{align}\label{eq:Priors_As_convention}
    A^{1/2} = (A_s/A_s^\mathrm{Planck})^{1/2}~, \qquad A_s^\mathrm{Planck} = 2.0989 \times 10^{-9}  ~.
\end{align}  

Yet another difference concerning nuisance parameters is the convention for treating shot noise. While the EFT implementation uses a Gaussian prior on the difference between the shot noise with respect to Poisson shot noise $\Delta P_\mathrm{noise} = P_\mathrm{noise}-P_\mathrm{Poisson}$, we implement a flat prior on the fractional difference $A_\mathrm{noise}$, the amplitude of the Poisson-like, scale independent shot noise contribution.
We have tested that this does not make any difference in the posterior distributions.

Considering these differences in model assumptions between \textit{ShapeFit} and FM fit we adopt two different nuisance parameter choices represented by a minimum freedom (``min") and a maximum freedom (``max") choice. The ``min" convention is oriented towards the fiducial setup of most classic RSD analyses, where the non-local bias parameters are fixed by the local Lagrangian (``lag.") prediction \cite{Baldauf_2012,Saito-ml-2014qha},
\begin{align} \label{eq:lagrangian_prediction}
    b_{s2} = -\frac{4}{7} (b_1 - 1), \qquad b_{3\mathrm{nl}} = \frac{32}{315} (b_1-1)~.
\end{align}
In the maximum freedom case $b_{s2}$, $b_{\rm 3nl}$ and $b_1$ are treated as independent parameters. 
We employ these relations also in the FM ``min" case and also fix the counterterm $c_0$ to zero, in order to match the \textit{ShapeFit} configuration. However we keep varying the counterterms $c_2, c_4$, as they are related to non-linear RSD, which in the \textit{ShapeFit} is parametrized by $\sigma_P$ as described above. The ``max" convention is oriented towards the fiducial setup of \cite{2020JCAP...05..042I}, where all counterterms are varied and the Lagrangian relations are relaxed. However, the third order non-local bias $b_{3\mathrm{nl}}$ is set to zero in the EFT implementation, because it is very degenerate with the monopole counterterm $c_0$. We do the same here, since we try to stick to the default configuration of \cite{2020JCAP...05..042I} as close as possible. However, we vary $b_{3\mathrm{nl}}$ in the {\it ShapeFit}, in order to compensate for the fact, that $c_0$ is not an ingredient of our model. Further tests of the {\it ShapeFit} concerning modelling choices of non-local bias parameters are shown in sections~\ref{sec:results:cosmo}, \ref{sec:Nseries} and in appendix \ref{sec:Shapefitdependencies}.

Regarding the cosmological parameters, we choose a similar setup as in the baseline analysis of \cite{2020JCAP...05..042I} for the \textsc{Patchy} mocks varying the parameters given in the top rows of table~\ref{tab:priors}.
We fix the baryon density to the value of the simulation \colored{$\ob=0.02214$} and do not take into account a varying neutrino mass, since the \textsc{Patchy} mocks were run with massless neutrinos. 
Concerning the primordial fluctuation amplitude $A_s$, for the FM fit we adopted the convention of \cite{2020JCAP...05..042I} varying $A^{1/2}$ given in eq.~\eqref{eq:Priors_As_convention}. However, for the step of cosmological inference from the compressed {\it ShapeFit} results we adopted a flat prior on $\ln\left(10^{10} A_s\right)$. This different choice does not affect our cosmological results at all.

\section{Results} \label{sec:results}

The results of our fiducial analysis on the mocks described in section~\ref{sec:Practical} is presented in two parts. 
First, we present the results of the parameter compression step comparing the \textit{ShapeFit} with the classic RSD method (section~\ref{sec:results:compression}). Afterwords,   assuming a $\Lambda$CDM model, we compare the cosmological analysis of the compressed  \textit{ShapeFit} results to the model's parameter constraints obtained with the FM method (section~\ref{sec:results:compression}). 
We also show extensions to our fiducial analysis by adding more cosmological parameters. In particular, we compare the performance of \textit{ShapeFit} and FM fit when varying $\ob$ and $n_s$ in sections \ref{sec:results:ob} and \ref{sec:results:ns} respectively.

For the mock ``data" we always use the mean of 2048 \textsc{Patchy} ``ngc\_z3" mocks, where we rescale the covariance to the volume 100 times  one of these mocks. This represents a factor 10 times larger than most previous analyses, and significantly larger than the volume of any single tracer or sample of forthcoming surveys. As it will be clear below, by choosing to calibrate the covariance for such a large volume we will see systematic shifts in some parameters which would have gone otherwise unnoticed. These shifts highlight the limitations of the current \cred{modelling} of non-linearities (see section~\ref{sec:theory-galaxies-redshift-space}), nevertheless, they  are still  below the $1\sigma$ expected statistical uncertainty for forthcoming surveys.

\subsection{Parameter compression: \cred{Classic} RSD vs. \textit{ShapeFit}} \label{sec:results:compression}
\begin{figure}[t]
    \centering
    \begin{minipage}[h]{0.495\textwidth}
    \centering
    Minimal Freedom \\
    \includegraphics[width=\textwidth]{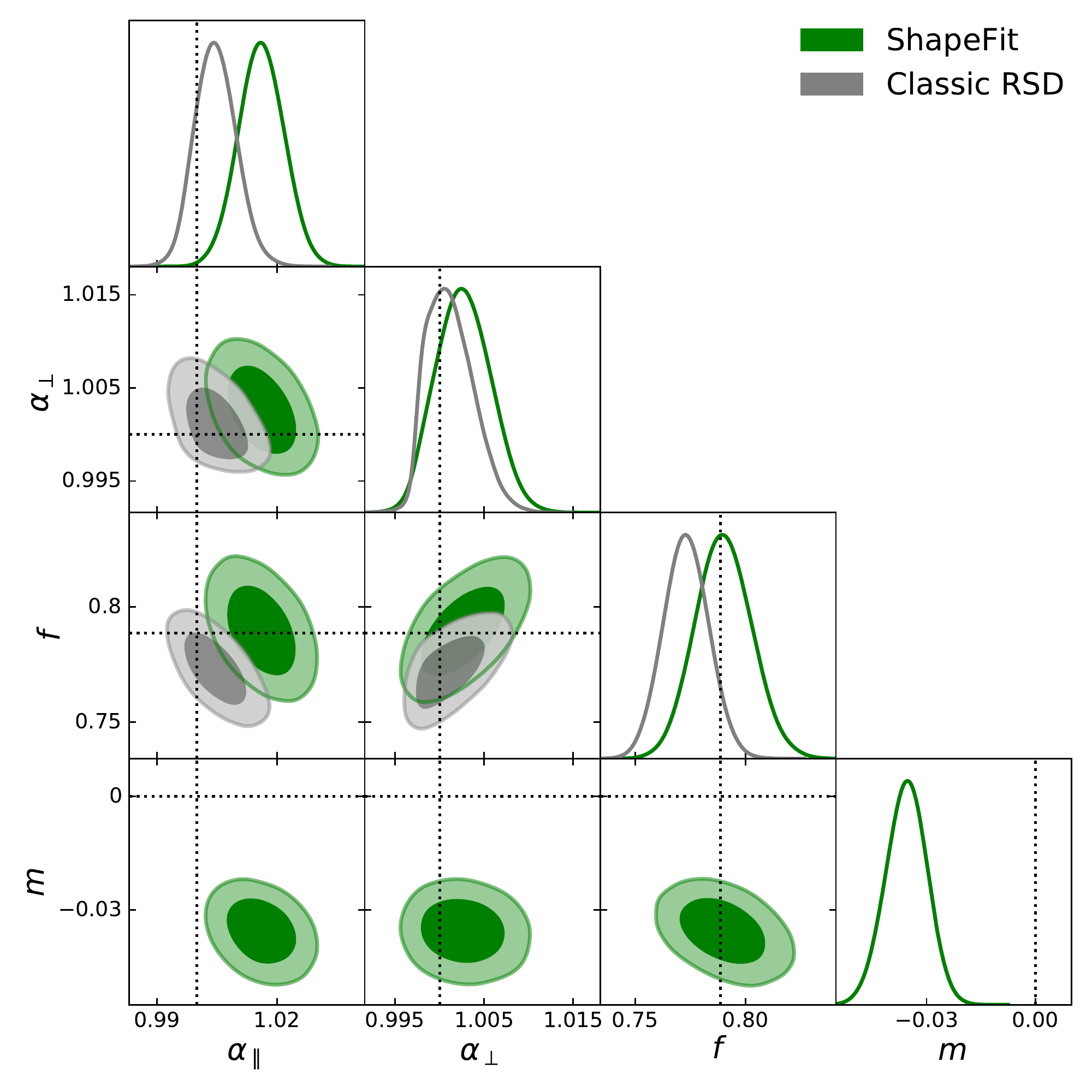}
    \end{minipage}
    \begin{minipage}[h]{0.495\textwidth}
    \centering
    Maximal freedom \\
    \includegraphics[width=\textwidth]{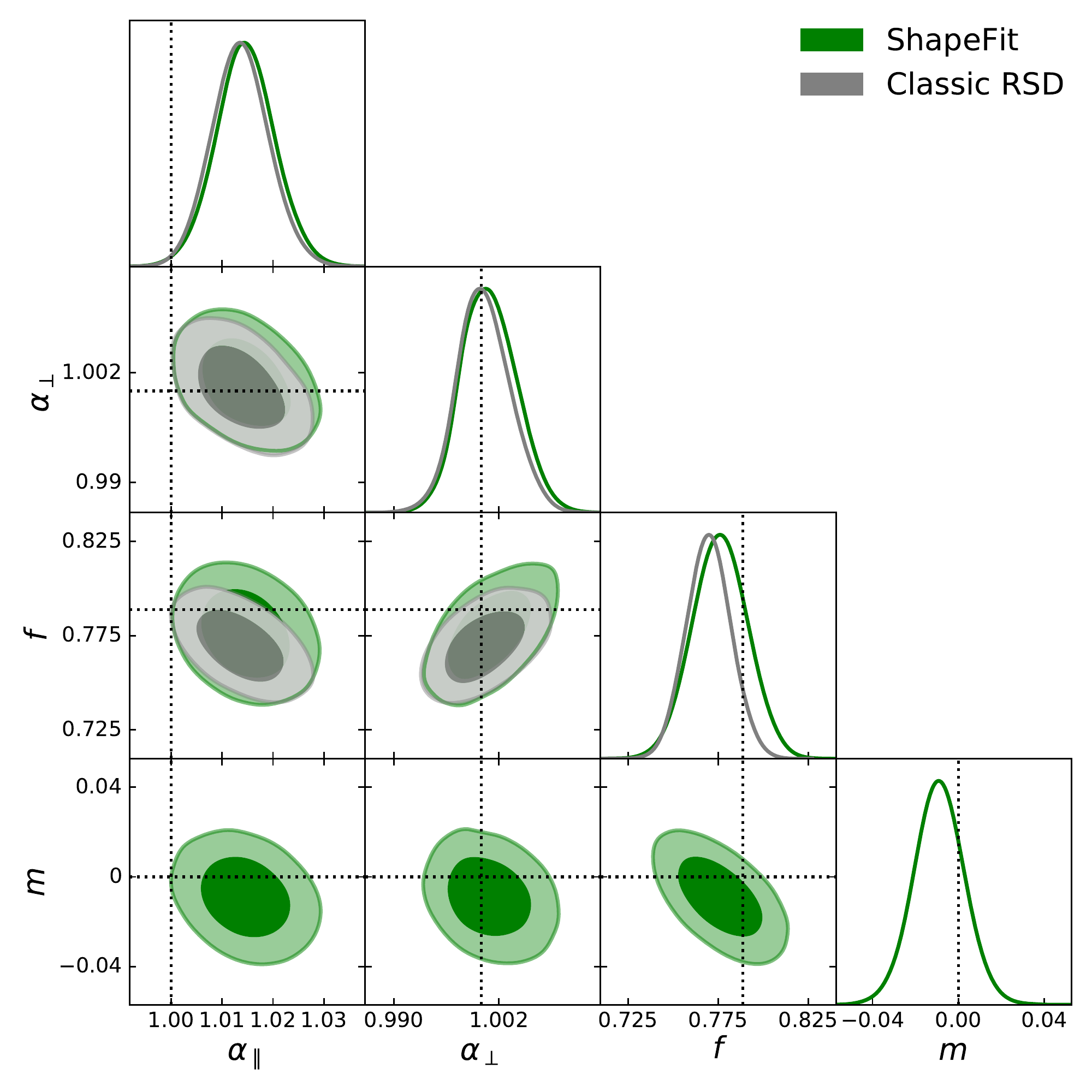}
    \end{minipage}
 \caption{Results of the classic RSD fit (grey) and the \textit{ShapeFit}  (green) applied to the \textsc{Patchy} ``ngc\_z3" sample, showing the ``min" (minimal freedom, non-local bias parameters  to follow the Lagrangian prediction) case on the left; and the ``max" (maximal freedom, fully free non-local bias parameters) case on the right panel. For the ``min" case we report a systematic deviation of the slope parameter from the expectation $m=0$, indicating that assumptions about the biasing scheme may introduce systematic shifts even at very large and linear scales. {\colored To make explicit and quantify biases in recovered parameter estimates, the error-bars are relative to an effective volume of $300\, [h^{-1}{\rm Gpc}]^3$.}}
    \label{fig:shapefit-result}
\end{figure}

We fit the classic RSD and \textit{ShapeFit} models to the mean of the \textsc{Patchy} ``ngc\_z3" mocks using the physical and nuisance parameters of table~\ref{tab:priors}. In both cases we use a template corresponding to the \textsc{Patchy} cosmology of eq.~\eqref{eq:patchy_cosmo}, where the slope parameter $m$ is varied in the \textit{ShapeFit} only, while it is fixed to $m=0$ in the classic RSD fit by definition. We perform the fits for the ``min" and the ``max" conventions, where the non-local bias parameters $b_{s2}$  and $b_{\rm 3nl}$ are either fixed to their Lagrangian prediction of eq.~\eqref{eq:lagrangian_prediction} or allowed to vary freely. The results for these cases are shown in figure~\ref{fig:shapefit-result} in the left and right panels, respectively, where grey contours correspond to the classic RSD Fit and green contours to the \textit{ShapeFit}. The dashed lines indicate the underlying parameter values of the simulation.

{\colored The uncertainties on $\alpha_{\parallel}$ and $\alpha_{\perp}$ are very similar, almost indistinguishable,  in the ``classic" and {\it ShapeFit} approaches. On the other hand, {\it ShapeFit} recovers slightly larger errors on $f$ than the classic approach. This can be understood by considering that while $m$ shows no significant correlation with the $\alpha$ parameters,  $m$ and $f$ are somewhat  correlated at least for a sample with ``ngc\_z3" features.} 
In the ``min" case (see figure~\ref{fig:shapefit-result} left panel) we observe that the classic RSD constraints are closer to the theoretical prediction than the \textit{ShapeFit} constraints. This is because when we enforce the non-local biases to follow the Lagrangian prediction, the constraint on $m$ experiences a systematic shift towards  $m=-0.036\pm0.006$, hence being  formally in $6\sigma$ tension with the expectation, $m=0$. This shift in $m$ appears to be a much better fit to the data (bestfit $\chi_{m\neq0}^2 = 51$) than in the classic RSD case, where $m$ is forced to zero (bestfit $\chi_{m=0}^2 = 95$). Once we allow the non-local bias parameters to vary freely (``max" case, right panel), we recover $m=0$ and the constraints on the other physical parameters show  very good agreement between classic RSD Fit and \textit{ShapeFit}. Indeed, the bestfit $\chi^2$ for both types of fit is very similar in the max case ($\chi_{\rm max}^2 \approx 44$) with a difference of only $\Delta \chi^2 = 0.02$ between classic RSD Fit and \textit{ShapeFit}.

This indicates that even seemingly reasonable and well-motivated assumptions about bias can induce systematic errors in recovered cosmological parameters by affecting  clustering even on very large, linear scales, and in particular  when the slope $m$ is used for cosmological interpretation.  
This finding highlights the importance of having a \cred{modelling} of bias as flexible as possible when interpreting the scale-dependence of clustering at all scales.
We speculate that  \textsc{Patchy} mocks may have some bias signature not fully consistent with the local Lagrangian bias scheme. This signal was not evident in the classic RSD analyses, but when we allow $m$ to vary this becomes important. We anticipate not finding such behaviour in the N-body galaxy mocks describing a similar set of galaxies (as we will see in section~\ref{sec:Nseries}).

In the ``max" case, there are remaining biases on $\alpha_\parallel$ (of order $2\sigma$) and $f$ (of order $1\sigma$), 
%
%
for an effective survey volume of $300 \,\mathrm{Gpc}^3$;  such \cred{a} large volume \cred{yields} statistical errorbars with similar size as the model uncertainty. 
%
%
%
The \cred{modelling} of non-linearities and redshift-space distortions will likely be improved before the on-going and future surveys are  completed and ready for cosmological interpretation. Still, our results indicate that one must be careful with model assumptions, in particular about the galaxy bias model, as its choice can have a significant impact on the measured slope $m$. We perform more tests on different sets of N-body simulations and galaxy mock catalogs in section~\ref{sec:additional_tests}.   {\colored Nevertheless the right panel of  figure~\ref{fig:shapefit-result} demonstrates that {\it ShapeFit} recovers the standard ``classic" compressed parameters with effectively the same uncertainties as the ``classic" approach. The degradation of the constraints on $f$ introduced by the extra parameter $m$ (due to a small degeneracy between $f$ and $m$), is minimal, $\sim 20\%$ for a volume of $300\, [h^{-1}{\rm Gpc}]^3$, which is expected to decrease for smaller, more realistic survey volumes (for a DESI-like volume of $30\, [h^{-1}{\rm Gpc}]^3$  the degradation decreases to $5\%$).} 

\subsection{Cosmological results: \cred{full modelling} vs. \textit{ShapeFit}}\label{sec:results:cosmo}
\begin{figure}[t]
    \centering
    \includegraphics[width=\textwidth]{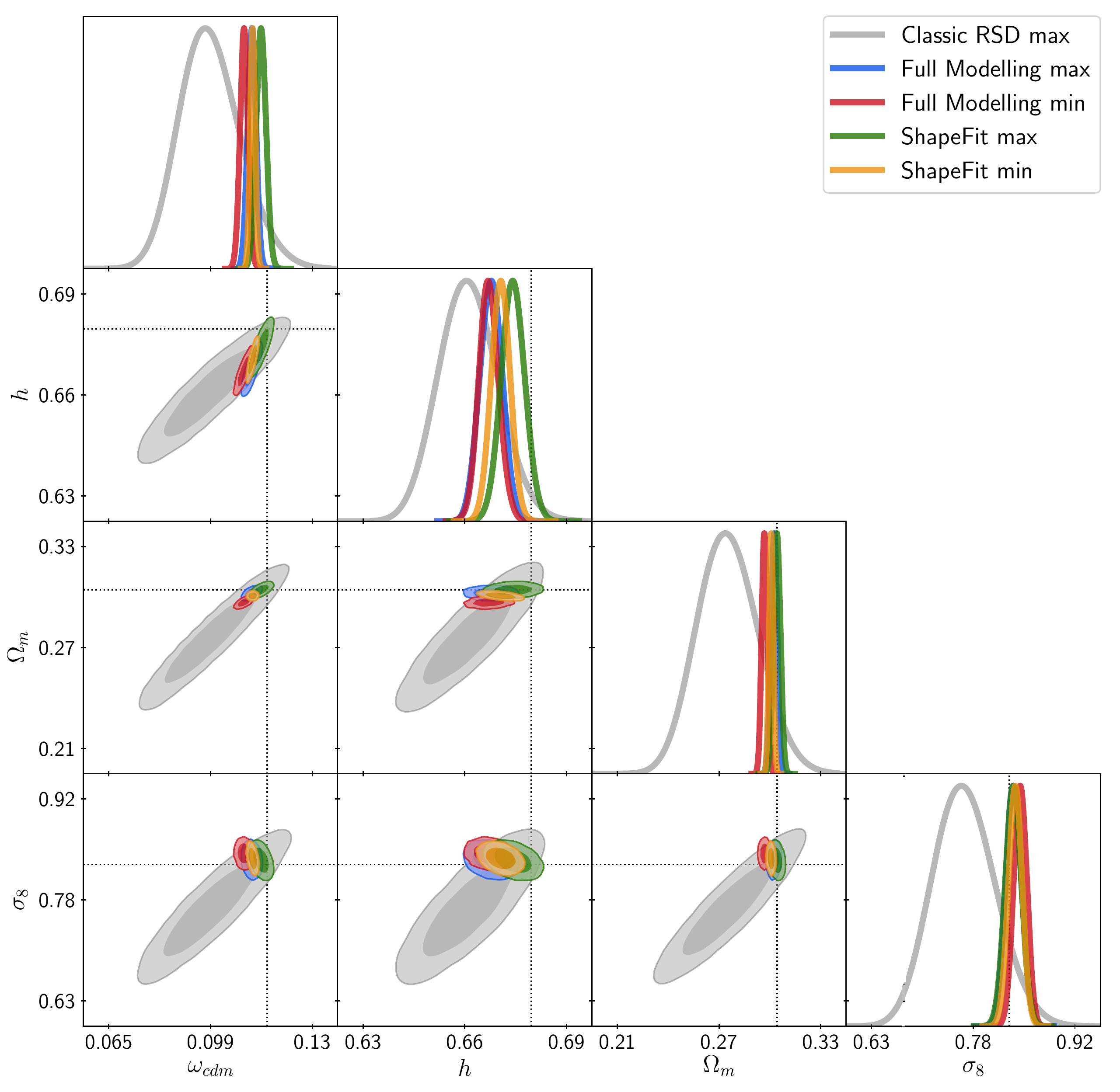}
 \caption{Results of the \textit{ShapeFit} and FM fits to the \textsc{Patchy} ``ngc\_z3" sample \cred{compared to the classic RSD "max" results for reference}. The size of the \cred{constraints} are very similar in the two cases, \cred{indicating} that \textit{ShapeFit} captures the bulk of cosmological information captured by FM. The systematic shifts associated to the FM contours are partly caused by neglecting the IR resummation correction, which modulates the BAO amplitude, but also broadens the constraints. \textit{ShapeFit} does not use any BAO amplitude information, and therefore does not need to crucially account for any IR resummation correction (although it could be easily incorporated). The fact that the size of the constraints is very similar in the two cases indicates that the cosmological information enclosed in the BAO amplitude is subdominant to the one enclosed in the large-scale shape of the power spectrum.
 }
    \label{fig:EFT-Shape-comparison-noIR}
\end{figure}

Figure~\ref{fig:EFT-Shape-comparison-noIR} displays the cosmological results on $\ocdm,\, h$ and the derived parameters, $\Om,\, \sigma_8$ obtained from interpreting the \textit{ShapeFit} results within a $\Lambda$CDM model as explained in section~\ref{sec:shapefit-cosmo}, as well as from the direct FM fit \cred{(for comparison we added the classic RSD results (grey contours) as well, but also see \cite{Brieden-ml-2021cfg})}. In this case we keep $\ob$ fixed at the mock expected value (see section~\ref{sec:results:ob} for results varying $\ob$). Again, we explore the effect of local-Lagrangian bias assumption through the ``min" and ``max" cases defined in table~\ref{tab:priors}. The FM fit results are displayed in blue (``max") and red (``min"), while the \textit{ShapeFit} results are represented by the green (``max") and orange (``min") contours. {\colored We also present the results in table~\ref{tab:results}, where we added the case of a more realisic, DESI-like-survey volume of $30\, [h^{-1}{\rm Gpc}]^3$ by scaling the covariance of one \textsc{Patchy} mock realisation by 10 (labeled $(V \times 10)$). As table~\ref{tab:results} indicates, considering the latter covariance} \textit{ShapeFit} recovers the cosmological parameters very well in the ``max" case \colored{(within $0.5\sigma$)}, while the results of the ``min" case are clearly biased \colored{(by up to $2\sigma$)}. In the ``max" case constraints are, unsurprisingly, weaker than in the ``min" case. The same degradation in cosmological parameter constraints between the ``min" and ``max" cases is observed for the FM fit, albeit it  shows larger biases in the recovered  parameter values.

\begin{table}[t]
\centering
\begin{tabular}{l|c|c|c|cc|cc}
\multirow{2}{*}{$\mathbf{\Omega}$} & \multirow{2}{*}{\textbf{Case}} & \multirow{2}{*}{\textbf{Fit}} & \multirow{2}{*}{\textbf{Mean}} & \multicolumn{2}{c|}{\textbf{Error, $V=3({\rm Gpc}/h)^3$}} & \multicolumn{2}{c}{\textbf{Bias/$\sigma_\mathbf{\Omega}$, $V=3({\rm Gpc}/h)^3$ }} \\
 & & & & $(V\times 100$) & ($V\times 10$) & ($V\times 100$) & ($V\times 10$) \\
\hline \hline \multirow{6}{*}{$\omega_{\mathrm{cdm}}$} & \multirow{3}{*}{min} & RSD & $0.1162$ & $0.0110$ & $0.0400$ & $-0.25$ & $-0.10$  \\
& & SF & $0.1139$ & $0.0009$ & $0.0027$ & $-5.57$ & $-1.86$ \\
& & FM & $0.1110$ & $0.0014$ & $0.0040$ & $-5.65$ & $-1.97$ \\ \cline{2-8}
& \multirow{3}{*}{max} & RSD & $0.0997$ & $0.0105$ & $0.0372$ & $-1.75$ & $-0.69$ \\
& & SF & $0.1167$ & $0.0017$ & $0.0045$ & $-1.24$ & $-0.50$ \\
& & FM & $0.1136$  & $0.0015$ & $0.0046$ & $-3.79$ & $-1.21$ \\ \hline \hline
\multirow{6}{*}{$h$} & \multirow{3}{*}{min} & RSD & $0.6751$ & $0.0079$ & $0.0271$ & $-0.35$ & $-0.12$ \\
& & SF & $0.6695$ & $0.0026$ & $0.0079$ & $-3.15$ & $-1.02$ \\
& & FM & $0.6665$ & $0.0027$ & $0.0079$ & $-4.00$ & $-1.38$\\ \cline{2-8}
& \multirow{3}{*}{max} & RSD & $0.6611$ & $0.0079$ & $0.0268$ & $-2.21$ & $-0.75$ \\
& & SF & $0.6729$ & $0.0032$ & $0.0102$ & $-1.50$ & $-0.49$  \\
& & FM & $0.6670$ & $0.0031$ & $0.0088$ & $-3.45$ & $-1.26$  \\ \hline \hline
\multirow{6}{*}{$\Omega_{\mathrm{m}}$} & \multirow{3}{*}{min} & RSD & $0.3031$ & $0.0174$ & $0.0620$ & $-0.24$ & $-0.08$  \\
& & SF & $0.3035$ & $0.0012$ & $0.0039$ & $-2.78$ & $-0.93$  \\
& & FM & $0.2996$ & $0.0017$ & $0.0049$ & $-4.42$ & $-1.57$ \\ \cline{2-8}
& \multirow{3}{*}{max} & RSD & $0.2783$ & $0.0178$ & $0.0623$ & $-1.70$ & $-0.54$  \\
& & SF & $0.3069$ & $0.0021$ & $0.0056$ & $-0.11$ & $-0.04$  \\
& & FM & $0.3052$ & $0.0019$ & $0.0057$ & $-1.01$ & $-0.33$ \\ \hline \hline
\multirow{6}{*}{$\sigma_{8}$} & \multirow{3}{*}{min} & RSD & $0.815$ & $0.044$ & $0.159$ & $-0.33$ & $-0.11$\\
& & SF & $0.838$ & $0.010$ & $0.029$ & $0.88$ & $0.30$ \\
& & FM & $0.845$ & $0.009$ & $0.029$ & $1.69$ & $0.55$  \\ \cline{2-8}
& \multirow{3}{*}{max} & RSD & $0.765$ & $0.044$ & $0.153$ & $-1.48$ & $-0.53$\\
& & SF & $0.835$ & $0.012$ & $0.033$ & $0.50$ & $0.18$  \\
& & FM & $0.836$ & $0.012$ & $0.032$ & $0.57$ & $0.21$ \\
 \end{tabular} \\
    \caption{\colored{This table shows parameter constraints for $\mathbf{\Omega} = \lbrace \ocdm, h, \Om, \sigma_8 \rbrace $ given by the corresponding mean $ \bar{\Omega}$, error $\sigma_\Omega$ and bias $(\bar{\Omega} - \Omega_\mathrm{Patchy})$ divided by $\sigma_\Omega$ with respect to the \textsc{Patchy} cosmological parameters $\mathbf{\Omega}_\mathbf{Patchy} = \lbrace 0.118911, 0.6777, 0.301175, 0.8288 \rbrace$. For the different bias model cases ``min" and ``max" we compare the results of our RSD, \textit{ShapeFit} (here abbreviated as SF) and FM fits. We carried out the fits using a covariance matrix corresponding to the volume of 100 stacked mocks $(V \times 100)$ and to 10 stacked mocks $(V \times 10)$, where $V = 3(h^{-1}\mathrm{Gpc})^3$. The mean values cited here are obtained from the $(V \times 100)$ runs, as these are more Gaussianly distributed. They do not necessarily coincide with the mean values of the $(V \times 10)$ runs due to non-Gaussianity, but we have checked that best-fits agree with each other. Also note that we present Gaussianized errors, although in fact they are slightly non-Gaussian, which is consistently taken into account for determining the bias. }}
    \label{tab:results}
\end{table}

The reason why the FM fit results do not recover the \textsc{Patchy} cosmology, is, at least in part, due to the systematic error in modelling the BAO wiggle amplitudes arising from neglecting IR effects.

From 1-loop Lagrangian PT it is well known that large scale bulk flows lead to a damping of the BAO amplitude. However, within Eulerian PT, which operates at the level of density field instead of displacement field, this effect is hard to model. The state-of-the-art attempt to model the large scale displacements within Eulerian PT,  known as IR resummation, is to phenomenologically damp only the BAO wiggles, while leaving the broadband unchanged. The magnitude of this damping effect is inferred from theory and is, hence, highly model-dependent, \colored{which may lead to an underestimation of the error bars as shown in \cite{Hinton-ml-2019nky}.} This is the reason why we designed the \textit{ShapeFit} in such a way that it only extracts cosmological information from the BAO position, the overall power spectrum normalization, and the power spectrum slope; but not from the BAO amplitude. 
In order to have significant cosmological information on the amplitude of the BAO, we require a survey volume significantly larger than $300\,{\rm Gpc}^3$, and this is not expected to be available in the next decade. Nevertheless, one could extend the \textit{ShapeFit} by an additional parameter governing the BAO amplitude; we leave this for future work. 

In appendix~\ref{sec:IR} we test the FM fit including the appropriate IR resummation correction finding that this helps to recover the expectation values of $\ocdm, h$ and $\Om$ within 1-$\sigma$ and $\sigma_8$ within  2-$\sigma$, albeit broadening the resulting constraints.
However it is important to recall that the \textsc{Patchy} mocks are not N-body and  simulate non-linearities in an approximate way. Hence, it is not guaranteed that the IR resummation scheme appropriately "describes" these mocks. 

From \colored{table \ref{tab:results}} we  notice that the size of the posterior constraints on the relevant cosmological parameters are very similar: \textit{ShapeFit} captures the same bulk of the cosmological information  extracted by the FM fit. This shows that the extra information captured by FM compared to the classic approach is concentrated on large-linear scales and is indeed the  early-time physics imprint left on the matter transfer function. 

\subsection{Impact of varying $\omega_\mathrm{b}$} \label{sec:results:ob}
In the previous sections we have fixed the $\ob$ parameter to its \textsc{Patchy} expected value, \colored{$\ob=0.02214$}. This is motivated by the fact that usually the constrains from LSS alone on this parameter are not competitive with CMB or BBN ones. However, in some cases we may want to perform an analysis with no external prior constrains. In this section we present results obtained considering $\ob$ as a parameter free to vary with a uniform prior  in the range $[0.005,0.04]$. 

\begin{figure}[t]
    \centering
    \begin{minipage}{0.495\textwidth}
    \includegraphics[width=\textwidth]{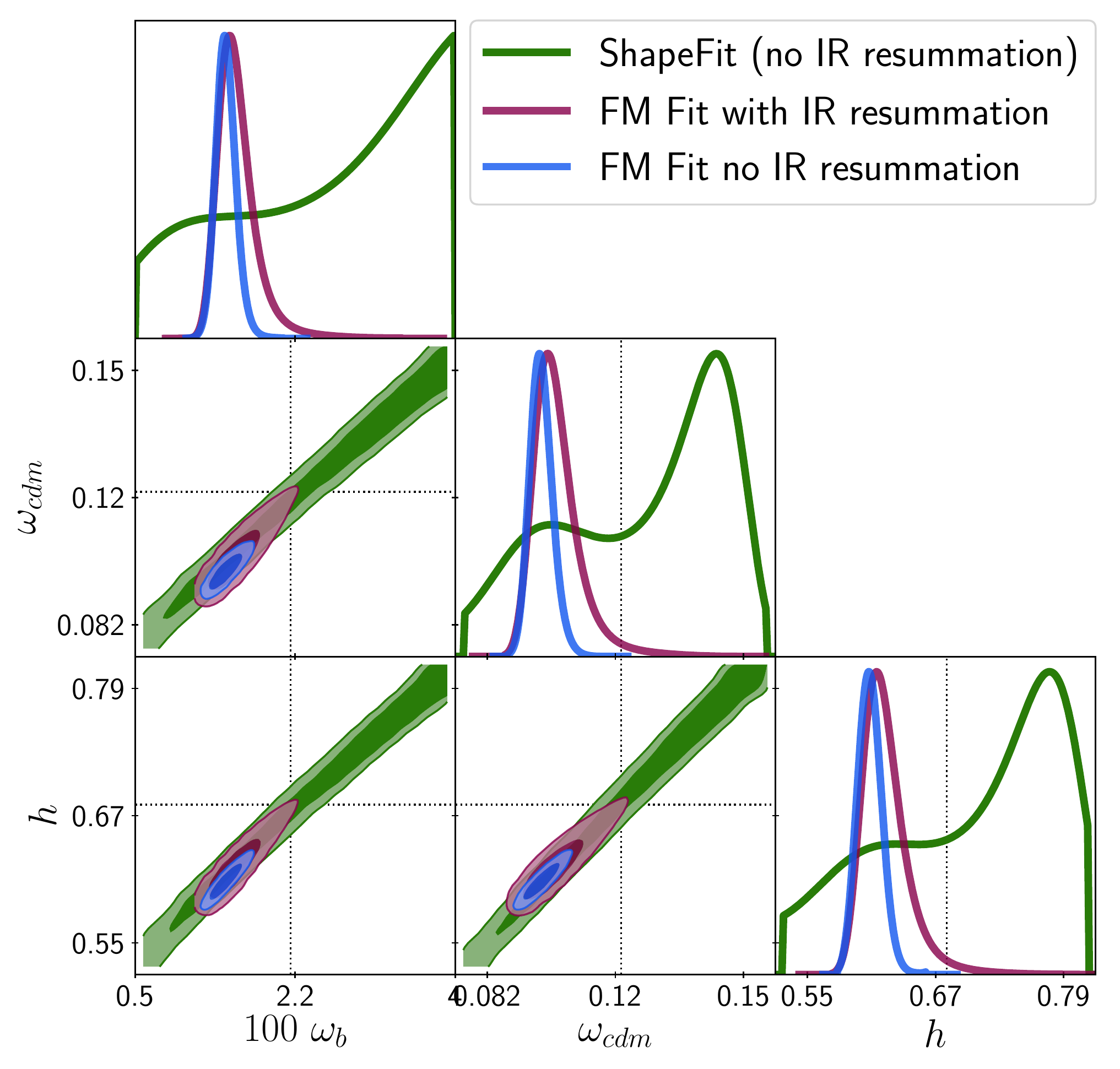}
    \end{minipage}
     \begin{minipage}{0.495\textwidth}
    \includegraphics[width=\textwidth]{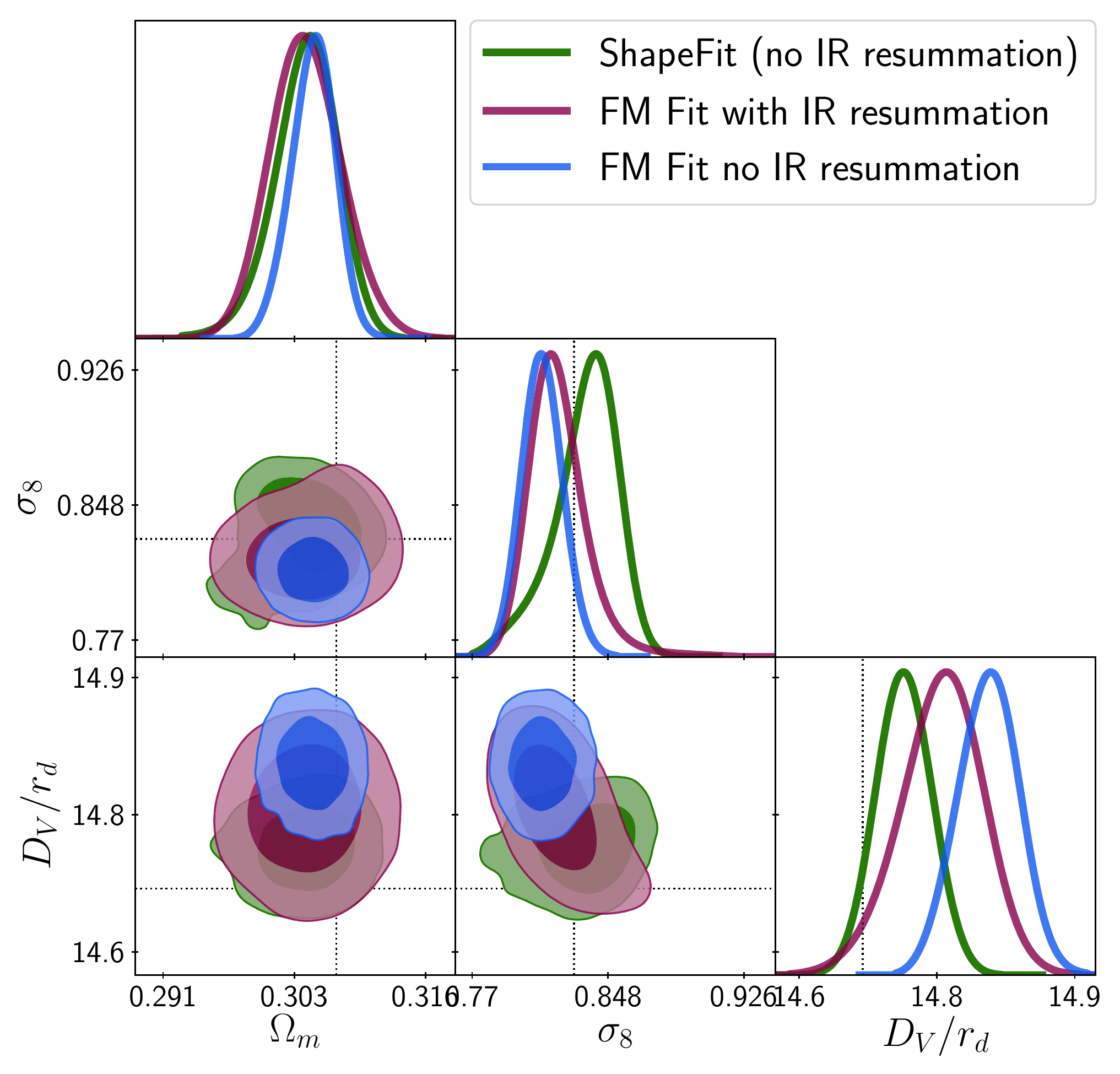}
    \end{minipage}   
 \caption{Results when \cred{analysing} the \textsc{Patchy} ``ngc\_z3'' sample with free $\ob$ within a flat prior range of $[0.005,0.04]$. The green contours display the results from {\it ShapeFit}, whereas the blue and purple contours correspond to the FM fit case, with (purple) and without (blue) the IR resummation correction (see text for more details). Since {\it ShapeFit} does not compress any BAO amplitude feature, it is not able to break the degeneracies between $\omega_{\rm cdm},\,\omega_{b},\,$ and $h$ \colored{(left panel)}, and in this case only cred{constrains} individually $\Omega_m$, $\sigma_8$ \colored{and $D_V/r_d$ (right panel)}. On the other hand, \cred{the} FM fit does use the BAO amplitude and therefore is able to break such degeneracies. However, for both cases explored here, the inferred \cred{constraints result} biased with respect to the expected values. This suggest that the BAO amplitude feature is not yet a reliable probe to be used in LSS analyses. } \label{fig:varying-ob}
\end{figure}

Figure~\ref{fig:varying-ob} shows the \textit{ShapeFit} (green) and FM fit (blue) results focusing on the ``max" case only. Since the slope $m$ is degenerate with $\ob$ and $\ocdm$, \textit{ShapeFit} does not constrain them individually, resulting in fully degenerate bands. Also the constrain on $h$  depends  strongly on the $\ob$-prior, as $h$ can only be measured from the BAO position once $\ocdm$ is fixed by the slope. Therefore, for \textit{ShapeFit} \colored{the parameters shown in the left panel} $\ob,\ocdm$ and $h$ are effectively unconstrained. Instead, it constrains the (derived) parameters $\Om$, $\sigma_8$ \colored{and $D_V/\rd$ shown in the right panel}.  

\begin{figure}[t]
    \centering
   \includegraphics[width=\textwidth]{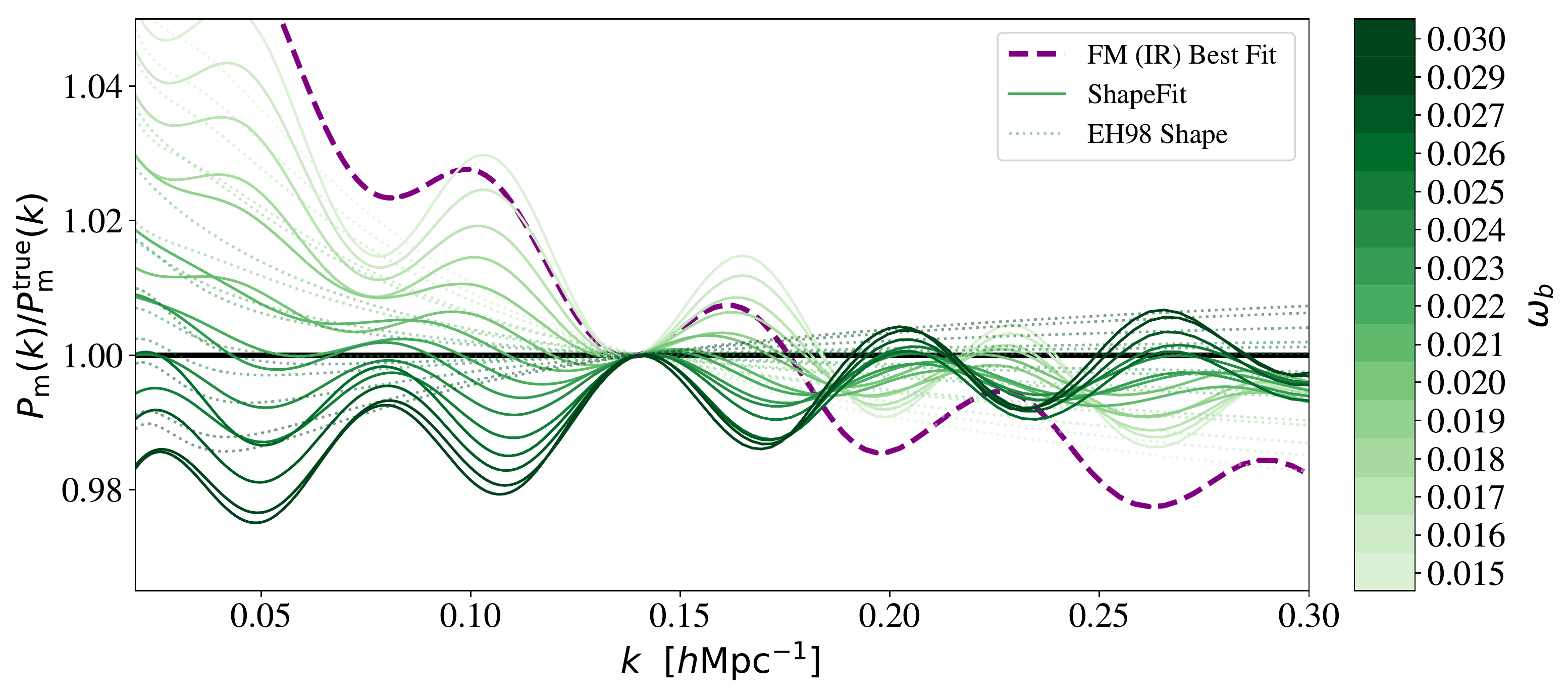}
 \caption{\colored{Linear matter power spectrum ratio $P_\mathrm{m}(k)/P_\mathrm{m}^\mathrm{true}(k)$ on scales $0.02<k\,[h\mathrm{Mpc}^{-1}]<0.3$ for different models: The dashed purple line corresponds to the FM bestfit (purple contours in figure~\ref{fig:varying-ob}), green lines are obtained from the green contours for different values of $\ob$. Green dotted lines show the corresponding EH98 approximation. For improved visibility, the curves are rescaled in amplitude to match $\sigma_8$.}}
 \label{fig:Pklin-ocdm-h-degeneracy}
\end{figure}

The additional constraining power of the FM fit comes from the BAO amplitude only, which breaks the degeneracy between $\ob$ and $\ocdm$.
Once the prior on $\ob$ is relaxed, the constraints on $\ob$ are
driven by the BAO amplitude, which depends linearly on $\ob/\ocdm$. 

\colored{To show this explicitly, we consider the linear matter power spectrum obtained from parameter combinations that follow the $\ob-\ocdm$ degeneracy of {\it ShapeFit} in figure~\ref{fig:varying-ob}. If the power spectra obtained along this degeneracy show any difference, this means that there is some information loss induced by the {\it ShapeFit} compression, that needs to be investigated. Figure \ref{fig:Pklin-ocdm-h-degeneracy} displays the linear power spectra for different values of $\ob$ (green lines), where all other cosmological parameters are read from the {\it ShapeFit} chain such as to represent the bestfits for each value of $\ob$. As expected, nearly all curves share the same values of $\Om$, $h \times\rd$ and hence also $D_V/\rd$, such that their BAO wiggle positions overlap with each other. 
However, there is a remaining difference in tilt between the curves as can be seen from the EH98 (dotted) lines. This is, because the fitting formula introduced in eq.~\eqref{eq:shapefit-m} is not optimized for varying $\ob$. As shown in figure~\ref{fig:m-param-dependence} our method reproduces the $\ob$-behaviour with $\sim 5\%$ precision for the range $0.015 < \ob < 0.03$ (which corresponds to a $40$-$\sigma$ region considering the BBN measurement of $\ob$).  
}

\colored{Nevertheless, it is obvious from figure~\ref{fig:Pklin-ocdm-h-degeneracy} that most of the differences between the green curves are encoded in the BAO amplitude.} It is important to note that the BAO amplitude is an early-time physics imprint which, however, is heavily processed by late-time effects (e.g., non-linearities, mode-coupling, bias). This is the reason why the FM fit delivers biased results when these late-time effects are not taken properly into account (see the no-IR modelling case represented by blue contours). The effect of including the IR resummation correction in the FM fit (purple contours) is to broaden the posteriors, but not changing their peaks maxima. As a result the posteriors are still biased, but the broadening reduces the tension with the expected values at $2\sigma$. We also observe that the inclusion of IR resummation correction is to broaden the contours precisely along the degeneracy direction given by the green ones. Therefore, we conclude, constraints on $\ob, \ocdm$ and $h$ from FM alone, without any prior from early-time measurements, are not reliable at better than 25\% for $\ob$, 10\% for $\ocdm$ and 10\% for $h$, when obtained with any of the state-of-the-art methods and \cred{modelling} explored here.

\subsection{Impact of varying $n_s$}\label{sec:results:ns}
\begin{figure}[t]
    \centering
   \includegraphics[width=\textwidth]{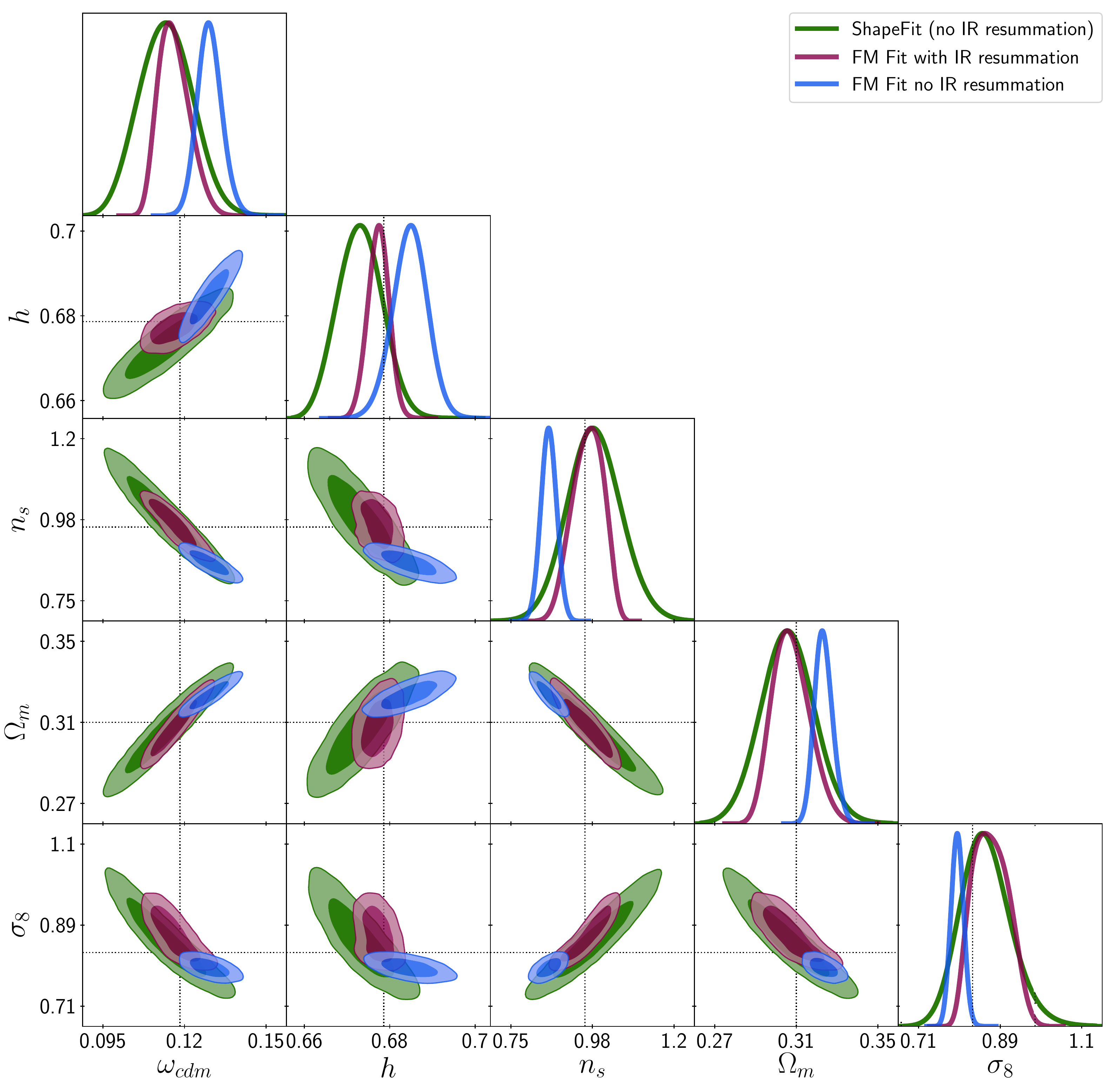}
 \caption{Derived posteriors for a flat-$\Lambda$CDM model (for a fixed $\omega_b$) for \textit{ShapeFit} (green contours) and the FM fit to the \textsc{Patchy} ``ngc\_z3'' sample. The FM results are shown for the IR resummation correction turned on and off, in violet and blue contours, respectively. The differences between {\it ShapeFit} and FM posteriors are due to the extra constraining power from the BAO-peak amplitude (not implemented in {\it ShapeFit}) which helps to break degeneracies along $\omega_{cdm} - n_s$. The BAO-damping effect due to non-linear bulk flows, which IR resummation describes within FM, greatly reduces the BAO-amplitude-based constraining power, and hence broadens the $n_s$ posteriors. }
 \label{fig:varying-ns}
\end{figure}

The constraining power of \textit{ShapeFit} when adding the scalar tilt $n_s$ as an additional free parameter in comparison with the FM fit is described here. In this case, \textit{ShapeFit} needs to be run with both slope parameters, $m$ and $n$, introduced in section~\ref{sec:shapefit-pk}. Since the scale-independent and the scale-dependent slopes have a similar effect on the power spectrum, $m$ and $n$ are strongly anti-correlated. See the full parameter degeneracies when varying $m$ and $n$ in figure~\ref{fig:full_triangle_for_Nseries_mn} of appendix~\ref{sec:Shapefitdependencies} for the Nseries mocks for reference. Here, in the case of the \textsc{Patchy} mocks, we do not show the \textit{ShapeFit} results on physical parameters for conciseness and focus on the cosmological results.

The green contours of figure~\ref{fig:varying-ns} display the posteriors derived from {\it ShapeFit} for the parameters of a flat-$\Lambda$CDM model. As in section~\ref{sec:results:ob}, the results for \cred{the} FM fit are shown for the cases where IR resummation correction is included (violet contours), and when \cred{it} is not (blue contours). Both \textit{ShapeFit} and FM + IR resummation  approaches recover the \textsc{Patchy} cosmology very well (marked with black dotted lines). On the other hand, the FM fit results without IR resummation are clearly biased, suggesting that the inclusion of IR resummation is crucial for the FM fit. 
 This can be understood by considering that
 $n_s$ and $\ocdm$ have similar effects on the power spectrum slope  and thus for \textit{ShapeFit} these two parameters are highly correlated. The FM fit uses the additional information provided by  BAO-peak amplitude  in order to break the degeneracy between $n_s$ and $\ocdm$\cred{, hence} the blue and violet contours are much narrower than the green contours. \colored{IR resummation affects the amplitude of the BAO signal, hence it is crucial to extract unbiased constraints from this signal. {\it ShapeFit} makes no use of the amplitude of the BAO \cred{and is hence} insensitive to this.}

Similar to the case when varying $\ob$, we conclude that FM fit constraints on $n_s$ depend on \cred{the} non-linear BAO damping model (late-time physics), while $\textit{ShapeFit}$ provides a conservative alternative, where the BAO amplitude is not used, such that the constrains on $\ocdm$ and $n_s$ are purely driven by the slope (early-time physics).
We note, that in this \cred{case we} fixed $\ob$ to its underlying value. If we allow to vary both $\ob$ and $n_s$ for the FM fit, the bias in $\ob$ observed in figure~\ref{fig:varying-ob} propagates into a biased result on $n_s$, even when the IR resummation correction is applied.

\section{Additional systematic tests on N-body catalogues} \label{sec:additional_tests}
\colored{There is no perfect  suite of mocks, all mocks are in some way an idealization of the survey and/or introduce some approximations. It is important to check whether results found on mocks are robust and not due to  approximations introduced e.g., in the mock generation.}  
\textsc{Patchy} mocks explored in section~\ref{sec:results} are not full N-body mocks, hence they are not appropriate for determining the systematic error budget of classic RSD analyses, and  even less  suited for the  newly introduced shape parameter $m$. \colored{In other words, the number of available simulations and the total effective volume they cover, makes  the \textsc{Patchy} mocks an invaluable resource to estimate things like error-bars. However, to investigate potential residual systematics of the proposed {\it ShapeFit} implementation, we prefer to resort to N-body mocks. }

In this section we present further tests of {\it ShapeFit} using two set of full N-body simulations. We want to  investigate further  the potential systematic errors on $m$, seen in the \textsc{Patchy} mocks,  in particular those under galaxy bias conditions such as local-Lagrangian assumptions (e.g., the left panel of figure~\ref{fig:shapefit-result}).

\subsection{Systematic tests on dark matter particles in real space: \cred{geometric} effects}\label{sec:DM}
We first focus on \cred{analysing} the simplest possible case of a set of dark matter particles in real space without survey geometry or  selection function. We use a N-body suite of 160 simulations with a flat $\Lambda$CDM cosmology consistent with the Wilkinson Microwave Anisotropy Probe bestfit cosmology (WMAP cosmology, \citep{2013ApJS..208...19H}), with a box size $L = 2.4\, {\rm Gpc}h^{-1}$ and a total number of $N_p=768^3$ particles. The initial conditions have been generated at $z=49$ by displacing the particles according to the second-order Lagrangian PT from their initial grid points. We use the output at the three snapshots, $z=0.5$, $1.0$ and $1.5$. Further details on these simulations can be found in section 3.1 and table 1 of \citep{gil-marin_dark_2014}. Although we have the velocity information for each of the simulated particles, we do not apply in this case any redshift-space distortion displacement for simplicity. 

We obtain the data-vector of each of these 160 realizations at each redshift bin, consisting of its monopole, quadrupole and hexadecapole signals between $0.02\leq k\,[{\rm Mpc}h^{-1}]\leq 0.15$, sampled in bins of $\Delta k=0.01\,h{\rm Mpc}^{-1}$ size, and with a total number of $13\times3$ elements. We take the average of the 160 data-vector realizations to form the mean data-vector  to use as our dataset. The total associated volume of this data-vector corresponds to $6,448\,{\rm Gpc}^3$. We make use of the 160 realizations to estimate the covariance, following the same corrections as described in section~\ref{sec:Practical-mocks}. \colored{In this section, covariance (and errors on the figures) are rescaled to be those corresponding to the full effective volume available. In fact the goal is to explore small systematic shifts, which we want to uncover and quantify with the maximum precision afforded by the simulations available, independently, for now, of the statistical power of specific surveys.}  We fit the redshift-space distortion model of eqs.~\eqref{eq:Pmodel} and \eqref{eq:TNS} with non-local bias parameters and  with the local-Lagrangian conditions of \eqref{eq:lagrangian_prediction}. We set $n=0$ for simplicity and only focus on exploring the posteriors of $\{\alpha_\parallel,\,\alpha_\perp,\,f,\,m\}$, when the rest of four nuisance parameters are also marginalized, $\{b_1,\,b_2,\,\sigma_P,\,A_{\rm noise} \}$. Note that since we are fitting the data-vector corresponding to dark matter particles in real space, and we will be using the reference template at the true own cosmology, we expect to recover $\alpha_{\parallel,\,\perp}=1$, $f=0$ and $m=0$. 

\begin{figure}[t]
    \centering
    \includegraphics[width=\textwidth]{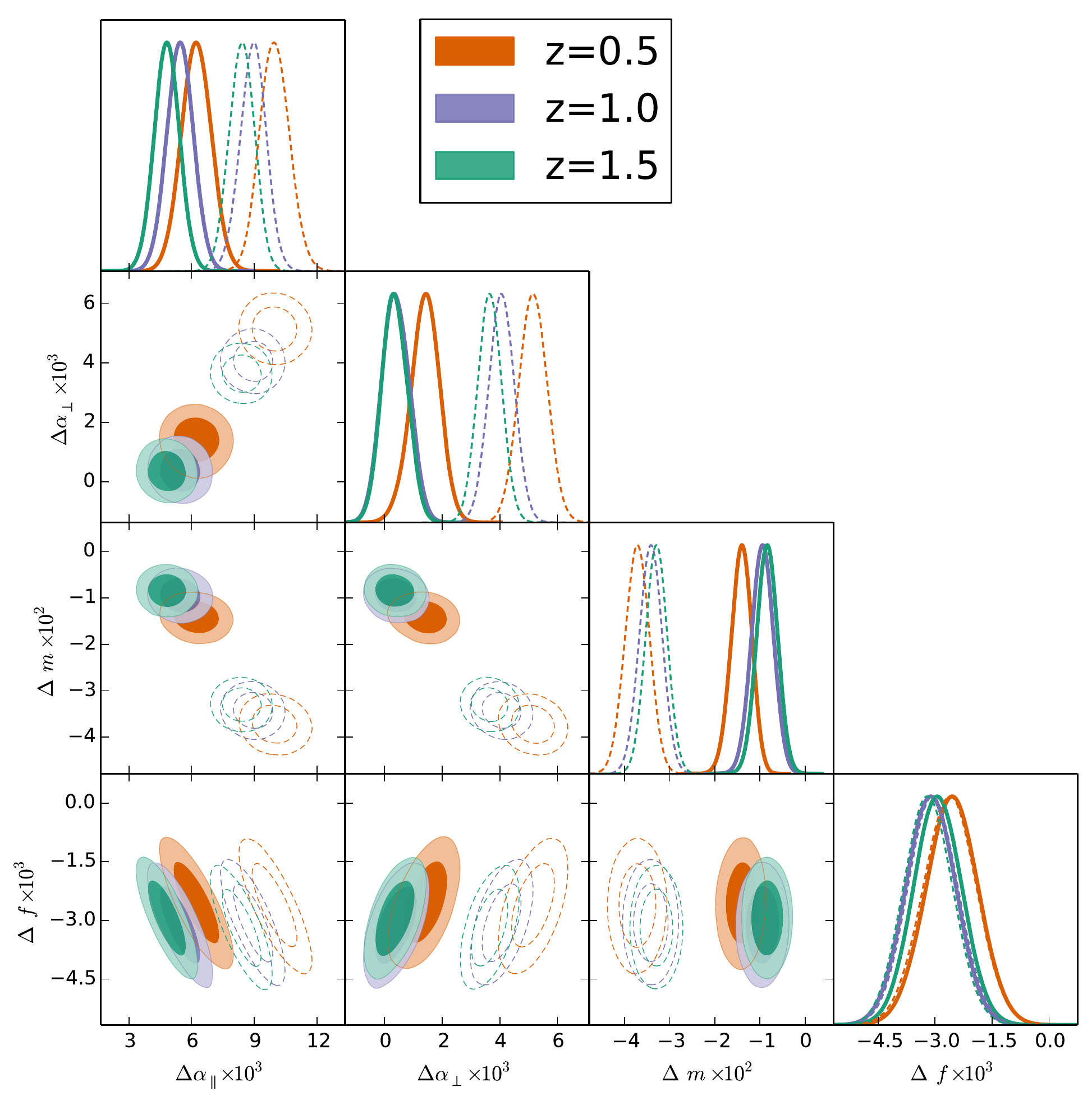}
  \caption{Posteriors derived when fitting the mean of 160 full N-body dark matter realization in real space. {\colored The covariance has been rescaled to correspond to an effective volume of $2200\, [h^{-1}{\rm Gpc}]^3$}. We have used the model described in eqs.~\eqref{eq:Pmodel} and \eqref{eq:TNS} and the local-Lagrangian bias scheme of eq.~\eqref{eq:lagrangian_prediction}. In the plot we only show the physical parameters ${\bf p}=\{\alpha_\parallel,\,\alpha_\perp,\,f\,,m\}$, although the remaining four nuisance parameters are also varied (see text). The empty-dashed contours display the results of fitting the data-vector without any geometric correction (as usually done), whereas the filled contours account for the geometric correction through eq.~\eqref{eq:Gcorrection}. For all ${\bf p}$-variables the expected value is $\Delta {\bf p}=0$. We observe that $m$ is especially sensitive to the geometric correction with a shift of about 0.03 towards positive values, for the specific set of geometric choices: $\Delta k=0.01\,h{\rm Mpc}^{-1}$, size of the periodic box, $L=2.4\,h^{-1}{\rm Gpc}$, and the $k$-range fitted, $0.02\leq k\,[h{\rm Mpc}^{-1}]\leq 0.15$.} 
    \label{fig:sys_m}
\end{figure}

The dashed-empty contours of figure~\ref{fig:sys_m} show the posteriors of such analysis, for the 3 redshift bins in different colors, as labeled. We only show the difference between the measured and the expected value for the 4 relevant physical parameters, accordingly scaled as indicated by the legend for visualization purposes. 
We notice that even in this highly idealized, simple case systematic shifts are present:  
$\Delta_{\alpha_\parallel}^{\rm sys}\simeq 0.01\, (1\%)$, $\Delta^{\rm sys}_{\alpha_\perp}\simeq 0.005\, (0.5\%)$, $\Delta^{\rm sys}_f\simeq0.003$ and $\Delta^{\rm sys}_m\simeq-0.04$.
Being redshift-independent  indicates that 
these shifts  are not related to a theoretical limitation of the PT-model, or any biasing model assumption, but likely a geometric effect.

In fact, as it is standard procedure, for each $k_i$-bin (defining the power spectrum band-power), the model is evaluated at the effective $k$-vector of that $k$-bin, $P^{\rm model}(k_{i\,,{\rm eff}})$, where $k_{i\,,{\rm eff}}\equiv\langle {\bf k} \rangle_i$ is the ensemble average over all possible directions of the ${\bf k}$-vector within the $i$-bin. However, the data-vector is measured by taking the average of $P({\bf k}_i)$ across all ${\bf k}$-directions, $P^{\rm data}(k_i) = \langle P^{\rm data}({\bf k}) \rangle_i $. It is clear that $P^{\rm data}(k_i)$ and $P^{\rm model}(k_{i\,,{\rm eff}})$ are not representing the same quantity: $P(\langle k\rangle)\neq \langle P(k)\rangle$. 
\colored{It is mode discreteness that generates this effect. For \cred{a} sufficiently large box the mode discreteness would be negligible, it is the survey geometry that introduces it and fully specifies it, hence the name "geometric effect"}.

Ignoring this effect can generate spurious signals, especially for the large-scale modes, where the number of modes per bin is small. Ideally we would like to evaluate the model's prediction for each $\bf k$ within the bin and  take the  average in the same way as when  measuring the data-vector, $\langle P^{\rm model}( {\bf k}_i)\rangle$. However this is too computationally expensive to be adopted in a MCMC.

We propose here an approximation to account for this effect  which is fast and sufficiently accurate for our purposes: we add the resulting ``mean" effect to the data.

We construct a new data-vector, $P^{{\rm data,}\, G^{-1}}$:
\begin{equation}
   P^{{\rm data,}\, G^{-1}}(k)= G^{-1}(k)\times P^{\rm data}(k),
\end{equation}
where the geometric factor, \cred{$G$, is defined as}
\begin{equation}
    G(k) \equiv \frac{\langle P^{\rm model}({\bf k}) \rangle}{P^{\rm model}(\langle {\bf k} \rangle)},
    \label{eq:Gcorrection}
\end{equation}
and where the ensemble average is taken over all directions of the ${\bf k}$-vector. 
Naturally $G(k)$ depends slightly on the parameters at which the model, $P^{\rm model}$ is evaluated,\footnote{Note that one could also leave the data-vector intact and apply this inverse correction into the model: $P^{{\rm model,}\, G}=G(k)\times P^{\rm model}(k)$ at each MCMC step, where $G$ would be pre-computed at a given fiducial model in order to save computational time. This approach may seem more physically motivated but is mathematically identical to what we follow.} so $G$ is obtained  through several  iterations, until we observe convergence (in practice  one or two iterations suffice). We start by fitting the data-vector without any correction ($G(k)=1$), evaluating $G(k)$ at the best fitting model (first trial correction), and first trial $ P^{{\rm data},\, G^{-1}}(k)$. We repeat this process until we observe convergence in the derived posteriors. We have found convergence is reached by the second iteration.

Following this two-iteration process approach we produce the posteriors displayed in figure~\ref{fig:sys_m} in solid contours. We notice how the systematic offsets observed initially (empty-dashed contours i.e., $G(k)=1$) are significantly reduced \colored{(by a factor 2 to 4 depending on the variable)} when we account for the geometric correction: $\Delta^{\rm sys}_{\alpha_\parallel}\simeq0.005\, (0.5\%)$, $\Delta^{\rm sys}_{\alpha_\perp}\simeq0\, (0\%)$, $\Delta^{\rm sys}_f=0.003$ and $\Delta^{\rm sys}_m=-0.01$. The geometric correction is particular important for $m$,  neglecting it induces a systematic  shift of $\Delta m\simeq -0.03$. The geometric effect is less important for  $\alpha_{\parallel\,,\perp}$-- neglecting it induces a  systematic shift of  $0.5\%$--
 and negligible for $f$. 
\colored{ We estimate that the residual bias on $m$ after the proposed approximate geometric correction becomes comparable to the statistical error  for an effective volume greater than  $\sim 400 {\rm Gpc}^3$. Without our proposed mean correction the bias would become comparable to the statistical error for volumes $\sim 25{\rm Gpc}^3$. Hence this correction is important and sufficient for on-going and forthcoming surveys.}
We conclude that such type of ``geometric" corrections may be important when doing precision cosmology, especially for signals  on large scales (i.e.,  $m$-derived quantities and \cred{the} FM fit approach).
It is important to keep in mind  that the systematic shifts reported in figure~\ref{fig:sys_m}  depend on the size of the chosen $k$-bin - the smaller the size of the bin, the smaller the required $G$-correction - as well as the size of the box in which the sample is embedded - the larger the box\cred{,} the smaller the $G$-correction - (see next section). In general, these shifts only set a floor for the type of systematics we expect in real-life applications. Inaccuracies when modelling galaxy bias schemes and redshift space distortions, may increase the systematic errors reported in figure~\ref{fig:sys_m}. We address these \cred{effects} in the following sub-section.

\subsection{Systematic tests on Nseries LRG mocks}\label{sec:Nseries}

\begin{figure}[t]
    \centering
    \includegraphics[scale=0.32, trim={400 0 50 0}, clip=false]{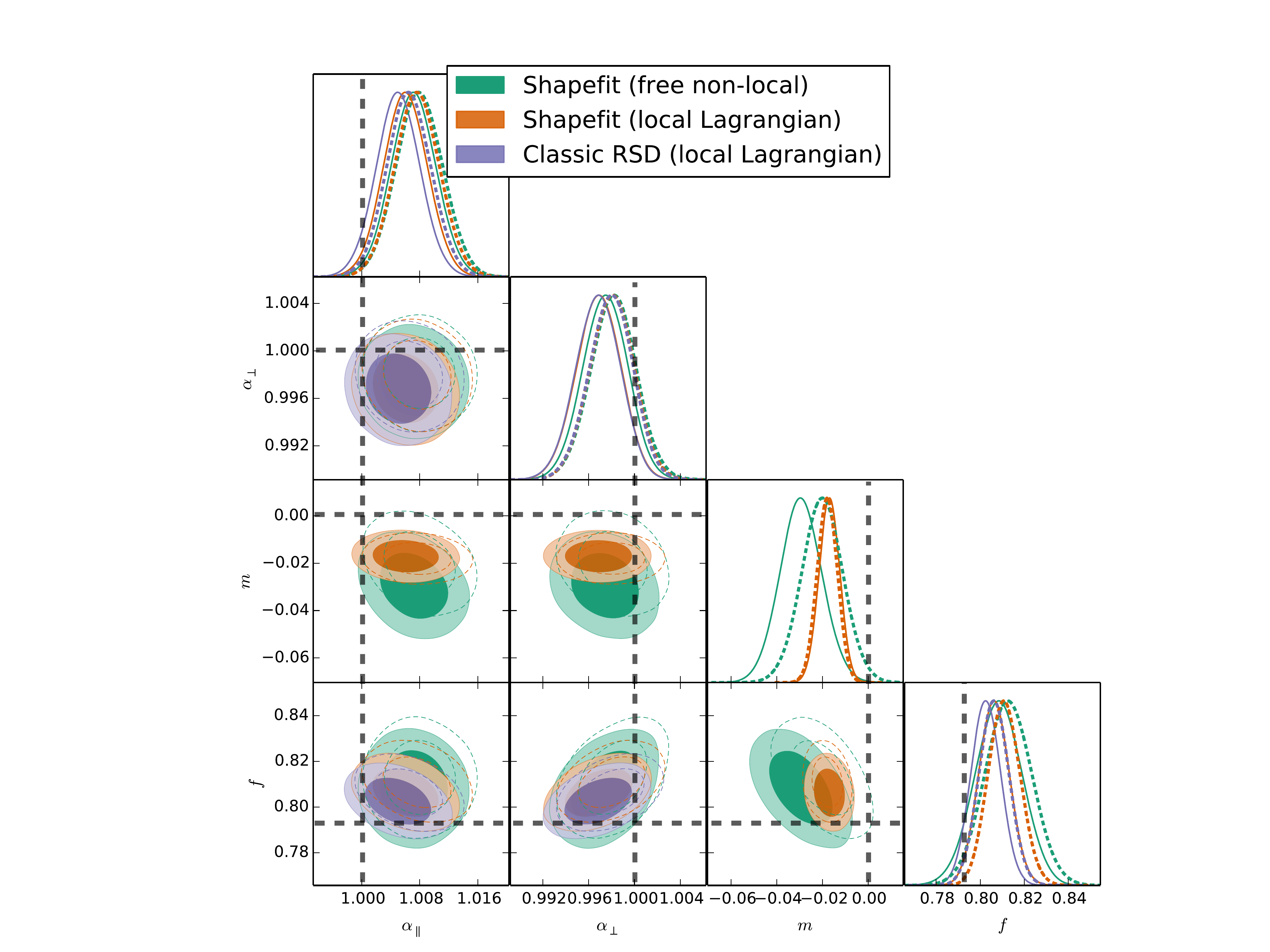}
     \includegraphics[scale=0.4, trim={50 0 100 0}, clip=false]{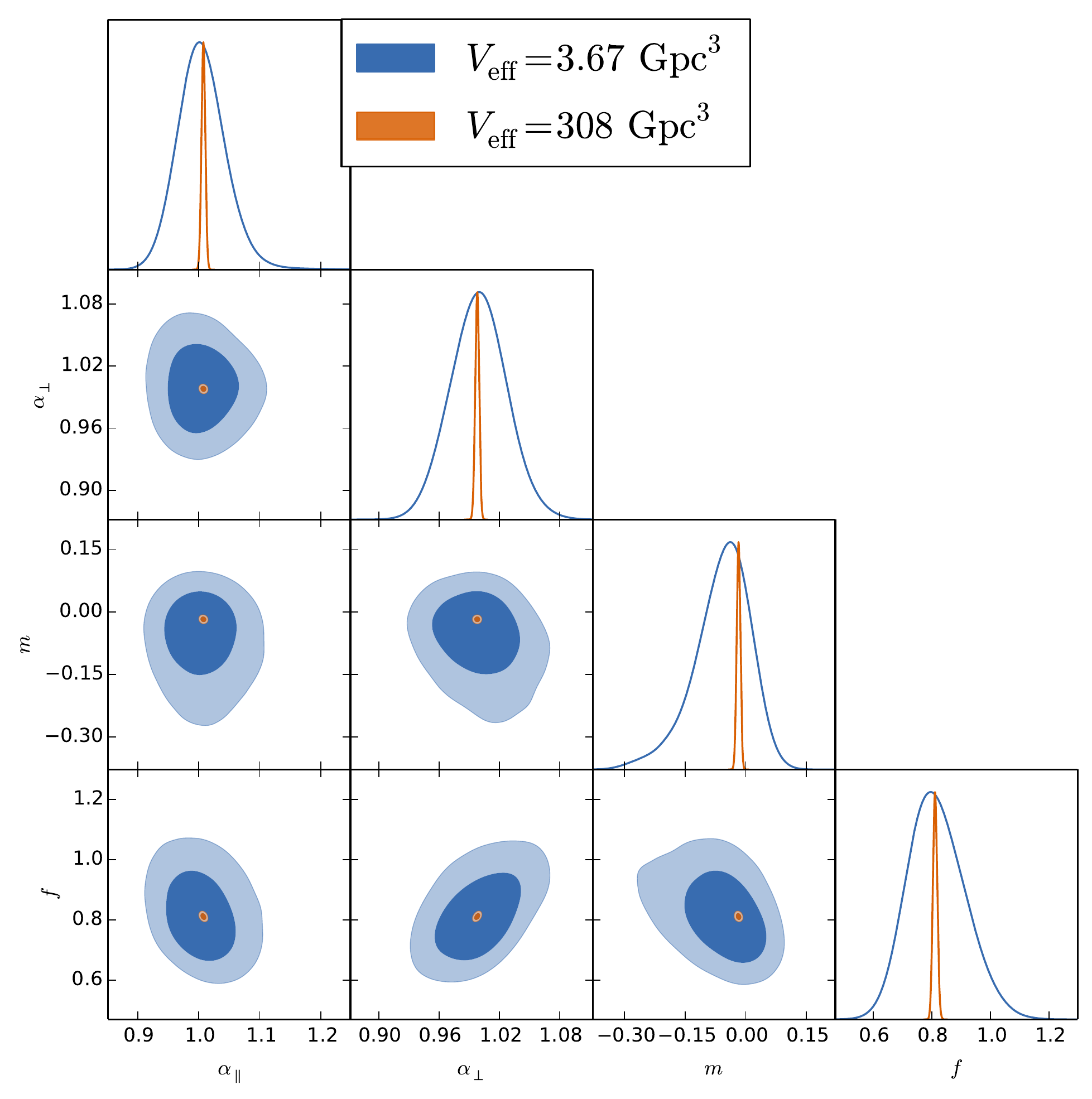}

 \caption{\cred{Posteriors} from Nseries N-body mocks using $P^{(0,2,4)}$ at $0.02\leq k\,[{\rm Mpc}h^{-1}]\leq0.15$. In the left panel we display the comparison of classic RSD (purple) and {\it ShapeFit} (orange), both with local-Lagrangian bias parametrization of eq.~\eqref{eq:lagrangian_prediction}, and {\it ShapeFit} with free non-local biases (green). All cases correspond to the fit of the mean power spectra of the 84 independent realizations with an associated effective volume of $V_{\rm eff}=308\,{\rm Gpc^3}$). Horizontal black-dashed lines represent the true expected value for each compressed variable. The empty-dashed contours display the results without any geometric correction ($G=1$) and the filled-solid contours with this correction included (see eq.~\eqref{eq:Gcorrection} and text for details). The right panel displays the case of {\it ShapeFit} with the local-Lagrangian bias assumption when fitting the mean of the 84 realizations (orange, same to the left panel) and when fitting the averaged-posteriors of all 84 individual realizations (blue contours), which in this last case represents an effective volume similar to BOSS CMASS NGC, $3.67\,{\rm Gpc}^3$. We note how the small systematic errors detected in the left panel, are negligible compared to the real-life statistical errors derived when fitting the realistic case of BOSS CMASS with a volume of $3.67\,{\rm Gpc}^3$ shown in blue in the left panel. }  
 \label{fig:sys2_m}
\end{figure}

We now explore how the compressed parametrization of {\it ShapeFit} performs when it is applied to full N-body galaxy mocks. We employ the Nseries galaxy mocks which have been  used for determining the modelling systematic error budget in BOSS \cite{alam_clustering_2017} and eBOSS \citep{eboss_collaboration_dr16} official RSD and BAO analyses.

The Nseries mocks have been generated out of 7 independent periodic boxes of $2.6\, h^{-1}{\rm Gpc}$ side, projected through 12 different orientations and cuts,  to extract, in total, 84 pseudo-independent realizations with the sky geometry similar to the northern galactic cap of CMASS DR12 data, for $0.43<z<0.7$, resulting in   
an effective redshift of 0.56. 
The  mass  resolution  is $1.5\times10^{11}\,M_\odot h^{-1}$, with $2048^3$ dark matter particles per box. The identified haloes are populated with galaxies following a halo occupation distribution model
tuned to match the clustering of LRGs observed by BOSS.
As in the standard procedure adopted by the  BOSS collaboration, the covariance is estimated from the 2048 realizations of NGC CMASS DR12 \textsc{Patchy} mocks catalogues. Additionally, we rescale all the estimated covariance terms by a 10\% factor based on the ratio of particles, as the \textsc{Patchy} mocks have $10\%$ fewer particles than the Nseries mocks due to veto effects on the DR12 CMASS data. The underlying cosmology of the Nseries mocks is close to the WMAP one and they sample a total effective volume of $84\times3.67\, [{\rm Gpc}]^3$. Further information and details on the Nseries mocks can be found in section 7.2 of \cite{eboss_collaboration_dr16}, as well as section 2.2.2 of \cite{Gil-Marin-ml-2020bct}. 

Each of the Nseries mocks  cover a larger physical volume than the  dark matter simulation, so we have embedded them in  a box of  $4\,{\rm Gpc}h^{-1}$ side length. We use the  same $k$-binning and data-vector entries as in  section~\ref{sec:DM}, but the larger box-size implies that the mode-sampling is much denser; this is expected to reduce the geometric effect seen in section~ \ref{sec:DM}.

The left panel of figure~\ref{fig:sys2_m} displays the posteriors resulting from fitting the mean power spectrum multipoles from the 84 Nseries mock realizations. 
The dashed-empty contours show the posteriors drawn when fitting the data without any geometric-correction factor applied, whereas the solid-filled contours are when the $G$-factor is applied (see eq.~\eqref{eq:Gcorrection} for reference). The different color schemes represent different types of fit or bias schemes, as labeled. The local-Lagrangian galaxy bias conditions of eq.~\eqref{eq:lagrangian_prediction} are applied to the classic RSD fit (purple contours) and to the {\it ShapeFit} (orange contours). Additionally, we also show the {\it ShapeFit} when the locality in Lagrangian space is relaxed (green contours). While only the physical parameters are shown,  the nuisance parameters are included in the fit and  marginalized over. For the {\it ShapeFit} parametrization we have set $n=0$ for simplicity. The black-dashed lines indicate the expected values for the underlying cosmology of the Nseries mocks. 

As expected, the geometric correction described in section~\ref{sec:DM} has a much smaller effect for the Nseries mocks. 

This is due to a combination of two effects:  the statistical errors for Nseries are larger than for the dark matter simulations, because of the smaller effective volume, $308\,{\rm Gpc}^3$ for Nseries and $6,448\,{\rm Gpc}^3$ for dark matter particles boxes. Moreover the larger size of the Nseries box, $L=4\,{\rm Gpc}h^{-1}$, yield a finer sampling of $k$-modes than for the dark matter simulation, $L=2.4\,{\rm Gpc}h^{-1}$, reducing the net effect. 

We note that for the 3 studied cases, classic RSD with the local-Lagrangian bias assumption, and {\it ShapeFit} with and without the local-Lagrangian bias assumption, the expected parameters are recovered very well. We only detect a systematic shift on $m$ of order $0.01-0.02$, towards negative values, similar to the one reported in section~\ref{sec:DM}. Note that  for the  Nseries mocks the recovered non-local bias parameters are very consistent with the local-Lagrangian prediction.

Additionally, letting  $n$ and $m$ to be simultaneously free for the local-Lagrangian bias case, yields   results  very consistent with the $n=0$ case (orange contours), and therefore with the expected value (see appendix \ref{sec:Shapefitdependencies}).

The right panel of figure~\ref{fig:sys2_m} illustrates the effective volume-effect for the {\it ShapeFit} case with the non-local bias set to their local-Lagrangian prediction. The orange contour displays the fit to the mean of the 84 Nseries mocks (same as in the left panel), whose associated effective volume is $308\,{\rm Gpc}^3$. The blue contours are  the resulting posterior from averaging the individual 84 posteriors, each of them with an associated effective volume of $3.67\,{\rm Gpc}^3$. For most of the parameters of interest the maxima of the posterior of the mean (orange) lies in the same position as the maxima of the mean of individual posteriors (blue). We observe a small displacement for $f$ and $m$, indicating some non-Gaussian behaviours on the tails of the distribution. We also note that the size of the systematic reported errors on the left plot, of order $\Delta_m^{\rm sys}\simeq 0.01-0.02$, are very sub-dominant with respect to the statistical errorbars of $m$ associated to a real-life volume, of $3.67\,{\rm Gpc}^3$, in the right panel in blue.

We conclude that for the {\it ShapeFit} analysis on Nseries mocks, with or without the local-Lagrangian bias assumption, the errors associated to modelling systematics are negligible for $\alpha_{\parallel\,,\perp}$ and $f$, and of 0.01-0.02 on $m$ towards negative values. We have not identified the source of such systematic shift, but we conclude that it  does not have any significant impact when fitting actual datasets, whose statistical errorbars on $m$ tend to be of order of 5-10 times larger. We leave  a more detailed study of such systematic effect and its mitigation in the next-generation of galaxy surveys for future work.

\section{Conclusions}\label{sec:conclusions}
The standard (classic) approach to \cred{analyse} galaxy redshift clustering, (\colored{that we refer to as} BAO and RSD analyses), is  conceptually different from the way, for example, CMB data are interpreted.  With the help of a  fiducial template of the power spectrum,  the clustering  data are compressed  into few physical observables which are sensitive only to late-time physics, and it is these observables that are then interpreted in light of a cosmological model.  
There has been a \colored{ renewed} effort \colored{ recently} to  \cred{analyse} galaxy redshift clustering in a similar way as CMB data: by comparing directly the observed power spectrum, including the BAO signal, the RSD signal, as well as the full shape of the broadband power to the model's prediction. In this case,  the model has to be chosen {\it ab initio}. We refer to this approach as full modelling (FM). 
The resulting posterior constraints on cosmological parameters  of the $\Lambda$CDM model, or its simple extensions, are significantly tighter than in the classic  analysis \colored{in a broad parameter space with no Planck constraints}.

In this paper we have provided a  full physical understanding  of where the additional constraining power afforded by the second approach arises,  and in doing so we  have bridged the  classic and new analyses in a transparent way. 

The compressed physical variables of the classic approach represent the universe's late-time dynamics;   they depend only on the geometry, expansion history and growth rate of the Universe in a model-independent way and  they can be in turn interpreted in light of the cosmological model of choice. These variables do not capture and are insensitive to  other physics relevant to processes at play at a different epoch in the Universe evolution such as equality scale, sound horizon scale, primordial power spectrum or other quantities that enter in  the matter transfer function.

\cred{However,} there is additional information in the clustering signal\cred{. Beside} the primordial tilt,  the broadband of the power spectrum is shaped by the matter transfer function encoding the evolution (scale and time dependence) of the initial fluctuations from inflation until the time of decoupling of the photon-baryon fluid, which in a $\Lambda$CDM model, depends on the physical baryon and matter densities $\om, \ob$ and  $h$.  This, we have shown, is located mostly on large scales,  and to a smaller extent in the amplitude of the BAO wiggles.

In the FM approach the parameter dependence of the transfer function and the geometry are not kept separated, in this way the information carried  by the shape of the transfer function, improves constraints on cosmological parameters that are usually interpreted as purely geometrical. This can be seen as an ``{\it internal model prior}".  The classic fixed template methods do not invoke a prior of that kind, as they do not establish this link.

We have thus extended the classic analysis with a single additional phenomenological parameter, that   captures the bulk of this extra information (section~\ref{sec:shapefit}).
We refer to this approach as {\it ShapeFit}.
\colored{The physical understanding we provide is rooted on landmark works on the matter transfer function e.g., \cite{BBKS, BondSzalay83,EH_TransferFunction}. We are aware that a single parameter like $\Gamma \sim \Omega_mh$ is insufficient to correctly describe the data given the statistical power of state-of-the-art galaxy surveys \cite{EH_TransferFunction}}. {\it ShapeFit} \colored{ introduces instead an effective parameter, $m$, the slope of the matter power spectrum at a specific pivot scale. 
The {\it ShapeFit} extension of the classic \cred{methodology}} captures the same  \colored{broadband shape} information as the FM fit by upgrading the classic RSD fit \cred{and} at the same time retaining the power of compressing the two point statistics into well understood  and model independent physical numbers that still  disentangle early from late-time physics. The `{\it internal model prior}' is not needed until the very last step of interpreting the physical variable in light of a model. 

In summary\cred{,} {\it ShapeFit}:
\begin{itemize}
\item preserves the model-independent nature of the compressed physical variables of the classic approach,
\item disentangles early-time from late-time physical information,
\item matches the constraining power of the FM approach when interpreted  within the cosmological model parametrization of choice (see figure~\ref{fig:EFT-Shape-comparison-noIR}),
\item is a simple addition to the established  “classic" codes and procedures with a simple physical interpretation,
\item the computational time is effectively indistinguishable from that of the classic approach and $\sim30$ times faster than the FM approach (at the level of cosmological inference),
\item reduces the (already small) template-dependence of the classic approach.
\end{itemize}

In passing we have  presented (section~\ref{sec:shapefit-s8}) a new  definition and interpretation of the physical parameter describing amplitude of velocity fluctuations which further reduces the model-dependence of the traditional RSD analysis. We recommend adopting it in classic analyses even without {\it ShapeFit} extension.

{\it ShapeFit} does not include the additional information enclosed in the BAO amplitude for two reasons: {\it i)} this early-time  information is relevant only if no CMB or BBN prior  is adopted for $\ob$, {\it ii)}  even though the BAO amplitude  is an early-time physics imprint it is however heavily processed by late-time effects (e.g., non-linearities, mode-coupling, bias) and therefore, we argue,  not robust (see figure~\ref{fig:varying-ob}).

Given the level of sophistication and the systematic control of the classic approach,  we have performed a battery of tests on mock surveys to quantify possible subtle systematic effects for {\it ShapeFit}.
These are presented in section~\ref{sec:Nseries} and the appendices.  We have highlighted a few systematic effects  and proposed and tested mitigation strategies  well suited to present and forthcoming surveys. The take-home message is that the power spectrum broadband shape is very sensitive to bias assumptions,  even on large,  linear scales. Therefore,  we advocate to always allow maximal freedom for the bias and nuisance parameters in forthcoming data analyses, especially when  FM  and {\it ShapeFit} are  used for cosmological interpretation.

We envision that the transparent physical interpretation offered by  the simple extension of the classic approach proposed here will be useful in \cred{analysing} and interpreting the clustering signal of current and forthcoming surveys in a robust  way.  

\acknowledgments

\begin{sloppypar}
H.G-M. and S.B. acknowledges the support from ‘la Caixa’ Foundation (ID100010434) with code LCF/BQ/PI18/11630024.
L.V. acknowledge support of European Unions Horizon 2020 research and innovation programme ERC (BePreSySe, grant agreement 725327).  Funding for this work was partially provided by project PGC2018-098866- B-I00 MCIN/AEI/10.13039/501100011033 y FEDER “Una manera de hacer Europa”, and the “Center of Excellence Maria de Maeztu 2020-2023” award to the ICCUB (CEX2019- 000918- M funded by MCIN/AEI/10.13039/501100011033).
The massive production of all MultiDark-\textsc{Patchy} mocks for the BOSS Final Data Release has been performed at the BSC Marenostrum supercomputer, the Hydra cluster at the Instituto de Fısica Teorica UAM/CSIC, and NERSC at the Lawrence Berkeley National Laboratory. We acknowledge support from the Spanish MICINNs Consolider-Ingenio 2010 Programme under grant MultiDark CSD2009-00064, MINECO Centro de Excelencia Severo Ochoa Programme under grant SEV- 2012-0249, and grant AYA2014-60641-C2-1-P. The MultiDark-Patchy mocks was an effort led from the IFT UAM-CSIC by F. Prada’s group (C.-H. Chuang, S. Rodriguez-Torres and C. Scoccola) in collaboration with C. Zhao (Tsinghua U.), F.-S. Kitaura (AIP), A. Klypin (NMSU), G. Yepes (UAM), and the BOSS galaxy clustering working group.
\end{sloppypar}

\appendix

\section{Impact of IR resummation on FM results}\label{sec:IR}
\begin{figure}[t]
    \centering
    \includegraphics[width=\textwidth]{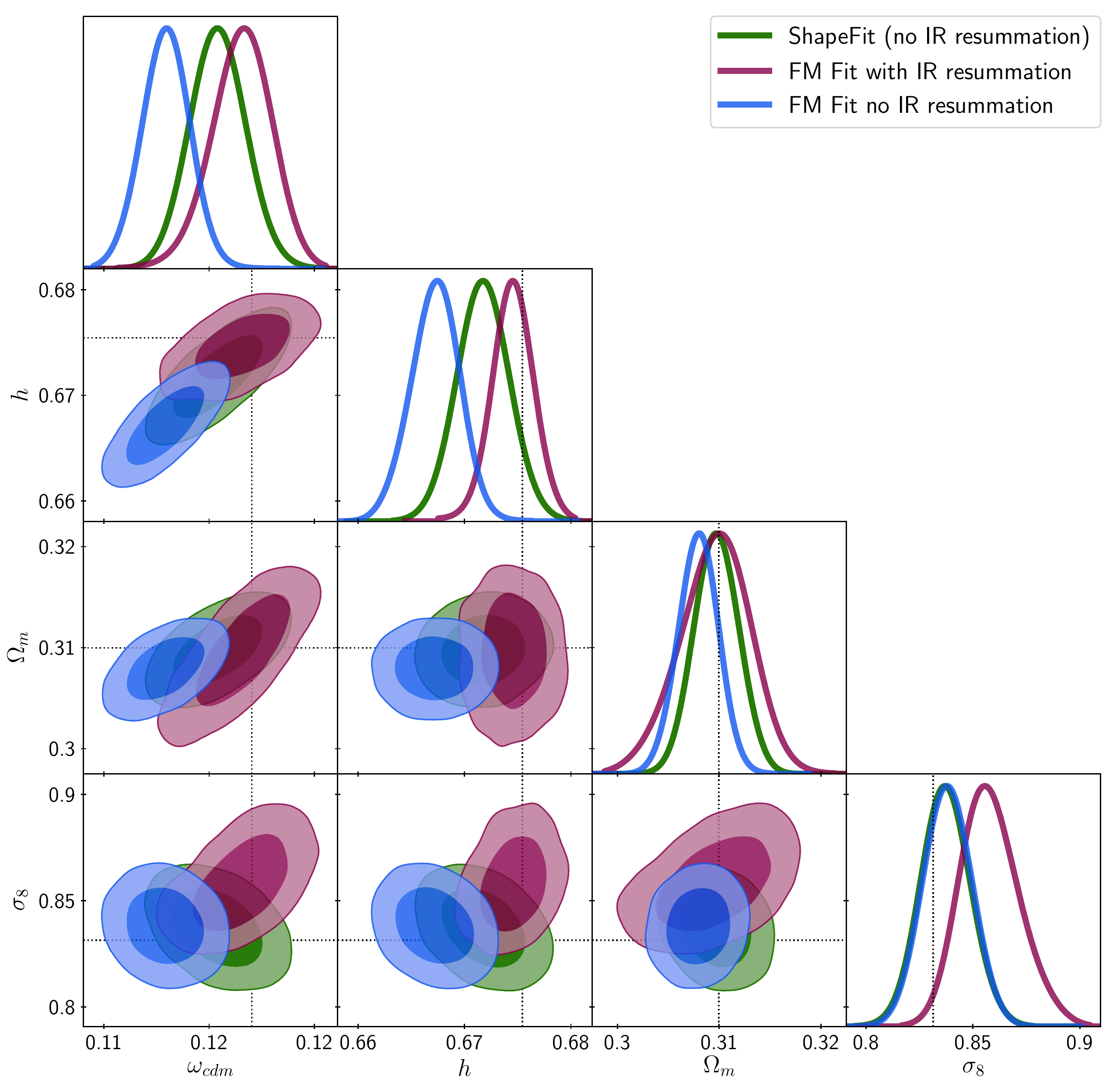}
\caption{Posteriors resulting from fitting the mean of 2048 \textsc{Patchy} ``ngc\_z3'' mocks with a covariance corresponding to 100 times the size of one single realization, and with an associated effective volume of $\sim300\,{\rm Gpc}^3$. For this case $\omega_b$ has been kept fixed to its true expected value (see figure~\ref{fig:varying-ob} for the case where $\omega_b$ is kept as a free parameter) and the ``max'' choice been made for the Lagrangian bias treatment (see section~\ref{sec:Practical-priors}). Blue and purple contours display the prediction from FM fit when IR resummation correction is ignored or accounted, respectively. The {\it ShapeFit} inferred contours are shown in green, where the BAO amplitude information is not used, and therefore the IR resummation correction has no effect.}
    \label{fig:EFT-Shape-comparison-withIR}
\end{figure}

The Infrared (IR) resummation effects are not included in the baseline analysis shown in the main text. This is motivated by the fact that  \textit{ShapeFit} does not include this effect in the PT model used (in this case 1-loop SPT). The effect of IR resummation in the FM fit is to damp the  amplitude of the BAO feature in the power spectrum model due to late-time physics effects (bulk flows), and therefore to broaden the likelihoods of those parameters sensitive to the amplitude of the BAO features. For this reason we opted for the non-IR option in FM fit as a baseline choice, for a fairer comparison between {\it ShapeFit} and FM fit likelihood's shapes.

However, since the IR resummation is an integral part of EFT power spectrum modelling, and therefore of FM fit, here we show how the likelihoods for these cases compare to each other when fitting the \textsc{Patchy} mocks, in the case of $\omega_b$ being fixed to its true value. Previously, in figure~\ref{fig:varying-ob} we have already shown how they compare for the case of a uniform and wide prior on $\omega_b$.

Figure~\ref{fig:EFT-Shape-comparison-withIR} is the analogous plot of figure~\ref{fig:EFT-Shape-comparison-noIR}, for the ``max'' case only (see text in section~\ref{sec:Practical-priors} for a full description of this choice), keeping the same color-code: blue for FM fit without IR resummation correction and green for {\it ShapeFit}. In purple we show the contours for FM fit with the IR resummation correction included. We remind the reader that {\it ShapeFit} does not use any BAO amplitude information, and therefore is, by construction, insensitive to IR resummation effects in $P(k)$.

Figure~\ref{fig:EFT-Shape-comparison-withIR} shows how the inclusion of IR resummation correction help the contours of FM fit to shift towards the expected  parameter values, at the expenses of broadening the contours. We stress that this shift only happens when the $\omega_b$ parameter is anchored to its true value. In the case $\omega_b$ is set to be free, the IR resummation correction does not produce the required shift towards the correct position, and only broadens the contours (see figure~\ref{fig:varying-ob}). The {\it ShapeFit} posteriors are naturally unaffected by ignoring the IR resummation correction and recover  the expected  parameters' values of the \textsc{Patchy} mocks cosmology. 

In real-life applications (i.e., for effective survey volumes of $\lesssim100\,{\rm Gpc}^3$) the IR resummation correction does not have a significant impact in the derived cosmological parameters. This is because the BAO amplitude information is dominated by other probes different to LSS, such as CMB- or BBN-based analyses, and getting it right from the LSS does not add any significant information to the combined analysis. Also, for those studies doing an integral LSS-alone analysis, the amplitude of BAO is not yet a reliable feature we should be trusting. The reason is that the BAO damping is highly model dependent and involves non-linear physics (including galaxy formation) that we do not understand 
at the level required for precision cosmology today. A practical proof of that are the contours of figure~\ref{fig:varying-ob}, where when LSS data is \cred{analysed} alone without any strong $\omega_b$ prior, even the IR resummation approach returns biased likelihoods for $\{h,\omega_b,\omega_{cdm}\}$.

\section{Investigating the \textit{ShapeFit} template dependence}
\label{sec:appendixtempldep}

 \begin{table}[h]
    \centering
    \begin{tabular}{c||c|c|c|c|c|c|c||c|c}
        Cosmology & $\om$ & $\ob$ & $h$ & $\sigma_8$ & $n_s$ & $M_\nu\,[\mathrm{eV}]$ & $N_\mathrm{eff}$ & $\Om$ & $r_{\rm d} \,[\mathrm{Mpc}]$ \\ \hline
       Planck & 0.1417 & 0.022  & 0.676 & 0.8288 & 0.9611 & 0.06 & 3.046 & 0.31  & 147.78\\
       \textsc{Patchy} & 0.1411 & 0.022  & 0.678 & 0.8288 & 0.9611 & 0.0 & 3.046 & 0.307  & 147.64\\
       Nseries & 0.1401 & 0.023  & 0.700 & 0.82 & 0.96 & 0.06 & 3.046 & 0.286  & 147.15\\
       X & 0.1599 & 0.022  & 0.676 & 0.814 & 0.97 & 0.056 & 3.046 & 0.35  & 143.17\\
       Y & 0.1599 & 0.022  & 0.676 & 0.814 & 0.97 & 0.056 & 4.046 & 0.35  & 138.77\\
       Z & 0.2053 & 0.037  & 0.75 & 0.9484 & 0.96 & 0.0 & 3.046 & 0.365  & 123.97\\
       Om-high & 0.1417 & 0.022  & 0.595 & 0.7349 & 0.97 & 0.0 & 4.046 & 0.4  & 142.85\\
       Om-low & 0.1417 & 0.022  & 0.913 & 0.7983 & 0.97 & 0.0 & 4.046 & 0.17  & 142.85\\
    \end{tabular}
    \caption{List of cosmological models for reference template potential dependence.}
    \label{tab:cosmoparams-templates}
\end{table}

Both the classic RSD Fit and the \textit{ShapeFit} are template-based fitting methods: they measure physical parameters related to late-time dynamics of the universe given a fixed template set by early-time physics. Keeping the template fixed and only varying it according to late-time effects is an effective way to decouple the early-time dependence of cosmological parameters from the late-time observations. This degree of model-independence goes at the expense of introducing a certain ``modelling systematic", coming from the fact that the template used for the analysis may not correspond to the underlying linear matter power spectrum of the universe.\footnote{As a side-note, this modelling systematic given by the template dependence does not exist in the FM approach, where the template is varied consistently at each step given the model, by definition. However, avoiding this systematic goes at the expense of model dependence, as internal model priors need to be adopted.} It is therefore important to quantify this systematic by studying the impact of different templates on physical parameter results and this is what this appendix is dedicated to. 

There are two questions we would like to address:
\begin{itemize}
    \item How do the classic RSD Fit and the \textit{ShapeFit} compare in terms of template independence for results on the traditional parameters $\alpha_\parallel, \alpha_\perp$ and $f\sigma_8$.
    \item What is the degree of template-dependence for the new  \textit{ShapeFit} parameter $m$.
\end{itemize}

\begin{figure}[t]
    \centering
    \includegraphics[width=\textwidth]{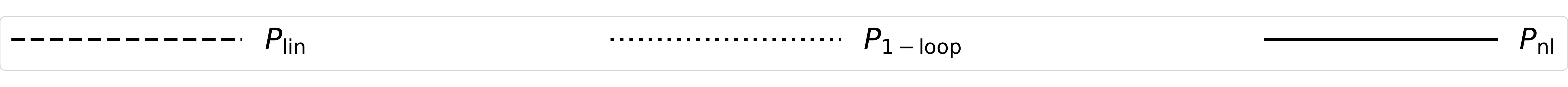}
    \includegraphics[width=\textwidth]{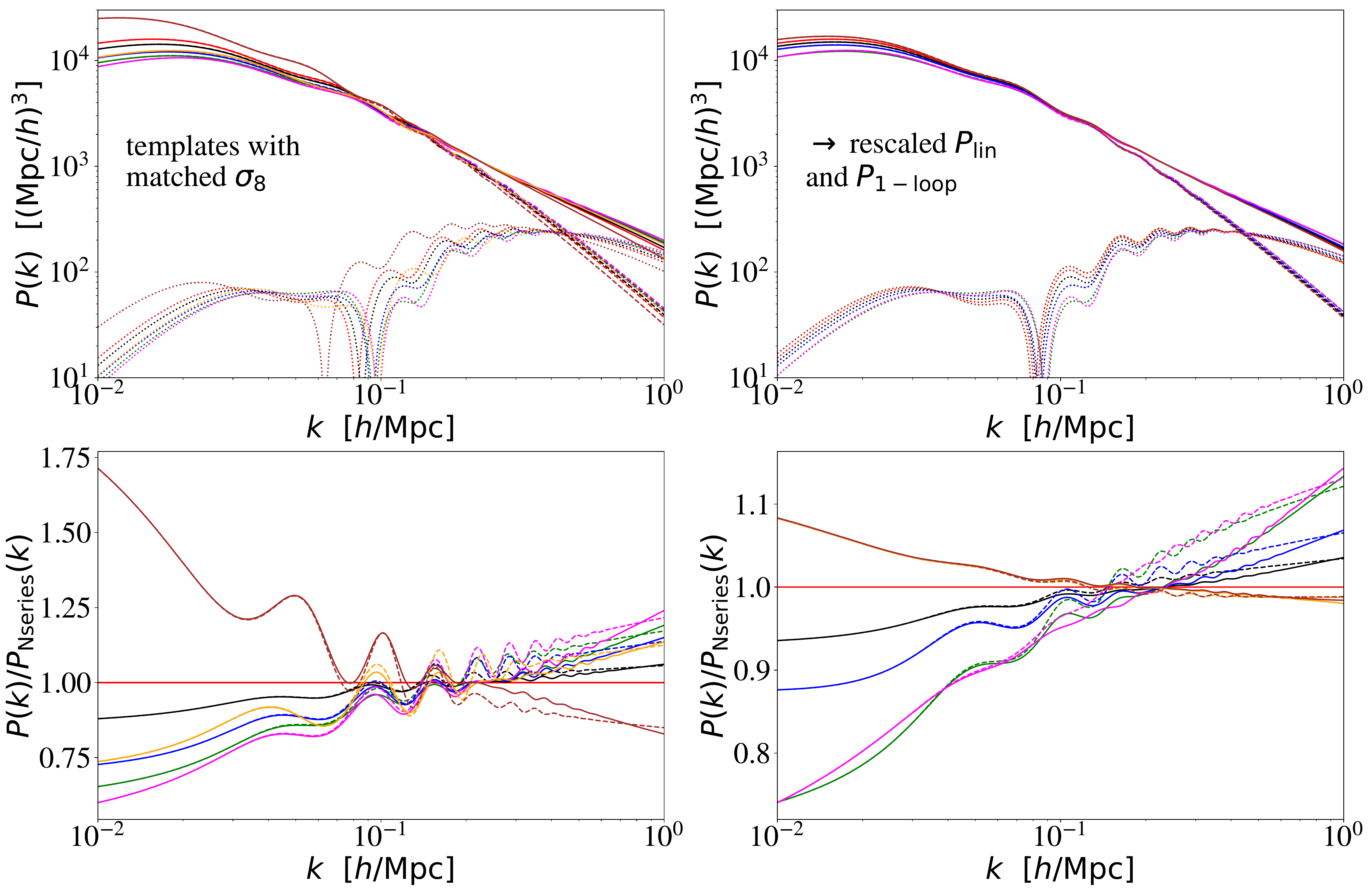}
    \includegraphics[width=\textwidth]{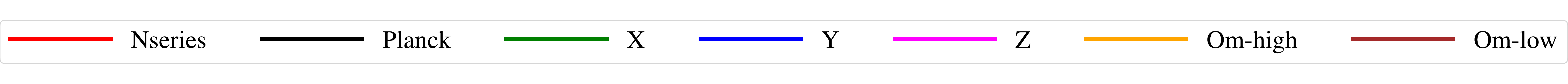}
 \caption{Power spectrum templates corresponding to the cosmologies of table \ref{tab:cosmoparams-templates} (without ``Patchy" cosmology, as it is very similar to ``Planck"). In the top row, we show the linear power spectra in dashed, the 1-loop corrections in dotted and their sum, the non-linear power spectra, in solid lines. The bottom row shows the power spectrum ratios with respect to the ``Nseries" cosmology for either $P_\mathrm{lin}$ or $P_\mathrm{nl}$. Left panels show the spectra rescaled in amplitude by $\sigma_8$ for better visibility and for the right panels we applied the early-time rescaling (by the sound horizon ratio) to the spectra, such that their BAO positions match.}
    \label{fig:templates}
\end{figure}

\begin{figure}[t]
    \centering
    \includegraphics[width=\textwidth]{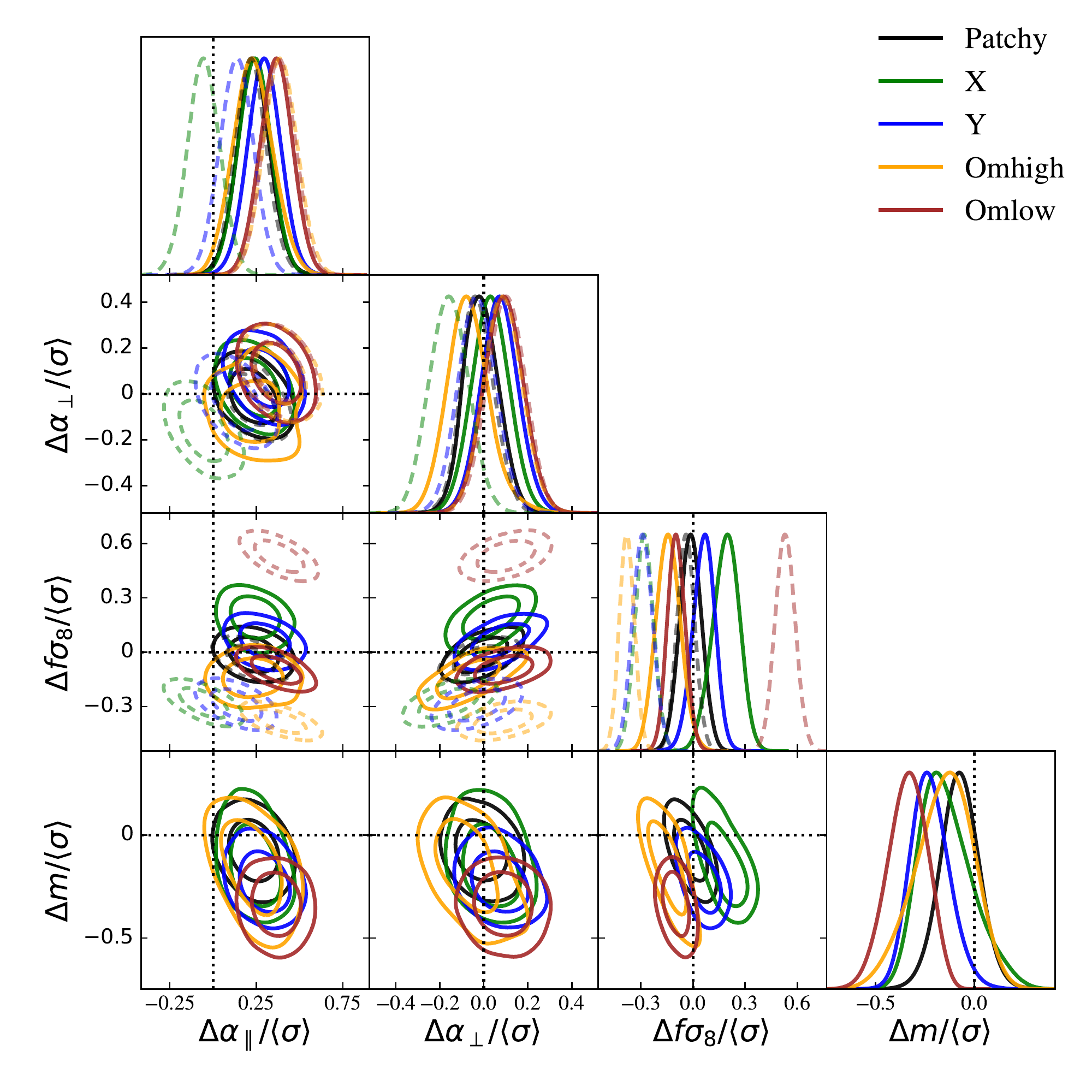}
 \caption{Comparison between templates used for the classic RSD Fit (dashed contours) and for the \textit{ShapeFit} (solid contours) on the mean of 2048 \textsc{Patchy} mocks with covariance for the volume of 100 realizations. We show posterior results for the physical parameters subtracted by the expectation, relative to the error normalized to the volume of a single \textsc{Patchy} mock realization, $\sim3\,{\rm Gpc}^3$.
 For all cases the non-local bias parameters are varied corresponding to the ``max" case (see section~\ref{sec:Practical-priors} for details)}
 
    \label{fig:triangle-template-dependence}
\end{figure}

To answer these questions we perform the classic RSD Fit and the \textit{ShapeFit} for a set of 8 different template cosmologies presented in table~\ref{tab:cosmoparams-templates}. The ``Planck" and ``\textsc{Patchy}" cosmologies are very similar (close to the cosmology preferred by Planck analysis) and have been introduced already in the main paper,  as the ``Nseries" parameters corresponding to the WMAP cosmology.
We also use the ``X", ``Y", and ``Z" templates, that correspond to $\om$-values extremely different from the ``Planck" reference, and  a different value of the effective number of neutrino species $N_\mathrm{eff}$ in the ``Y" case. All these templates have also been used to study the template dependence of eBOSS results in \cite{Gil-Marin-ml-2020bct}. In addition, we use templates generated from the ``Om-high" and ``Om-low" cosmology, that share the same value of $\om$ as ``Planck", but extremely different Hubble parameters $h$, leading to a very high and a very low value of $\Om$ respectively.

The linear power spectrum templates as well as the 1-loop corrections and the full non-linear templates are also shown in figure~\ref{fig:templates} for all cosmologies except for the ``Patchy" cosmology, as it is very similar to ``Planck". From the left panels one can see that the templates show deviations of up to 50\% on large and 20\% on small scales. After rescaling them via the early-time scaling given in eq.~\eqref{eq:Theory_Pmodel_rescaling}  to match the BAO positions (right panels), the deviations reduce to 25\% and 10\%, respectively. One can appreciate, even by eye, that the remaining disagreement between the templates after rescaling is well described by a slope. This is precisely the additional degree of freedom \cred{that} \textit{ShapeFit} delivers via the parameter $m$, which is missing in the classic RSD fit.

\begin{figure}[t]
    \centering
    \includegraphics[width=\textwidth]{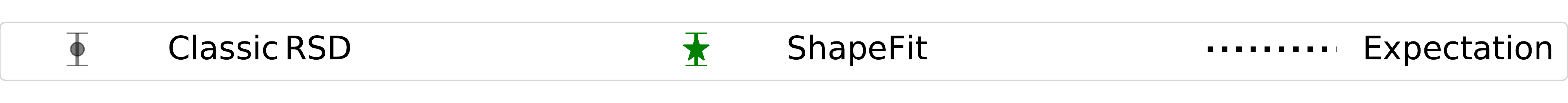}
    \includegraphics[width=\textwidth]{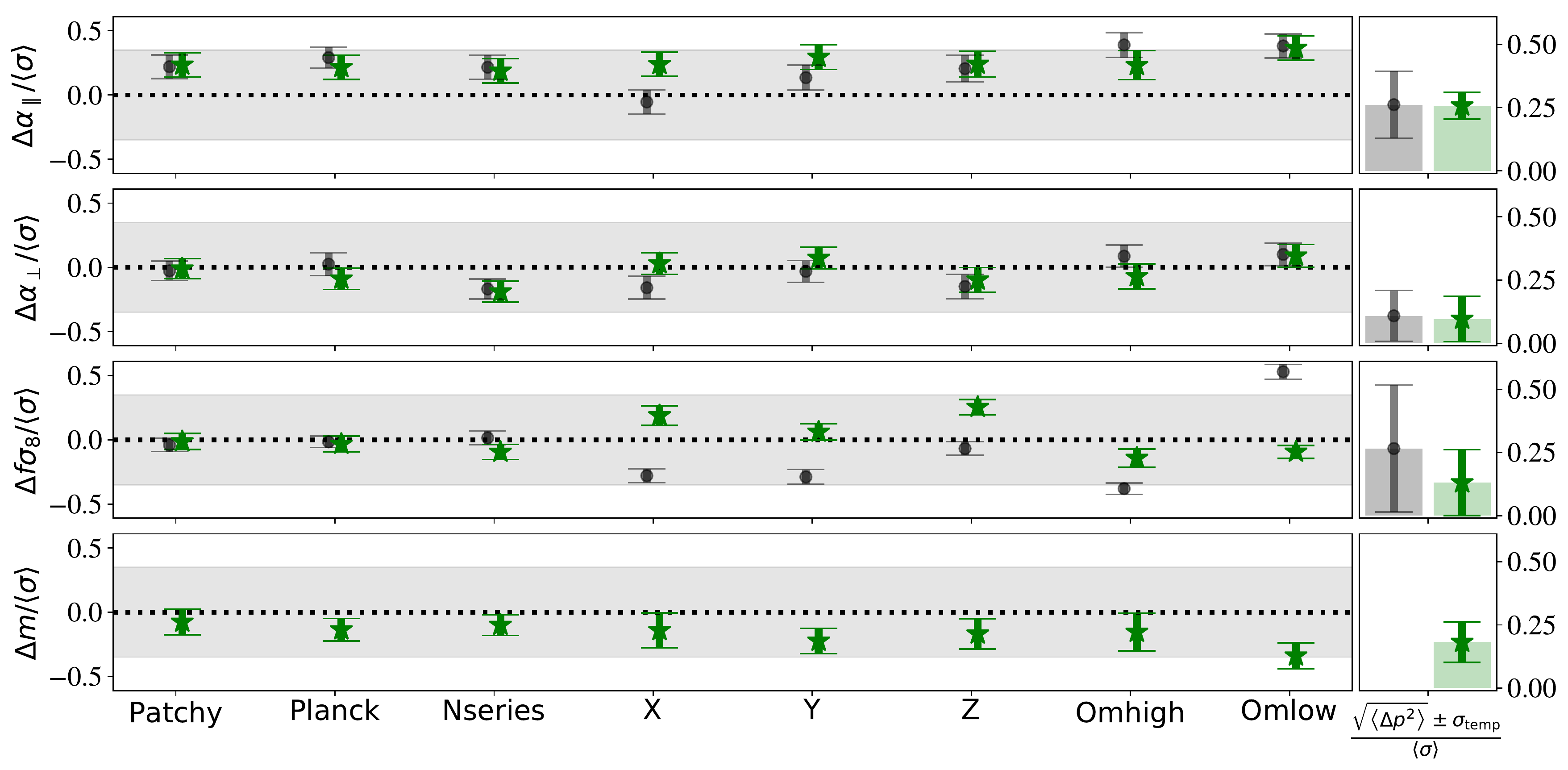}
 \caption{Results of compressed variables for all templates (see table~\ref{tab:cosmoparams-templates}) used for the classic RSD Fit and the \textit{ShapeFit}, in both cases allowing non-local bias parameters to vary (``max" case in table \ref{tab:priors}). For each template we show \colored{the} deviation \colored{$\Delta p$} from the expectation of \colored{$p$} for \colored{$\Delta p \in \left\lbrace \Delta \alpha_\parallel, \Delta \alpha_\perp, \Delta f\sigma_8, \Delta m \right\rbrace$} divided by the corresponding errors $\left<\sigma\right>$ of each parameter when fitting the BOSS ``ngc\_z3" \cred{sample} using the covariance of a single realization \colored{corresponding to a volume of $\sim 3\, {\rm Gpc}^3$}. \colored{The classic RSD fit barely exceeds $0.5 \left<\sigma\right>$ while {\it ShapeFit} is bound below $0.35 \left<\sigma\right>$ deviations indicated by the grey band.} \colored{The last column shows, again in units of $\left<\sigma\right>$, the template-averaged deviation from the truth $\sqrt{\left<\Delta p^2\right>}$ (represented by vertical histogram-bars) over-plotted with the intrinsic scatter among all templates $\sigma_\mathrm{p,temp}$ (represented by error bars) for each parameter $p$. It is important to note that the error bars do {\it not} indicate the error on the error, but the statistical spread, while the histogram bar represent the cumulative systematic bias. Hence, in case there is no systematic deviation, the error bar size is equal to the histogram height, (as it is the case for $\Delta \alpha_\perp$ and $\Delta f\sigma_8$.) }}
    \label{fig:means-template-dependence}
\end{figure}

Figure~\ref{fig:triangle-template-dependence} shows the posteriors for the northern ``ngc\_z3'' \textsc{Patchy} mocks  \cred{analysed} using {\it ShapeFit} (solid contours) and the classic RSD (dashed contours) using some of the different templates (displayed in different colours) listed in table~\ref{tab:cosmoparams-templates} as the reference cosmology. 
In all cases we set $n=0$ and we allow the non-local galaxy bias parameters to vary freely.
For each case the data-vector has been constructed from the mean monopole and quadrupole signals of the 2048 realizations, and the associated covariance correspond to the volume of 100 \textsc{Patchy} mocks. Each physical parameter, $p$, is displayed with its expected value, $p^{\rm exp}$ (different for each reference template) subtracted in such a way that the expected value coincides with 0, $\Delta p = p-p^{\rm exp}$. Additionally each $\Delta p$ is divided by the statistical error corresponding to one single realization of these mocks. \colored{As already discussed above, the choice of reporting the results for an effective volume of $\sim 3 {\rm Gpc}^3$ is motivated by the fact that \cred{the \textsc{Patchy} mocks' accuracy} in reproducing the observed clustering properties is not guaranteed much beyond the limit afforded by the statistics of a single realization. Moreover we follow the procedure for template sensitivity presented in  \cite{Gil-Marin-ml-2020bct}.} Using {\it ShapeFit} over the classic RSD method helps to bring the measured value of $\alpha_\parallel$ and $f\sigma_8$ close to the expected value ($\Delta p=0)$, removing a weak systematic residual associated with the reference template choice, which is present for the classic RSD analysis.  $\alpha_\perp$ is unbiased for both classic RSD and {\it ShapeFit}. 
The new shape parameter $m$ does not show any significant bias neither. The deviation with respect to the expectation remains well below one half of the statistical error-bars expected for a volume of about $3\,{\rm Gpc}^3$ even for the extreme case of the templates `X', `Y' and 'Z'.

Figure~\ref{fig:means-template-dependence} presents the displacement of the same physical variables shown in figure~\ref{fig:triangle-template-dependence} in 1-dimensional panels, for the additional cosmologies of table~\ref{tab:cosmoparams-templates}, `Planck', 'Nseries' and `Z'.

\colored{In addition, the last column provides a ``summary statistic" of all templates to facilitate evaluating the overall {\it ShapeFit} performance. The colored bars represent the ``least squared" deviation $\sqrt{\left<\Delta p^2\right>}$ from the truth, where we averaged over all $N^\mathrm{temp}$ templates 
\begin{align}
    \left<\Delta p^2\right> = \frac{\sum_i^{N^\mathrm{temp}} \Delta p_i^2/\sigma_{p_i}^2 }{\sum_i^{N^\mathrm{temp}} 1/\sigma_{p_i}^2}~,
\end{align}
and the error bars show the intrinsic scatter of the bestfit values with template $\sigma_\mathrm{temp}$. We can see that the overall effect of {\it ShapeFit} on $\sqrt{\left<\Delta p^2\right>}$ is rather mild for $\alpha_\parallel$ and $\alpha_\perp$, but very strong (factor 2 improvement) for $f\sigma_8$. On the other hand, {\it ShapeFit} shows significant improvement concerning the scatter $\sigma_\mathrm{temp}$ for $\alpha_\parallel$ and $f\sigma_8$, but not for $\alpha_\perp$.}
\colored{Ideally the cumulative systematic bias (histogram-bars) should not be larger than the statistical scatter (error-bars). Clearly for $\alpha_\parallel$ this is not the case, but is also not a problem with the adopted compression.  We interpret it as an indication of insufficient accuracy in the  adopted theoretical  \cred{modelling} at the level  below or $\sim 0.25$ of one standard deviation. We can thus roughly estimate that, if the \cred{modelling} is not improved,  such bias may become a concern for volumes of the order of $\sim 40$ Gpc$^3$. }
\colored{Considering the individual templates} we observe a similar trend as in figure~\ref{fig:triangle-template-dependence}\cred{:} \colored{the visible bias of the classic RSD approach for $f\sigma_8$ in the X and Y cosmologies as well as the Om-high and Om-low cases, has puzzled and worried the experts for a while. } {\it ShapeFit} helps to reduce the already small template dependence of the classic RSD analysis, even for  extreme cosmologies \colored{(such as X, and Y)}, completely ruled out by CMB observations; 
even the parameter $m$ does not show a significant template dependence, although we note that the error on $m$ may increase for an inappropriate choice of template. We argue however, that this feature does not have a significant impact on future data analysis, as the extreme templates studied here are used for testing purposes only.

\begin{figure}[h]
    \centering
    \includegraphics[width=\textwidth]{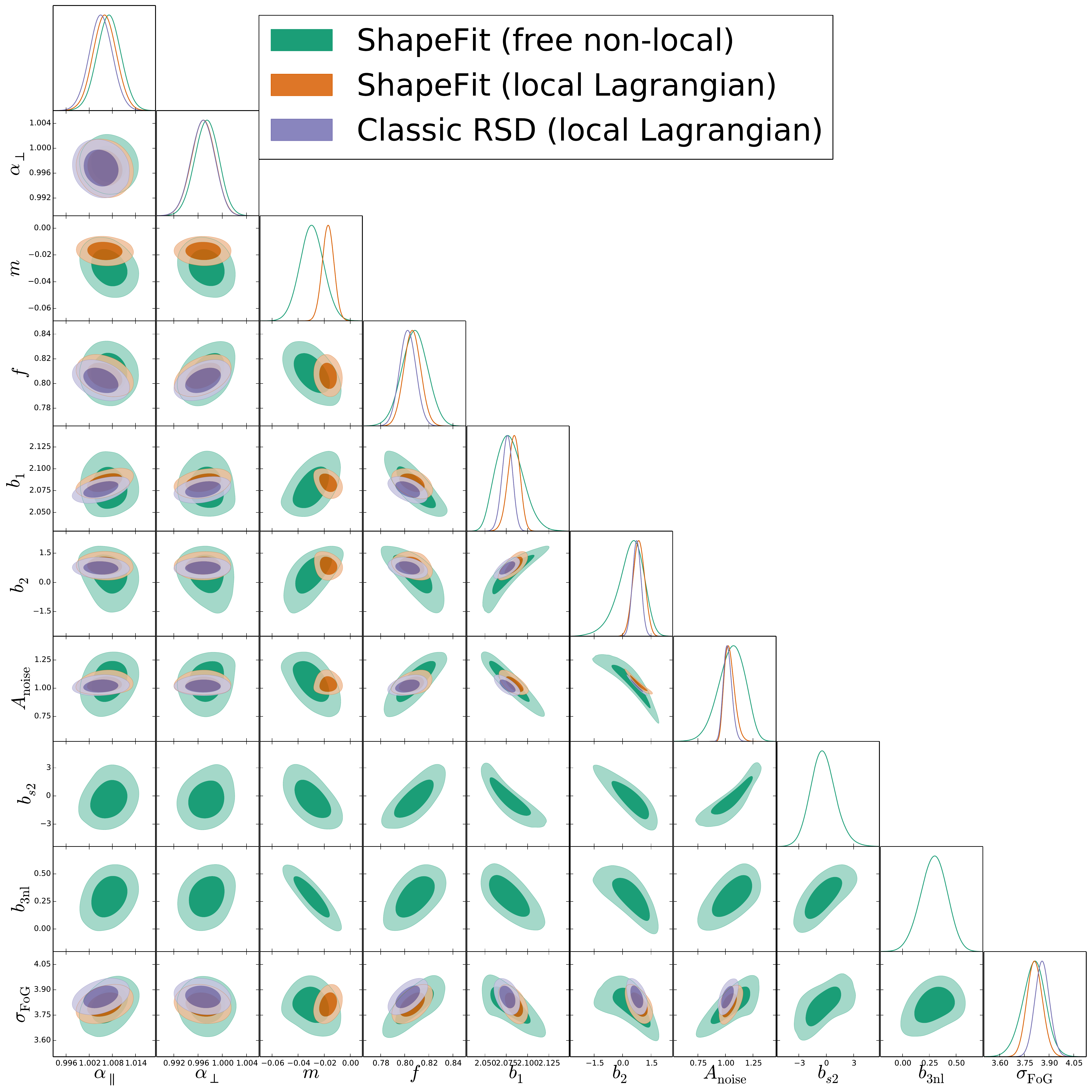}
    \caption{Posterior distribution for the mean of the Nseries mocks corresponding to what is shown in the left panel of figure~\ref{fig:sys2_m}, but in this case explicitly displaying the dependencies on nuisance parameters, including the non-local biases. Note the strong correlation between the shape parameter $m$ and the non-local biases $b_{s2}$ and $b_{3{\rm nl}}$.}

    \label{fig:full_triangle_for_Nseries}
\end{figure}

\section{Full parameter-dependencies for \textit{ShapeFit}}\label{sec:Shapefitdependencies}

\begin{figure}[t]
    \centering
    \includegraphics[width=\textwidth]{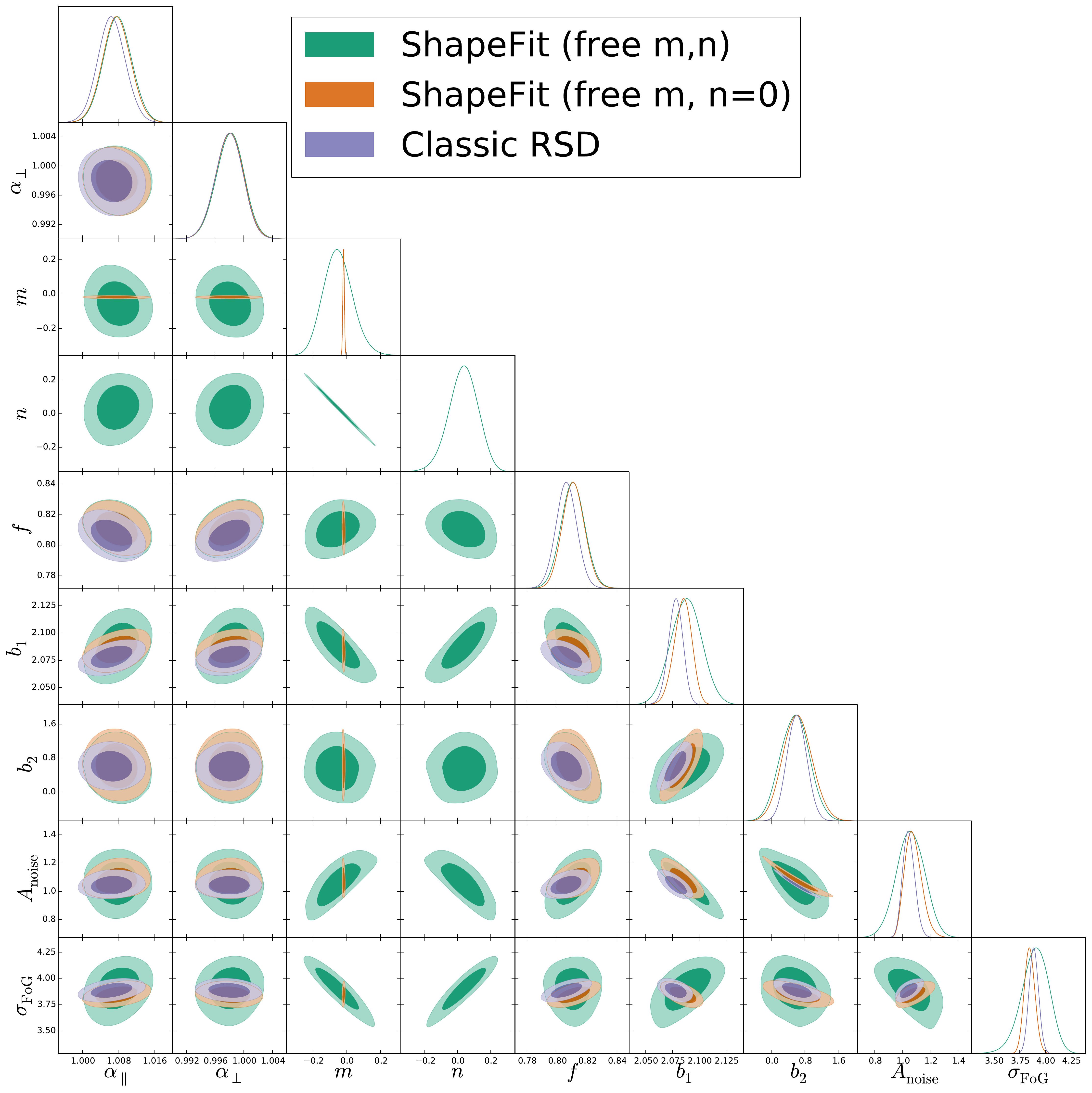}
    \caption{Posterior distribution for the mean of the Nseries mocks, corresponding to the ``min'' case shown in  figure~\ref{fig:sys2_m} (i.e., when locality in Lagrangian space is assumed), for {\it ShapeFit} with only $m$ varying (in orange contours), and when also $n$ is varied simultaneously to $m$ (green contours). For comparison, the classic RSD is also shown in purple contours. We note the strong correlation between $n$ and $m$ parameters.}
    \label{fig:full_triangle_for_Nseries_mn}
\end{figure}

For clarity, the main text did not show the full posteriors including both physical and nuisance parameters. It is however important to study possible correlations  between the shape parameter $m$ and the non-local biases: correlations between nuisance parameters and the physical parameters of interest may induce systematic biases in cosmological inference if the \cred{modelling} of nuisance effects is incorrect or if unsuitable priors are imposed on the nuisance parameters.  This is studied  in this appendix. 

In figure~\ref{fig:full_triangle_for_Nseries} we show the full correlations in all fitted parameters, both physical and nuisance, for the Nseries case with $m$ free, for the cases of local-Lagrangian (``min''case in orange contours) and free non-local biases (``max'' case in green). For comparison we also show in purple the classic RSD case for local-Lagrangian. 

On the other hand, in figure~\ref{fig:full_triangle_for_Nseries_mn} we show the dependencies for the ``min'' case when both  $m$ and $n$ are freely varied within {\it ShapeFit} (green contours). We note the high correlation between $m$ and $n$. This has to do with the intrinsic degeneracy between $\ob$, $\om$ and $n_s$ through the slope, which can only be broken by modelling the BAO amplitude and imposing a strong prior on $\ob$.

The full parameter degeneracies figures make a crucially important point (as already anticipated in section \ref{sec:Nseries}): 
 the  power spectrum  broadband shape, and hence the slope $m$, is very sensitive to bias assumptions, even on large, linear scales. Therefore, we advocate to always  allow  maximal freedom for the bias and nuisance parameters in forthcoming data analyses, especially for FM fits and when the slope $m$ is used for cosmological interpretation. Of course, this slows down  MCMC chains convergence, but \textit{ShapeFit} has an advantage over the FM fit, as in the former the fit only needs to be done once, while for the latter it has to be repeated for any model of choice.

\section{Impact of rescaling the non-linear template.}\label{sec:rescaling}

We investigate the approximation of  factorizing the parameter $m$ and $s$ outside the loop-integral corrections. This approximation is particularly  useful because it allows us  to pre-compute all loop-correction terms at a given reference cosmology, but varying the slope, $m$, and the BAO-template peak position, $s$, at each MCMC step.  Note that the approximation  involving $s$ has been extensively used in all the `classic' RSD methods using the `fixed-template' approach.

\begin{figure}
    \centering
    \includegraphics[width=\textwidth]{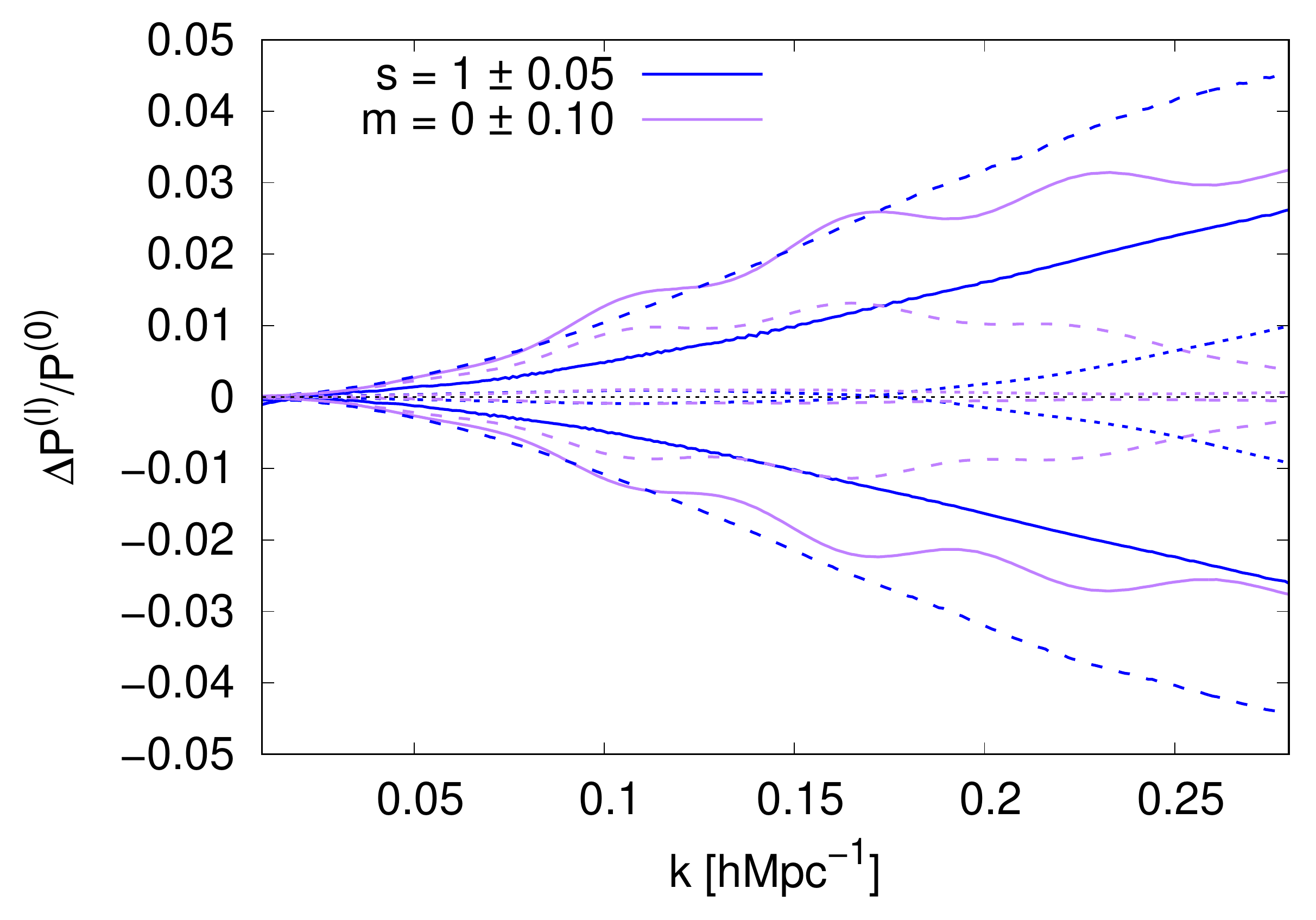}
    \caption{Systematic errors produced by rescaling the BAO-peak position $s$ and the shape parameter $m$ in the fixed-template implementation, $\Delta P^{(\ell)}\equiv P^{(\ell)}_{\rm rescaled}-P^{(\ell)}_{\rm exact}$, relative to the amplitude of the monopole, for reference. Blue lines display the effect of rescaling $s$ by $s=1\pm0.05$ and purple lines for $m$ by $m=0\pm0.1$. Solid, dashed and dotted lines display the effect for the monopole, quadrupole and hexadecapole, respectively. For the $k_{\rm max}=0.15\,h{\rm Mpc}^{-1}$ used in this paper, the systematic error stays always below 2\%.}
    \label{fig:rescaling}
\end{figure}

 Figure~\ref{fig:rescaling} displays the difference between the non-linear 1-loop SPT (taking into account the non-linear bias and TNS terms corrections) exact evaluation of the power spectrum multipoles
 and the corresponding rescaling
 of 
 a reference template evaluated at different values of $s$ ($\pm 0.05$, in blue) and $m$ ($\pm 0.1$, in purple), as it would be
 used in an actual MCMC run.
 Solid/dashed/dotted lines show the difference for the monopole/quadrupole/hexadecapole, relative to the amplitude of the monopole.  The rest of nuisance parameters have been set to values close to the best-fitting case for the \textsc{Patchy} and Nseries mocks. 
 Of course marginalizing over the nuisance parameters will absorb some of these differences (see below).

We see that for both $m$ and $s$ the approximation is better than 2\% for $k\leq 0.15\,h{\rm Mpc}^{-1}$, and  3\% for $k\leq 0.20\,h{\rm Mpc}^{-1}$. These are actually comparable to  the absolute typical errors of the model adopted in this paper (1-loop-SPT, TNS model, 1-loop bias corrections). The errors made by factorizing $m$ outside the loop integrals are of the same order as  those introduced by  the scaling of the BAO-template peak position, for shifts of $m\pm0.1$ and $s\pm0.05$, respectively. These small systematics errors are partially absorbed by nuisance parameters, such as $b_2$ and $\sigma_{P}$, and not affecting in any significant way the cosmological parameters inference, as it can be seen from appendix~\ref{sec:appendixtempldep}.

We conclude that the `fixed-template' implementation is a valid approach for both $s$- and $m$-rescaling, and produces systematic errors well within the current systematic error budget, as they are  of the order of systematic errors associated to the theory model itself. 
\colored{Should  the  maximum $k$  be  pushed so that  more non-linear scales are (reliably) included and constraints  shrink significantly compared to the cases considered in this paper,  this approximation may need to be improved. We leave  to future work how to do this without representing a computational bottleneck}.

\end{document}